\pdfoutput=1
\pdftrailerid{}
\documentclass[letterpaper,twocolumn,10pt]{article}
\usepackage{usenix}

\usepackage{relsize}
\usepackage{fontawesome5}
\usepackage[most]{tcolorbox}
\usepackage{amsmath,amssymb,amsfonts}
\usepackage{graphicx}
\usepackage{textcomp}
\usepackage{enumitem}
\usepackage{tikz}
\usetikzlibrary{arrows.meta, shadows.blur, positioning, fit, calc, backgrounds, shapes.geometric, decorations.pathreplacing}
\newcommand{\auditSchemes}{7}%
\newcommand{\auditRuns}{3}%
\newcommand{\auditRatings}{63}%
\newcommand{\auditUnanimous}{56}%
\newcommand{\auditUnanimousPct}{88.9\%}%
\newcommand{\auditDisagree}{7}%
\newcommand{\auditIndAgree}{195}%
\newcommand{\auditIndTotal}{231}%
\newcommand{\auditIndAgreePct}{84.4\%}%
\newcommand{\auditMatch}{39}%
\newcommand{\auditMatchPct}{61.9\%}%
\newcommand{\auditCodeBetter}{21}%
\newcommand{\auditCodeWorse}{3}%
\newcommand{\auditBetterNoExec}{10}%
\newcommand{\auditSThreeFull}{6}%
\newcommand{\auditSchemesTotal}{15}%
\newcommand{\auditNoCode}{7}%
\newcommand{\auditUnsandboxed}{2}%
\newcommand{\auditPersistStore}{4}%
\newcommand{\auditPickle}{2}%
\newcommand{\auditMissingComponent}{4}%
\newcommand{\auditMatchStrict}{31}%
\newcommand{\auditIndPerScheme}{33}%

\newcommand{\vtwoNoneRobustCells}{39}
\newcommand{\vtwoRobustCells}{60}
\newcommand{\vtwoFullSecurityCells}{9}
\newcommand{\vtwoSecurityCells}{75}
\newcommand{\vtwoNoneToPartial}{10}
\newcommand{\vtwoNoneToFull}{3}
\newcommand{\vtwoPartialToFull}{8}
\newcommand{\vtwoPartialToNone}{3}
\newcommand{\vtwoNewsSchemes}{14}
\newcommand{\vtwoPriceSchemes}{15}
\newcommand{\vtwoFilingSchemes}{12}
\newcommand{\vtwoGptFourFamily}{12}
\newcommand{\vtwoMultiAgentSchemes}{10}

\newcommand{\vtwoOpusCriticalFanout}{10}
\newcommand{\vtwoOpusOutbreakTwenty}{85\%}

\newcommand{\vtwoGeminiOutbreakTen}{97\%}
\newcommand{\vtwoGeminiCriticalFanout}{4}

\newcommand{\probeCalls}{324}
\newcommand{\lunaSellNews}{77\%}
\newcommand{\dsSellNews}{100\%}
\newcommand{\lunaTradeNoNews}{0\%}
\newcommand{\dsSellDropUnder}{89\%}

\newcommand{\gemSellNews}{100\%}
\newcommand{\gemBuyDipMax}{67\%}
\newcommand{\gemSellDropUnderMax}{33\%}

\newcommand{\dsBuySize}{180}
\newcommand{\gemBuySize}{109}
\newcommand{\crashFakeHundred}{2\%}
\newcommand{\fakeCrashGap}{8}
\newcommand{\crashFakeOneFifty}{30\%}
\newcommand{\crashFakeTwoHundred}{92\%}
\newcommand{\crashSellTwoHundred}{32\%}
\newcommand{\crashSellThreeHundred}{94\%}
\newcommand{\severeSellEightHundred}{98\%}

\newcommand{\pnlFakeEightHundred}{10.5\%}
\newcommand{\pnlFakeFourHundred}{5.2\%}

\newcommand{\ddSellBase}{2.6\%}
\newcommand{\ddFakeBase}{0.1\%}

\newcommand{\crashSellGemSixHundred}{58\%}
\newcommand{\crashSellGemEightHundred}{92\%}

\newcommand{\ddOutBase}{9.2\%}
\newcommand{\ddOutDsTwoHundred}{13.8\%}
\newcommand{\ddOutDsFourHundred}{17.3\%}
\newcommand{\ddOutDsEightHundred}{23.7\%}
\newcommand{\ddOutGemEightHundred}{11.6\%}

\newcommand{\crashSeeds}{50}

\newcommand{\sensOutSixty}{35.9\%}
\newcommand{\sensOutTwoForty}{17.3\%}
\newcommand{\sensOutSixtyCtl}{9.2\%}

\newcommand{\holdOutDsFourHundred}{6.9\%}
\newcommand{\soldLowDsFourHundred}{66\%}

\newcommand{\crashFakeLimits}{6\%}

\newcommand{\soldLowFakeFourHundred}{63\%}

\newcommand{\robProbeCases}{27}
\newcommand{\robSellMinDs}{25}
\newcommand{\robSellMinLuna}{20}
\newcommand{\robTraderNoNewsSellMax}{5}
\newcommand{\robTraderNoNewsSellMin}{4}

\newcommand{\resDsSellLo}{40\%}
\newcommand{\resDsSellHi}{100\%}

\newcommand{\resDsOutDdLo}{15.8\%}
\newcommand{\resDsOutDdHi}{18.1\%}

\newcommand{\resGemSellLo}{0\%}
\newcommand{\resGemSellHi}{100\%}

\newcommand{\resPolicies}{30}

\newcommand{\robOutAddMin}{3.5}

\newcommand{\robGapMax}{1.1}
\newcommand{\robNoThinSell}{4.3\%}

\newcommand{\robDeepSell}{2.2\%}

\newcommand{\luldFakeRuns}{300}
\newcommand{\luldFakePausedRuns}{0}

\newcommand{\luldFakeDd}{8.2\%}

\newcommand{\luldFakePnl}{4.8\%}

\newcommand{\luldOutBase}{9.2\%}
\newcommand{\luldOutBasePauses}{2}
\newcommand{\luldOutDs}{12.9\%}

\newcommand{\luldOutDsCtl}{9.2\%}

\newcommand{\calmGemVolNone}{9.8\%}
\newcommand{\calmGemHaltsThree}{2.4}
\newcommand{\calmGemVolThree}{4.1\%}
\newcommand{\calmGemHaltsFive}{1.5}

\newcommand{\calmDsHaltsThree}{0.6}
\newcommand{\calmDsHaltsFive}{0.2}
\newcommand{\calmGemVolCapSix}{9.8\%}
\newcommand{\fiveSigmaGapMax}{1.5}

\newcommand{\capSixDiffMax}{0.1}

\newcommand{\personaProbeCalls}{1{,}944}
\newcommand{\personaSellSharedLuna}{77\%}
\newcommand{\personaSellMixLuna}{53\%}

\newcommand{\personaOutLuna}{12.7\%}

\newcommand{\personaSellCrashLuna}{100\%}
\newcommand{\personaSellSharedDs}{100\%}
\newcommand{\personaSellMixDs}{67\%}
\newcommand{\personaFakeTwoHundredDs}{4\%}

\newcommand{\personaOutBase}{9.2\%}

\newcommand{\personaSellCrashSharedLuna}{0\%}

\newcommand{\revDepthHiMin}{96\%}
\newcommand{\revDepthHiMax}{100\%}
\newcommand{\revDepthLoMin}{0\%}
\newcommand{\revDepthLoMax}{24\%}
\newcommand{\revDepthFive}{100\%}

\newcommand{\revStopRuns}{450}
\newcommand{\revStopCollapse}{400}
\newcommand{\revStopFastOutRuns}{150}
\newcommand{\revStopFastOutCollapse}{150}
\newcommand{\revTrueDrop}{8\%}
\newcommand{\revTrueOverMin}{9.6\%}
\newcommand{\revTrueOverMax}{11.1\%}
\newcommand{\revTrueBehindMin}{4.2}
\newcommand{\revTrueBehindMax}{5.9}

\newcommand{\revTrueCtlOverMax}{0.1\%}

\newcommand{\liveTwoHundred}{9 of 10}
\newcommand{\liveTwoHundredMeasured}{9 of 10}
\newcommand{\liveFourHundred}{5 of 5}
\newcommand{\liveFourHundredMeasured}{5 of 5}
\newcommand{\liveCalls}{10{,}003}
\newcommand{\cycSixHundred}{97\%}
\newcommand{\cycFourHundred}{13\%}

\newcommand{\splitTwoHundredTwoMin}{13\%}
\newcommand{\splitTwoHundredTwoMax}{17\%}
\newcommand{\splitTwoHundredFour}{0\%}
\newcommand{\splitTwoHundredOneMin}{87\%}
\newcommand{\splitTwoHundredOneMax}{93\%}

\newcommand{\splitOutDsTwo}{13.2\%}
\newcommand{\splitOutDsFour}{11.2\%}
\newcommand{\leakQuarterEightMin}{80\%}
\newcommand{\leakQuarterEightMax}{100\%}

\newcommand{\leakHalfFourMin}{88\%}

\newcommand{\histSellAllEmuLuna}{100\%}

\newcommand{\histCrashEmuLunaTwoHundred}{92\%}

\newcommand{\histSellLillyLuna}{0\%}

\newcommand{\histSellApDs}{99\%}

\newcommand{\histSellPentDs}{99\%}

\newcommand{\histSellLillyDs}{25\%}

\newcommand{\histSellApGem}{96\%}

\newcommand{\histSellPentGem}{90\%}

\newcommand{\histSellLillyGem}{97\%}

\newcommand{\histSellApVerbGem}{28\%}

\newcommand{\histCtlDdMax}{1.3\%}
\newcommand{\histLuldPausedRuns}{0}
\newcommand{\histLuldRuns}{1{,}500}
\newcommand{\histLunaHolds}{318}
\newcommand{\histLunaTwo}{324}
\newcommand{\histLunaHoldsConfirm}{307}
\newcommand{\histGemVerbOther}{116}
\newcommand{\histGemVerbKnown}{98}
\newcommand{\histCalls}{2{,}430}

\newcommand{\frameworkShort}{FARSIGHT}
\newcommand{\frameworkLong}{Financial Agent Robustness and Security Investigation and Global Holistic Taxonomy}
\usepackage{forest}
\usepackage{xcolor}
\usepackage{listings}
\usepackage{booktabs}
\usepackage{multirow}
\usepackage{makecell}
\usepackage{pifont}
\usepackage{array}
\usepackage{tabularx}
\usepackage{colortbl}
\usepackage{wasysym}
\usepackage{bmpsize}
\usepackage{pgfplots}
\pgfplotsset{compat=1.18}
\usepgfplotslibrary{fillbetween, groupplots}
\usepackage{adjustbox}
\usepackage{pgf-pie}
\usepackage{caption}
\usepackage{ragged2e}
\usepackage{xurl}
\hypersetup{colorlinks=true, linkcolor=blue, citecolor=blue, urlcolor=blue}
\hypersetup{pdftitle={Trading Agents or Market Crashers? Dissecting Robustness and Security Failures in Academic Financial LLM Trading Schemes},
            pdfauthor={Mengxiao Wang, Nitesh Saxena}}
\def\BibTeX{{\rm B\kern-.05em{\sc i\kern-.025em b}\kern-.08em
    T\kern-.1667em\lower.7ex\hbox{E}\kern-.125emX}}

\definecolor{fsBlue}{RGB}{31,94,166}
\definecolor{fsBlueL}{RGB}{222,235,250}
\definecolor{fsOrange}{RGB}{214,104,24}
\definecolor{fsOrangeL}{RGB}{253,234,214}
\definecolor{fsGreen}{RGB}{40,135,80}
\definecolor{fsGreenL}{RGB}{220,242,226}
\definecolor{fsRed}{RGB}{190,45,45}
\definecolor{fsRedL}{RGB}{250,222,220}
\definecolor{fsPurple}{RGB}{110,70,160}
\definecolor{fsPurpleL}{RGB}{236,228,248}
\definecolor{fsGray}{RGB}{90,90,90}
\definecolor{fsGrayL}{RGB}{238,238,238}
\newtcolorbox{insightbox}[1][]{enhanced, colback=fsBlueL!60!white, colframe=fsBlue,
  boxrule=0.6pt, arc=2pt, left=4pt, right=4pt, top=3pt, bottom=3pt,
  fonttitle=\bfseries\small, coltitle=white, title={#1}}
\newcommand{\cmark}{\textcolor{fsGreen}{\ding{51}}}
\newcommand{\xmark}{\textcolor{fsRed}{\ding{55}}}
\newcommand{\pmark}{\textcolor{fsOrange}{$\sim$}}
\newcommand{\nmark}{\textcolor{fsGray!70}{--}}
\definecolor{okblue}{HTML}{0072B2}
\definecolor{okorange}{HTML}{E69F00}
\definecolor{okverm}{HTML}{D55E00}

\begin{document}

\date{}
\title{\Large \bf Trading Agents or Market Crashers? Dissecting Robustness and Security Failures in Academic Financial LLM Trading Schemes}

\author{
{\rm Mengxiao Wang}\\
Texas A\&M University\\
\texttt{jacksonwang@tamu.edu}
\and
{\rm Nitesh Saxena}\\
Texas A\&M University\\
\texttt{nsaxena@tamu.edu}
}

\maketitle
\begin{abstract}
Large language model (LLM) agents already trade real capital, and builders copy a few published designs that share models and news feeds, so one shock reaches many agents at once. We present \frameworkShort{} (\textit{\frameworkLong}), a systematization that rates trading agents on robustness to market stress and on security against source manipulation, agent compromise, and agents acting as attackers, and checks the ratings against released code. Across 15 academic schemes, no paper describes a halt that skips LLM inference or combines source provenance with cross-source corroboration, and the \auditSchemes{} code releases enforce no stop-loss or position limit outside the LLM. Three LLMs sell whole positions on a fake headline and on real hoaxes, including the fake press release behind the 2000 Emulex crash. In a simulated market driven by these decisions, one fake headline crashes the market in all runs with 400 agents, and 400 agents that sell when prices fall nearly double a 2010-sized crash. The pause that U.S. exchanges apply to large stocks starts in no headline run, whereas automatic halts and hard loss limits outside the model remove both effects. Restated as a domain-independent threat model, \frameworkShort{} maps the same gaps onto agents in healthcare, law, software, and crypto, where compromised agents spread harm as worms and as cascades. We derive 14 open problems and urge deployers to corroborate sources, keep interrupts that bypass the model, and enforce loss limits outside it.
\end{abstract}

\section{Introduction}
Autonomous large language model (LLM) agents increasingly invoke tools, move data, and take consequential actions, and their security is now a serious concern. Recent systematizations~\cite{sokagentic2026, raza_trism_2025} map agentic-AI attacks and defenses at large, but their domain-agnostic treatments cannot capture what changes once an agent operates in a \textit{high-stakes} domain. Financial markets~\cite{imf2024ai_markets} are the sharpest such setting: actions are financially irreversible, adversaries are directly profit-motivated, and one agent's output instantly becomes another's input. \textit{Financial LLM agents}~\cite{ding_large_2024} already manage real capital (e.g., RockAlpha~\cite{rockalpha2025}, Composer~\cite{composer2025}) and serve retail users (e.g., Robinhood~\cite{robinhood2026agentic}, Nof1.ai~\cite{nof1ai2025}), and because financial systems can themselves be weaponized~\cite{lin2016, mulder2022}, a compromised agent could pass its errors on to other market participants. This is no longer a research-only concern. In mid-2026 the Bank of England warned publicly that autonomous AI agents could trigger a market meltdown~\cite{boe2026breeden}, and a Wolters Kluwer survey found that 72\% of U.S. banks lack AI-model kill switches or failure-reporting protocols~\cite{wolters2026banking}. U.S. legislators have already escalated. In June~2026, House Financial Services Democrats sent SEC Chair Atkins 13 questions on ``correlated trading decisions'' and ``herding behavior'' by AI trading agents on registered brokerages~\cite{donachie2026lawmakers}.

\begin{figure*}[htp]
  \centering
  \includegraphics[width=0.8\linewidth]{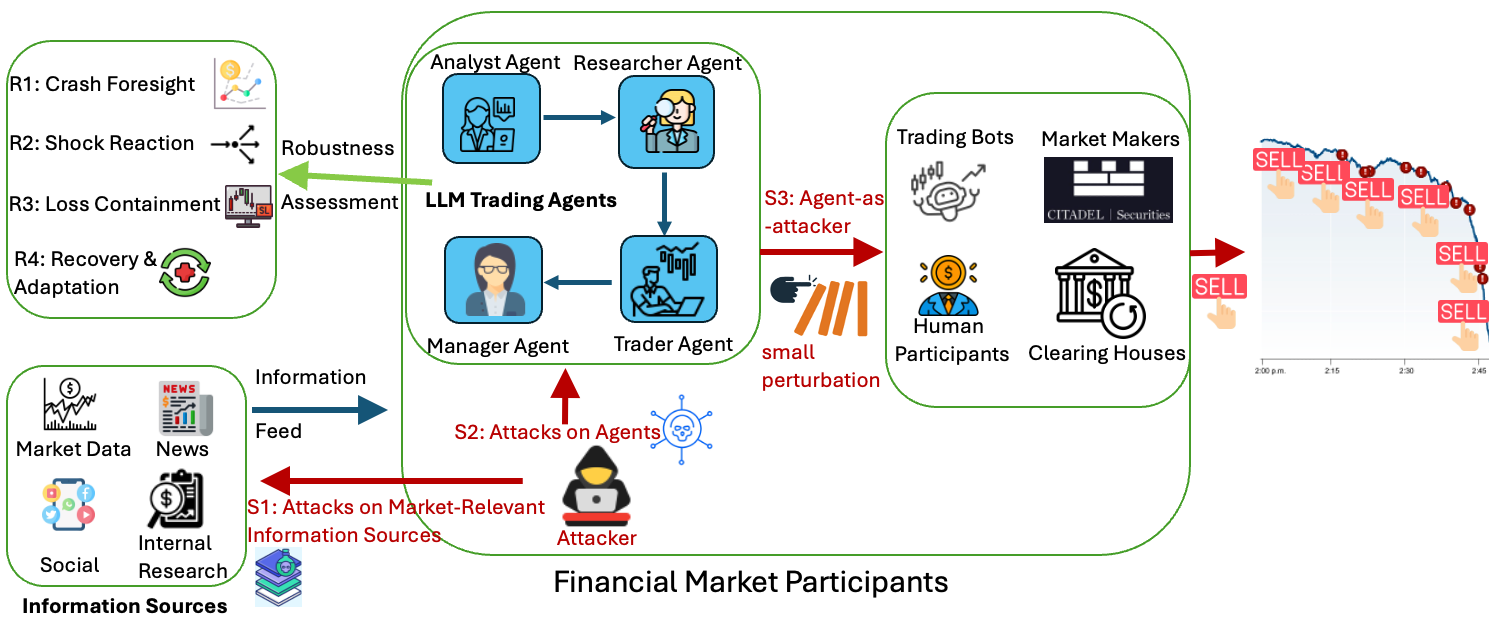}
  \caption{
  Overview of threat surfaces in financial LLM agent ecosystems.
  S1 denotes attacks on market-relevant information sources (e.g., data or news manipulation) that indirectly influence agent decisions;
  S2 represents direct attacks on LLM trading agents (e.g., prompt injection or model poisoning);
  and S3 captures scenarios where compromised agents act as attackers themselves: after being influenced via S1 or S2, they inject small perturbations into trading behaviors, which can propagate through market participants (bots, human traders, and market makers), collectively amplifying reactions and potentially triggering large-scale sell-offs or flash crashes.
  In addition to these security threat surfaces, our robustness assessment (R1–R4) evaluates whether agents can foresee market turbulence, react rapidly to shocks, contain cascading losses, and recover after stress events, reflecting the inherent risks of operating in volatile financial environments.
  }
  \label{fig:attack-types}
\end{figure*}

As Figure~\ref{fig:attack-types} shows, the systems we study follow recent multi-role designs~\cite{yu2025fincon, xiao_tradingagents_2025, li_can_2025, chen_mm-drex_2025}: analyst, researcher, manager, and trader agents that jointly process market data and execute trades. Yet their behavior under extreme conditions is underexplored, even though minor algorithmic reactions can cascade, as in the 2010 Flash Crash that briefly erased nearly one trillion dollars~\cite{kirilenko2017flash, cftcsec2010flash, itbrew2025flash}, and compromising even a small subset of the deployed LLM trading agents (e.g., the more than 150,000 agentic trading accounts that Robinhood reported by September 2026~\cite{robinhood2026agentic, godbole2026robinhood}) could deliberately induce the same effect through small perturbations in low-liquidity stocks. This scale of impact, where a fleet of compromised agents can destabilize an entire market, does not appear in the single-session threat models that generic agentic-security studies assume. Recent work further shows that LLM trading agents systematically inherit and amplify herding, which can amplify flash crashes~\cite{xia2026agentictradingllmagents}. This gap frames our central goal, which is \textit{to measure and ensure that financial LLM agents remain stable under normal market conditions and resilient under adversarial ones.} We approach it in two steps, a scheme-level assessment of published trading-agent designs that maps their attack surface before defenses are designed, and a simulated market that measures the cost of the missing controls once many agents trade together. A systematization fits this stage because deployment has run ahead of research. Products already let agents trade for retail users, and the most widely used academic design has over 100k GitHub stars, yet no common security standard has been applied to these designs.

Answering it starts with what makes these agents different. Their attacks tend to share five traits: they are \textit{Instant} and \textit{Inconspicuous}, rely on \textit{Unverified} inputs, remain \textit{Adaptive} across channels, and can set off \textit{Cascading} effects across the market. Generic prompt-injection and jailbreak taxonomies do not model this combination, which our framework is designed to capture.
We propose the \frameworkShort{} (\frameworkLong{}) framework to systematize agents along robustness and security. Unlike agent-security SoKs~\cite{sokagentic2026, raza_trism_2025}, which are domain-agnostic, and surveys of financial agents~\cite{ding_large_2024, xia2026agentictradingllmagents}, which catalog capabilities, \frameworkShort{} rates every published design against one codebook that treats market-stress robustness as a security property, and it checks these ratings against the released code. Robustness is measured through four metrics: \textit{crash foresight}, \textit{reaction}, \textit{loss containment}, and \textit{recovery}.  
As shown in Figure~\ref{fig:attack-types}, security includes three metrics: \textit{S1} (attacks on market-relevant information sources), \textit{S2} (attacks on agent logic and prompts, including prompt-injection, tool/Application Programming Interface (API) exploitation, and memory manipulation), and \textit{S3} (agent-as-attacker behavior).  

Across the 15 schemes, robustness and security controls are largely missing by design. Twelve of the 15 (80\%) receive no credit on at least one of the four robustness metrics. No paper describes a way to interrupt trading without waiting for LLM inference (R2), and none describes both source provenance and cross-source corroboration before acting on a signal (S1). The released code of the \auditSchemes{} schemes that publish it adds a second view. None of these releases contains an order-execution path or enforces a stop-loss or position limit outside the LLM, \auditUnsandboxed{} run model-written code on the host without a sandbox, and \auditMissingComponent{} lack a component that their paper or documentation describes, such as QuantAgent's risk agent (\S\ref{subsec:code-audit}). The recurring failures show that data, tools, and memory are attack surfaces, so safety must rest on cross-validated information, pre-committed limits, and containment that sit outside the model. The adversary here is financially motivated and the agent holds real execution authority, so once such an agent is deployed, a single planted feed or hijacked tool could move real capital. Robustness here is itself a security concern. In a reflexive, zero-sum market, an attacker can deliberately trigger the shock that a non-robust agent mishandles, so robustness and security cannot be assessed apart.

Ratings alone cannot show the cost of these gaps, because the harm appears once many agents trade the same stock. We therefore measure how three LLMs, gpt-5.6-luna, deepseek-v4-flash, and gemini-3.5-flash, trade one position, and we let hundreds of agents that decide as they do trade in a simulated market (\S\ref{sec:market-experiment}). All three models sell the whole position after one fake headline and after the fake press release that cost Emulex \$2.2 billion in 16 minutes in 2000~\cite{sec2000litigation}.

As Figure~\ref{fig:crash-threshold} shows, the headline then crashes the market in all runs with 400 agents, because each agent sells its whole position in one order and the orders of all agents arrive within one decision cycle. Without any attacker, 400 agents whose model sells when prices fall deepen a 2010-sized crash from \ddOutBase{} to \ddOutDsFourHundred{}. Limit Up-Limit Down, the trading pause of U.S. exchanges~\cite{sec2012luld}, starts in \luldFakePausedRuns{} of \luldFakeRuns{} headline runs, whereas a fast path that freezes orders without an LLM call (R2) and hard loss limits outside the model (R3) remove both effects (Table~\ref{tab:crash-controls}).

More broadly, financial trading agents are a sharp instance of \textit{high-stakes agentic security}, because every property that makes a compromised agent costly, from irreversible actions to coupling with other actors, is at its maximum in markets. The missing controls recur outside finance. Restated as domain-independent properties, the \frameworkShort{} surfaces map onto clinical, legal, software, and crypto agents (\S\ref{sec:beyond}), and S3 ties agent security to market cascades and computer worms (\S\ref{sec:cascades}). Options agents raise every stake further (\S\ref{subsec:options}). With published injection rates, a channel that ten agents read lets one compromised agent start an outbreak even against the frontier model that best resists indirect injection in LivePI~\cite{zhao2026livepi}. To our knowledge, ours is the first work to jointly systematize market robustness and security for the academic trading-agent literature and to measure, with LLM decisions in a simulated market, what the missing controls cost.

\smallskip
\noindent \textbf{Our Contributions:} Our main contributions are outlined below:  

\noindent 1. \textit{\textbf{\frameworkShort{}: A Scheme-Level Framework for High-Stakes Agentic Security.}}
We present \frameworkShort{}, a systematization methodology (not a benchmark suite) that rates financial LLM agents on both robustness and security via a reproducible codebook (Appendix~\ref{subsec:metric-assignment}), with dedicated threat models for the execution-authority, reflexive, and zero-sum conditions that make these agents high-stakes.

\noindent 2. \textbf{\textit{Characterizing the Unique Attack Surface of Financial Trading Agents.}}
We systematically categorize financial LLM agents and identify five defining traits (\textit{Instant}, \textit{Inconspicuous}, \textit{Unverified}, \textit{Adaptive}, \textit{Cascading}) that widen their exposure and separate them from agents in law, healthcare, or education.

\noindent 3. \textit{\textbf{Systematic Robustness and Security Assessment of 15 Trading Schemes.}}
Applying \frameworkShort{} (codebook in Appendix~\ref{appendix:codebook}) to 15 representative schemes, we find that 12 of the 15 (80\%) receive no credit on at least one robustness metric and that none earns full credit on S1 or R2. A parallel LLM-based assessment disagreed with ours on 53\% of metric-level judgments (Appendix~\ref{appendix:llm_analysis}), so fine-grained scoring still requires human expertise. We further audit the released code of the \auditSchemes{} open-source schemes in three independent runs and find that code-level and paper-level ratings agree on \auditMatch{} of \auditRatings{} ratings, with gaps in both directions (\S\ref{subsec:code-audit}). 

\noindent 4. \textbf{\textit{Measuring What the Missing Controls Cost.}}
We show that three LLMs sell whole positions on a fake headline and on hoaxes that moved real markets, that a few hundred agents that decide as they do cause and amplify flash crashes in a simulated market that the exchange pause misses, and that the fast path and hard limits remove both effects across 15 robustness settings (\S\ref{sec:market-experiment}).

\noindent 5. \textbf{\textit{Lessons Beyond Trading and a Research Agenda.}}
We restate \frameworkShort{} as a domain-independent threat model, rate seven high-stakes domains on what makes their agents dangerous, classify six channels through which agents propagate harm, including agent worms, derive 14 open problems (\S\ref{sec:beyond}--\S\ref{sec:open}), and show why options agents are more dangerous (\S\ref{subsec:options}).

\smallskip
\noindent \textbf{Relevance to Security.}
Each \frameworkShort{} metric maps to a systems-security question with an established USENIX, S\&P, or CCS literature on non-financial substrates: source poisoning (S1), prompt, tool, and memory compromise (S2), agent-as-attacker containment (S3), and incident response under adversarial stress (R1--R4)~\cite{greshake2023indirect, liu2024formalizing, shen2024dan, carlini2024poisoningweb, chen2024agentpoison, ruan2024toolemu}. Like prior security SoKs that systematize artifacts from adjacent communities~\cite{papernot2018sok}, we carry this literature into a high-stakes domain that the domain-agnostic SoKs~\cite{sokagentic2026, raza_trism_2025} leave out.

\section{Background}

\noindent\textbf{Large language models and LLM agents.}
Large Language Models are transformer-based neural architectures~\cite{vaswani2017attention} trained on large text corpora to capture linguistic and factual patterns. 
Through large-scale pretraining, they generalize across diverse language tasks, achieving state-of-the-art performance in reasoning, summarization, and generation. 
Representative models include GPT-4~\cite{openai2023gpt4}, LLaMA~\cite{touvron2023llama}, Mixtral~\cite{jiang2024mixtral}, and Claude~\cite{anthropic2025claude}. 
Safety and alignment are typically enhanced via supervised fine-tuning and Reinforcement Learning from Human Feedback (RLHF)~\cite{ouyang2022training}, while Retrieval-Augmented Generation (RAG)~\cite{gao2023ragsurvey} improves factual grounding.
Building on these foundations, \textit{LLM agents} extend pretrained models with capabilities for autonomous reasoning, planning, and environment interaction. 
Unlike static LLMs, agents iteratively reason, act, and observe outcomes through feedback loops~\cite{yao2023reactsynergizingreasoningacting,wang2023voyager}, integrating modules such as \textit{planning}, \textit{memory}, and \textit{tool use}~\cite{langchain2023,autogpt2023}. 
Modern frameworks (e.g., LangChain, AutoGPT, ReAct) enable multi-step task execution, including web searches, code execution, and data analysis. 

\begin{figure*}[t]
\centering
\begin{tikzpicture}[
  font=\footnotesize,
  stage/.style={draw=#1, fill=#1!7, line width=0.8pt, rounded corners=4pt, text width=2.5cm, align=left, inner sep=3pt, minimum height=2.75cm, anchor=north west},
  head/.style={font=\footnotesize\bfseries, text=#1},
  num/.style={fill=#1, text=white, circle, inner sep=0.5pt, minimum size=0.38cm, font=\footnotesize\bfseries},
  flow/.style={-{Stealth[length=2.4mm]}, line width=1.0pt, draw=#1}
]
\def\W{2.66}\def\G{0.29}
\newcommand{\stagebox}[6]{%
  \node[stage=#2] (#6) at ({#1*(\W+\G)}, 2.9) {\textcolor{#2}{\rule{0pt}{1.25em}\smash{\Large #3}}\hfill\tikz[baseline=-0.6ex]\node[num=#2]{\the\numexpr#1+1\relax};\\[2pt]{\color{#2}\bfseries #4}\\[2pt] #5};}
\stagebox{0}{fsBlue}{\faBook}{Collect}{127 candidate papers, 40 read in full, 15 schemes selected (\S\ref{sec:taxonomy})}{s1}
\stagebox{1}{fsPurple}{\faListUl}{Codebook}{9 metrics, \auditIndPerScheme{} binary indicators, deterministic rules (App.~\ref{appendix:codebook})}{s2}
\stagebox{2}{fsGreen}{\faUserCheck}{Rate papers}{Two experts, 135 ratings, Cohen's~$\kappa{=}0.89$ (Table~\ref{tab:combined-eval})}{s3}
\stagebox{3}{fsOrange}{\faCode}{Audit code}{\auditSchemes{} releases, \auditRuns{} LLM runs each, pattern scan (\S\ref{subsec:code-audit})}{s4}
\stagebox{4}{fsRed}{\faTools}{Recommend}{Controls outside the model, mapped to Rule 15c3-5 and RTS~6 (\S\ref{sec:recommendations})}{s5}
\stagebox{5}{fsGray}{\faGlobeAmericas}{Generalize}{Other domains, cascades and worms, 14 open problems (\S\ref{sec:beyond}--\S\ref{sec:open})}{s6}
\foreach \a/\b/\c in {s1/s2/fsBlue, s2/s3/fsPurple, s3/s4/fsGreen, s4/s5/fsOrange, s5/s6/fsRed}{
  \draw[flow=\c] (\a.east) -- (\b.west);
}
\node[draw=fsGray, dashed, line width=0.7pt, rounded corners=4pt, fill=white, text width=5.4cm, align=left, inner sep=3pt, anchor=north] (llm) at ($(s3.south)+(0,-0.45)$) {\textcolor{fsGray}{\faRobot}\ \textbf{LLM-based assessment, comparison only:} disagrees with the experts on 24 of 45 cells (App.~\ref{appendix:llm_analysis})};
\draw[flow=fsGray, dashed] (llm.north) -- (s3.south);
\end{tikzpicture}
\caption{Overview of the \frameworkShort{} pipeline. Two human experts assign the paper-level ratings, an LLM-based assessment serves only as a comparison, and three independent audit runs check the ratings of the seven open-source schemes against their code. Sections~\ref{sec:beyond}--\ref{sec:open} carry the findings beyond trading.}
\label{fig:farsight-overview}
\end{figure*}

\noindent\textbf{Financial LLM agents.}
Ding et al.~\cite{ding_large_2024} define a \textit{financial LLM agent}, often instantiated as an \textit{LLM trading agent}, as an autonomous system that leverages large language model reasoning for trading tasks such as market prediction, portfolio optimization, and risk assessment. 
These agents integrate LLMs with financial data streams, market simulators, and trading APIs to enable data-driven investment decisions~\cite{lopez2025canllmtrade,saha_large_2025,mahdavi_integrating_2025}. 
Unlike conventional algorithmic trading systems, they exhibit adaptability and contextual understanding through natural language processing, supporting applications such as sentiment-based forecasting~\cite{zhang_when_2024}, portfolio construction~\cite{guo_mass_2025}, high-frequency trading (HFT)~\cite{xiong_quantagent_2025}, and multi-agent risk management~\cite{li_hedgeagents_2025}. 
Recent work further explores contest-based reinforcement~\cite{zhao_contesttrade_2025} and decision auditing in investment management~\cite{saha_large_2025}. 
Two main paradigms have emerged: \textit{Trader Agents}, which execute buy, hold, or sell decisions~\cite{yu_finmem_2023,yu2025fincon,xiao_tradingagents_2025}, and \textit{Alpha Miner Agents}, which identify predictive signals (\textit{alphas}) from textual and quantitative data~\cite{yuan_alpha-gpt_2024}. 
In practice, Alpha Miners often serve as signal generators for Trader Agents, and we treat them as a functional subset in this study.

\section{Related Work and Positioning}
\label{sec:related}

Six bodies of work surround this paper: surveys of financial LLM agents, security systematizations of LLM agents, attacks on trading agents, studies of correlated LLM decisions and simulated markets, studies of propagation among agents, and agent security in other high-stakes domains. Table~\ref{tab:related} positions \frameworkShort{} against representative studies of each along six dimensions: whether a study targets trading agents, builds a security threat model, treats robustness under market stress, considers spillover from compromised agents to other agents or the market, assesses each studied system against explicit criteria, and examines released code.

\begin{table*}[t]
\centering
\caption{Positioning of \frameworkShort{} against representative related work. \cmark: a central part of the study; \pmark: discussed, or limited to one attack surface or a few systems; \nmark: not covered. Rob.\ = robustness under market stress; Spillover = harm that spreads from a compromised agent to other agents or the market; Per-system = each studied system is assessed against explicit criteria; Code = released implementations are checked against the published designs.}
\label{tab:related}
\footnotesize
\renewcommand{\arraystretch}{1.08}
\setlength{\tabcolsep}{5pt}
\begin{tabular}{l l l c c c c c c}
\toprule
\rowcolor{fsGrayL}
\textbf{Group} & \textbf{Study} & \textbf{Type} & \textbf{Trading agents} & \textbf{Threat model} & \textbf{Rob.} & \textbf{Spillover} & \textbf{Per-system} & \textbf{Code} \\
\midrule
\multirow{7}{*}{\makecell[l]{Surveys of\\financial\\LLM agents}}
& Ding et al.~\cite{ding_large_2024} & Survey & \cmark & \nmark & \nmark & \nmark & \nmark & \nmark \\
& Dong et al.~\cite{dong2025agents} & Survey & \pmark & \pmark & \nmark & \pmark & \nmark & \nmark \\
& Saha et al.~\cite{saha_large_2025} & Survey & \pmark & \nmark & \pmark & \nmark & \nmark & \nmark \\
& Mahdavi et al.~\cite{mahdavi_integrating_2025} & Survey & \pmark & \nmark & \nmark & \nmark & \nmark & \nmark \\
& Hua et al.~\cite{hua2026agentic} & Survey & \cmark & \nmark & \nmark & \nmark & \nmark & \nmark \\
& Xia et al.~\cite{xia2026agentictradingllmagents} & Survey & \cmark & \pmark & \pmark & \pmark & \nmark & \nmark \\
& Gong~\cite{gong2026agents} & Survey & \pmark & \nmark & \nmark & \pmark & \nmark & \nmark \\
\midrule
\multirow{4}{*}{\makecell[l]{Security\\systematizations}}
& Kim et al.~\cite{sokagentic2026} & SoK & \nmark & \cmark & \nmark & \nmark & \nmark & \nmark \\
& Raza et al.~\cite{raza_trism_2025} & Survey & \nmark & \cmark & \pmark & \nmark & \nmark & \nmark \\
& Zeng et al.~\cite{zeng2026meets} & Survey & \pmark & \cmark & \pmark & \pmark & \nmark & \nmark \\
& Mao et al.~\cite{mao2026sokcommerce} & SoK & \pmark & \cmark & \nmark & \pmark & \nmark & \nmark \\
\midrule
\multirow{5}{*}{\makecell[l]{Attacks on\\trading agents}}
& Rizvani et al.~\cite{rizvani2026adversarial} & Attack & \cmark & \pmark & \nmark & \nmark & \pmark & \nmark \\
& Liu et al.~\cite{liu2026autoredtrader} & Attack & \cmark & \pmark & \nmark & \nmark & \pmark & \nmark \\
& Na et al.~\cite{na2026poisoning} & Attack & \cmark & \pmark & \nmark & \nmark & \pmark & \nmark \\
& Sua et al.~\cite{sua2026contagion} & Attack & \cmark & \pmark & \pmark & \nmark & \pmark & \nmark \\
& Yan et al.~\cite{yan2025tradetrap} & Benchmark & \cmark & \pmark & \pmark & \nmark & \pmark & \nmark \\
\midrule
\multirow{3}{*}{\makecell[l]{Correlation and\\market simulation}}
& Apte~\cite{apte2026trading} & Stress test & \cmark & \nmark & \pmark & \pmark & \nmark & \nmark \\
& Lopez-Lira~\cite{lopez2025canllmtrade} & Simulation & \cmark & \nmark & \nmark & \pmark & \nmark & \nmark \\
& Ross et al.~\cite{ross2026better} & Simulation & \cmark & \pmark & \nmark & \cmark & \nmark & \nmark \\
\midrule
\multirow{6}{*}{\makecell[l]{Propagation\\among agents}}
& Schroeder de Witt et al.~\cite{csdw2025open} & Agenda & \nmark & \cmark & \nmark & \cmark & \nmark & \nmark \\
& Hammond et al.~\cite{hammond2025multi} & Report & \nmark & \pmark & \nmark & \cmark & \nmark & \nmark \\
& Cohen et al.~\cite{cohen2024here} & Attack & \nmark & \cmark & \nmark & \cmark & \pmark & \nmark \\
& Gu et al.~\cite{gu2024agentsmith} & Attack & \nmark & \cmark & \nmark & \cmark & \nmark & \nmark \\
& Lee and Tiwari~\cite{lee2024promptinfection} & Attack & \nmark & \cmark & \nmark & \cmark & \nmark & \nmark \\
& Triedman et al.~\cite{triedman2025multi} & Attack & \nmark & \cmark & \nmark & \pmark & \cmark & \pmark \\
\midrule
\multirow{4}{*}{\makecell[l]{Other\\high-stakes\\domains}}
& Alber et al.~\cite{alber2025medical} & Attack & \nmark & \cmark & \nmark & \nmark & \nmark & \nmark \\
& Clusmann et al.~\cite{clusmann2025prompt} & Attack & \nmark & \cmark & \nmark & \nmark & \pmark & \nmark \\
& Magesh et al.~\cite{magesh2024hallucination} & Measurement & \nmark & \nmark & \nmark & \nmark & \cmark & \nmark \\
& Patlan et al.~\cite{patlan2025real} & Attack & \pmark & \cmark & \nmark & \nmark & \pmark & \pmark \\
\midrule
& \textbf{This work} & \textbf{Rating + simulation} & \cmark & \cmark & \cmark & \cmark & \cmark & \cmark \\
\bottomrule
\end{tabular}
\end{table*}

\noindent\textbf{Surveys of financial LLM agents.} Surveys of LLM agents in trading and finance catalog architectures, data inputs, tasks, benchmarks, and backtest performance~\cite{ding_large_2024,dong2025agents,saha_large_2025,mahdavi_integrating_2025,hua2026agentic}. They describe what trading agents can do, and security appears in them at most as a deployment challenge. Two recent surveys go further. Xia et al. note that open-web memory can be poisoned and call for deterministic risk guardrails~\cite{xia2026agentictradingllmagents}, and Gong discusses the systemic implications of agent adoption in financial markets~\cite{gong2026agents}. Neither defines a threat model or assesses individual systems against security criteria. In contrast, \frameworkShort{} turns these concerns into a threat model with three attack surfaces and four robustness phases, and it rates each of 15 schemes against them.

\noindent\textbf{Security systematizations of LLM agents.} Agent-security systematizations map attacks and defenses for LLM agents in general~\cite{sokagentic2026,raza_trism_2025,deng2025agents}. They are deliberately domain-agnostic, and their threat models do not capture an agent that holds execution authority over capital in a reflexive market, where one agent's output becomes another's input. Newer backbones do not retire these attacks, because indirect prompt injection still succeeds against GPT-5-class, Claude~4.x, and Gemini~3.x models~\cite{zhao2026livepi,abdelnabi2026agents}. Closer to finance, Zeng et al. organize seventeen attack subtypes across the lifecycle of financial AI, from data poisoning to prompt injection and deepfake fraud against identity checks~\cite{zeng2026meets}, and Mao et al. systematize the security of agents in agentic commerce, including payment, negotiation, and market-manipulation risks~\cite{mao2026sokcommerce}. Both work at the level of model lifecycles or commerce protocols, above the level of concrete trading agents, and neither treats robustness under market stress. In contrast, \frameworkShort{} adds a robustness axis with no direct analogue in these works (crash foresight, shock reaction, loss containment, and recovery) and instantiates the security axis with the two surfaces that are specific to trading: manipulation of market-relevant information (S1) and compromised agents that act on the market (S3).

\noindent\textbf{Attacks on trading agents.} A growing line of empirical work attacks trading agents directly. Rizvani et al. show that adversarial news headlines cut the profit of an LLM-driven trading system~\cite{rizvani2026adversarial}, Liu et al. generate finance-specific misinformation that misleads trading agents~\cite{liu2026autoredtrader}, Na et al. poison the inputs of four multi-agent trading architectures through role-specific adversaries~\cite{na2026poisoning}, Sua et al. trace how poisoned social-media posts spread inside multi-agent trading stacks~\cite{sua2026contagion}, and Yan et al. perturb four components of two classes of trading agents in backtests~\cite{yan2025tradetrap}. Each study tests one or a few attack surfaces on one or a few systems, most often the information surface. These results show that the attacks we systematize are practical. Our systematization complements them by showing how widely the corresponding controls are missing across 15 published schemes and in their released code (\S\ref{subsec:code-audit}).

\noindent\textbf{Correlated decisions and simulated markets.} Correlation turns the errors of single agents into a market problem. Regulators and economists warn that common models and data sources produce correlated behavior and that AI can amplify procyclicality~\cite{fsb2024financial, gensler2020deep, danielsson2022artificial}, and measurements now find this correlation in LLMs. Their errors are correlated across models~\cite{kim2025correlated}, shared prompts concentrate the actions of LLM trading agents in a microstructure stress test~\cite{apte2026trading}, stock-price predictions covary across eight LLMs~\cite{mcclellan2025financial}, and ten preference-tuned forecasters in a prediction market are worth about 1.4 independent ones~\cite{begin2026preference}. Ross et al. show that LLM traders that read the same misleading commentary misprice together in a simulated market, where more capable models act more alike~\cite{ross2026better}, and Ng sorts AI-driven flash crashes into herding, error-cascade, and disinformation crashes~\cite{ng2026herding}. Agent-based market models study such instability with fundamentalist and trend-following traders~\cite{brock1998heterogeneous}, support the leading explanation of the 2010 Flash Crash~\cite{paddrik2012agent}, and show that fake news lowers prices~\cite{mignot2024fake}. Recent versions use LLM traders~\cite{lopez2025canllmtrade}, which form bubbles on a simulated social network~\cite{noauthor_twinmarket_nodate} and switch trading styles only partly as behavioral finance predicts~\cite{li2026behavioral}. These works measure correlated decisions, classify crashes, or simulate traders. In contrast, we feed measured LLM decisions into such a market, find the number of agents at which correlation becomes a crash, and test the controls that market rules require and that no rated design implements (\S\ref{sec:market-experiment}).

\noindent\textbf{Propagation among agents.} Research agendas for multi-agent security taxonomize the threats that emerge when agents interact, including secret collusion, swarm attacks, and the spread of jailbreaks and poisoned data through network effects~\cite{csdw2025open,hammond2025multi}. Attack studies have made propagation concrete. Self-replicating prompts spread through RAG-based email assistants~\cite{cohen2024here} and through multi-agent systems~\cite{lee2024promptinfection}, one adversarial image jailbreaks a population of up to one million agents exponentially fast~\cite{gu2024agentsmith}, adversarial content hijacks multi-agent frameworks into executing arbitrary code~\cite{triedman2025multi}, and network topology shapes how far attacks spread~\cite{yu2024netsafe,huang2025resilience}. These studies examine agents that infect one another through messages. In contrast, we connect propagation to markets, where harm also spreads without compromise through the prices that every agent reads, and we use published per-hop success rates to estimate when a single compromised agent can start an outbreak (\S\ref{sec:cascades}).

\noindent\textbf{Agent security in other high-stakes domains.} Domain studies document the same surfaces one domain at a time. Medical models can be poisoned through a tiny fraction of their training data~\cite{alber2025medical} or their weights~\cite{han2023medical}, and medical vision-language models follow instructions hidden in images~\cite{clusmann2025prompt}. General-purpose models and commercial legal research tools still invent legal facts and citations~\cite{dahl2024legal,magesh2024hallucination}. Memory injection steers Web3 agents into unauthorized asset transfers~\cite{patlan2025real}, and autonomous science agents raise safety risks that call for safeguards over autonomy~\cite{tang2025risks}. Each study covers one domain and mostly one surface. In contrast, Section~\ref{sec:beyond} maps all \frameworkShort{} surfaces and resilience phases onto seven domains and rates them on the properties that make trading agents dangerous.

\noindent\textbf{Differences from prior work.} \frameworkShort{} differs from all six bodies of work along six dimensions at once. First, it targets trading agents specifically, unlike the domain-agnostic agent systematizations~\cite{sokagentic2026,raza_trism_2025} and the finance-wide surveys~\cite{dong2025agents,zeng2026meets}. Second, it builds a security threat model with three attack surfaces, which the capability surveys~\cite{ding_large_2024,saha_large_2025,mahdavi_integrating_2025,hua2026agentic} lack. Third, it treats robustness under market stress (crash foresight, shock reaction, loss containment, and recovery) as a security property, which none of the other works covers as a central part. Fourth, it covers spillover from compromised agents to the market (S3), whereas the trading attack studies measure harm to the attacked agent's own decisions or, at most, propagation inside one multi-agent stack~\cite{sua2026contagion}, and the propagation studies stop at agents that infect one another. We also measure this spillover, with three LLMs' decisions in a simulated market, where the simulation studies above measure correlation but not the controls that would stop it. Fifth, it rates every one of 15 schemes against the same deterministic codebook, whereas each attack study tests one or a few systems. Sixth, it checks these ratings against the released code of the seven open-source schemes, which no other work in Table~\ref{tab:related} does as a central part. Beyond the six dimensions, this version carries the framework to other high-stakes domains and to cascades between agents (\S\ref{sec:beyond}--\S\ref{sec:open}). To our knowledge, \frameworkShort{} is the first systematization of academic trading agents that combines all six.

\section{Our \frameworkShort{} Framework} \label{framework}
This section introduces our \frameworkShort{} (\textit{\frameworkLong}) framework, a structured methodology for systematizing the robustness and security of financial LLM agents.
\frameworkShort{} is a \textbf{scheme-level assessment framework}: each system is scored as a \textit{scheme}, the design described in its published paper, not as a running instance. It is not a benchmark suite with fixed datasets and automated test harnesses. It provides structured criteria, a codebook-based scoring methodology, and a reproducible assessment pipeline for expert review of agent architectures. This design choice reflects the heterogeneity of financial LLM agents: their diverse architectures, data sources, and deployment contexts make a single standardized benchmark insufficient for capturing the full spectrum of robustness and security concerns.
To ensure replicability and transparency, we include a detailed \textit{codebook} outlining the coding scheme, metric definitions, and annotation examples in Appendix~\ref{appendix:codebook}.

\subsection{Workflow of \frameworkShort{}}

Inspired by the SoK methodology presented by Akanda et al.~\cite{akanda2025sok}, we developed \frameworkShort{} to assess the state of financial LLM agents through literature-grounded assessment and metric-based analysis. 
Unlike prior frameworks that focus on implementation toolkits (e.g., LangChain, AutoGPT, ReAct) or benchmarking tasks (e.g., AgentDojo~\cite{debenedetti2025agentdojo}, StockBench~\cite{chen2025stockbenchllmagentstrade}), \frameworkShort{} introduces a unified and reproducible assessment pipeline that emphasizes robustness and security as primary concerns.

Section~\ref{sec:related} positions \frameworkShort{} against surveys of financial LLM agents, security systematizations of LLM agents, and attacks on trading agents.

As shown in Figure~\ref{fig:farsight-overview}, our study follows the \frameworkShort{} framework to assess the robustness and security of financial LLM agents.
We began by collecting a comprehensive corpus of academic works and real-world products related to financial LLM agents. 
From this corpus, we filtered and selected those directly relevant to trading and market-interacting systems, specifically agents that autonomously or semi-autonomously perform financial analysis, strategy generation, or trade execution. 
Each selected work was analyzed as an \textit{instantiated trading agent}, forming the foundation of our systematization.

Based on this dataset, \frameworkShort{} introduces two categories of metrics: (\textit{i}) \textit{Robustness Metrics}, which measure an agent’s capacity to maintain operational stability under volatile or flash-crash-like conditions, and (\textit{ii}) \textit{Security Metrics}, which assess resilience against three major threat dimensions: \textit{attacks on the information environment}, \textit{attacks on agents}, and \textit{agent-as-attacker} behaviors. 
The framework translates these metrics into a systematic assessment process that runs from literature collection through codebook-based rating to insight generation. 
We applied these metrics to 15 representative financial LLM agents, identifying common design weaknesses, overlooked risks, and insufficient consideration of real-world adversarial conditions. 
Our findings show that most existing systems fail to anticipate the impact of extreme market turbulence or targeted manipulation, leaving users and markets vulnerable to cascading financial losses. 
Finally, \frameworkShort{} provides actionable recommendations for building more robust and secure financial LLM agents, with robustness and security built in from the design stage.

\subsection{Robustness Metrics}
Robustness in \frameworkShort{} refers to a financial LLM agent’s ability to maintain stable, safe, and rational behavior under sudden and extreme market disruptions (e.g., flash crashes, black swan events).
We note that market volatility and price corrections are not inherently undesirable: classical economic theory recognizes rapid price adjustment as a feature of efficient markets~\cite{fama1970efficient}. Our robustness metrics do not target market volatility \textit{per se}, but rather \textit{agent-amplified instability}: scenarios where LLM agents, through delayed reaction, herding behavior, or cascading feedback loops, \textit{exacerbate} market disruptions beyond what would occur with human-only participation. The distinction is between natural price discovery and artificial destabilization caused by poorly designed autonomous systems~\cite{minsky1986stabilizing,hayek1931prices}.
We define four key robustness metrics that reflect different phases of a market crisis, from early detection to post-crash recovery. The phases mirror the incident-handling functions of the NIST Cybersecurity Framework~2.0~\cite{nist2024csf}: R1 maps to Detect, R2 and R3 map to Respond (R3 to its containment of incidents), and R4 maps to Recover.

\noindent\textbf{R1. Crash Foresight and Early Awareness.}  
This metric assesses whether an agent can anticipate extreme market volatility or potential crashes in advance by identifying anomalous signals (e.g., spikes in the Chicago Board Options Exchange (CBOE) Volatility Index (VIX), correlated asset drops, or sentiment shifts).  
It measures the agent’s ability to perform proactive risk sensing before catastrophic losses occur.

\noindent\textbf{R2. Rapid Reaction Capability under Market Shocks.}  
This metric evaluates the timeliness and appropriateness of an agent’s response during sudden market disruptions.  
A robust agent has a pre-committed path that halts risky orders, freezes trading, or switches to a safe mode without waiting for LLM inference, with a reaction time matched to the horizon of its strategy. The reaction time that suffices ranges from below 0.1~ms in high-frequency trading to about 10~s in daily trading, whereas TradingAgents takes over 240~s per decision (Table~\ref{tab:latency-comparison}).

\noindent\textbf{R3. Loss Containment and Stop-Loss Execution.}  
This metric measures whether an agent can effectively contain losses and prevent cascading risk once a market crash begins.  
It captures the design of automated stop-loss mechanisms, exposure caps, and cool-off intervals that limit systemic damage. None of the \auditSchemes{} released implementations enforces such a limit outside the LLM (\S\ref{subsec:code-audit}).

\noindent\textbf{R4. Recovery and Post-Crash Adaptation.}
This metric captures the agent’s ability to recover from an adverse event and restore stable operation.
It considers whether the agent can rebalance parameters, retrain on updated data, or adaptively recalibrate its strategy after a market crash.

To ensure reproducibility, each \frameworkShort{} metric is assigned using an \textit{evidence-based checklist} with deterministic scoring rules that leave no room for subjective judgment. We detail this assignment methodology (together with the inter-rater agreement between our two independent evaluators) in Appendix~\ref{subsec:metric-assignment}.

\subsection{Security Metrics and Underlying Threat Models}
This section presents the threat models associated with each security metric in \frameworkShort{}, outlining the underlying assumptions, adversary capabilities, attack flows, and the security properties each metric is intended to assess.

\noindent \textbf{S1: Attacks on Market-Relevant Information Sources.}

\noindent \textit{Threat Model:}
We assume the financial LLM agent continuously ingests market-relevant information (news feeds, sentiment data, social-media signals) without enforcing strong source authentication, provenance validation, or cross-source consistency checks. The attacker cannot access or modify the agent’s internal prompts or model parameters, but can compromise, forge, delay, or selectively manipulate upstream information sources that the agent relies on. By controlling these external data channels, the adversary can shape the agent’s perception of market conditions and influence its downstream trading decisions.

\noindent \textit{Attack Flow / Objective:}
The attacker first compromises an external information provider or feed (e.g., a news API, sentiment stream, or social media source).
Manipulated content (falsified news, selectively delayed signals, or biased sentiment) is then propagated through the agent’s subscribed channels.
The agent ingests these deceptive inputs, integrates them into its decision-making context, and executes trades or reallocations based on corrupted market signals.
The objective is to indirectly influence or steer the agent’s trading logic, causing misinformed actions, adversary-favorable outcomes, or broader systemic disruption, all without directly compromising the agent itself.

This metric examines whether the agent relies on trusted, cross-validated information sources and whether it detects or filters unreliable inputs.
In particular, we assess whether the agent validates news authenticity, aggregates signals across multiple feeds, or naively consumes highly manipulable sources such as social media, widely recognized as the easiest channel for coordinated misinformation.

\noindent \textbf{S2: Attacks on Agent Logic and Prompts.}

\noindent \textit{Threat Model:}
We assume the financial LLM agent operates in an interactive or API-accessible environment that accepts external inputs (chat requests, automated trading commands, tool-invocation APIs) without strict isolation between system prompts, user prompts, tool permissions, and persistent memory. The attacker can send crafted prompts or payloads to the agent, exploiting its instruction-following behavior, its ability to call tools or APIs, and its capability to write or retrieve long-term memory. The attacker aims to directly subvert the agent’s decision logic, induce unauthorized financial actions, or implant persistent malicious context that continues influencing behavior across sessions.

\noindent \textit{Attack Flow / Objective:} The attack proceeds by injecting crafted instructions that blend benign content with adversarial payloads. Once processed by the agent, these injections may override internal trading constraints, trigger unintended tool or API calls, or contaminate long-term memory with biased or adversarial instructions. \frameworkShort{} decomposes S2 attacks into the following categories:
\begin{itemize}[leftmargin=*]
  \item \textbf{S2.1 Prompt-Injection and Instruction Override:}
  Malicious text manipulates the agent’s reasoning chain or bypasses predefined trading rules through context hijacking or instruction override.

  \item \textbf{S2.2 Tool-Use and API Exploitation:}
  Compromised reasoning leads to unauthorized execution of tools or APIs, enabling unintended trades, fund movements, or sensitive data exfiltration.

  \item \textbf{S2.3 Memory or Context Manipulation:}
  The attacker injects persistent malicious content into memory, causing long-lived bias or harmful strategy drift that survives resets or retraining cycles.
\end{itemize}

This metric evaluates whether the agent enforces prompt isolation, input sanitization, tool-use sandboxing, API permission boundaries, and memory-integrity protections, critical safeguards required to prevent direct logic-layer compromise.

\noindent \textbf{S3: Agent-as-Attacker Behavior.}

\noindent \textit{Threat Model:}
We assume that one or more financial LLM agents have already been compromised through S1 or S2 and continue to operate in a multi-agent or market-simulation environment where other agents trust their outputs. The attacker fully controls these compromised agents and can use them to autonomously generate misinformation, propagate malicious prompts, or execute manipulative trades. The attacker aims to exploit the compromised agent’s trusted role to influence broader market dynamics and induce cascading or destabilizing effects.

\noindent \textit{Attack Flow / Objective:}
A compromised agent behaves as an active adversary rather than a passive victim. It may disseminate false signals, coordinate harmful trades with other controlled agents, or manipulate liquidity or volatility to trigger cascading reactions. Because financial agents often treat peer outputs as actionable signals, a single compromised agent can rapidly propagate harmful behavior across the ecosystem. The attacker’s objective is to amplify systemic volatility, trigger coordinated sell-offs or flash-crash-like events, or exploit arbitrage gaps created by the disruption.

This metric evaluates whether the system can detect malicious coordination, isolate compromised agents, and contain cascading impacts across interconnected agent networks, critical safeguards for preventing ecosystem-wide failures. It rates whether these controls are present, not whether the rated schemes have caused spillover.

\noindent\textbf{Trading-specific changes.} Table~\ref{tab:generic-vs-finance} states, for each surface, what changes once the agent trades. Two changes recur: the market pays the adversary, because a pre-positioned trade turns a manipulated decision into profit, and the harm is irreversible, because an executed order settles before any human review.

\begin{table*}[t]
\centering
\caption{Generic agent threats and their instantiation in trading agents. The last column states what the trading setting changes relative to the generic threat.}
\label{tab:generic-vs-finance}
\small
\setlength{\tabcolsep}{4pt}
\begin{tabular}{>{\raggedright\arraybackslash}p{0.9cm} >{\raggedright\arraybackslash}p{3.6cm} >{\raggedright\arraybackslash}p{4.5cm} >{\raggedright\arraybackslash}p{7.0cm}}
\toprule
\textbf{Surface} & \textbf{Generic agent threat} & \textbf{Instantiation in trading agents} & \textbf{What changes in trading} \\
\midrule
S1 & Indirect prompt injection via retrieved content~\cite{greshake2023indirect} & Forged headline, spoofed filing, or coordinated social-media posts in the agent's feeds & The payload is plausible news rather than an instruction, so injection filters miss it. The attacker profits from a position taken beforehand, as in the 2013 AP Twitter hack~\cite{karppi2015hackcrash}. \\
S2.1 & Direct prompt injection that overrides instructions~\cite{liu2024formalizing} & Injected text that overrides trading rules such as position limits or restricted symbols & A rule violation becomes an order that settles. A small numeric nudge suffices, which is harder to spot than an instruction override. \\
S2.2 & Unauthorized tool or API calls~\cite{ruan2024toolemu,debenedetti2025agentdojo} & Order submission, fund movement, or execution of model-written code (\S\ref{subsec:code-audit}) & Actions are irreversible, and the damage scales with account size and leverage, not with the number of calls. \\
S2.3 & Memory and retrieval poisoning~\cite{chen2024agentpoison} & Poisoned reflections or decision logs that bias later trades (\S\ref{subsec:code-audit}) & The bias compounds through self-evolving trading memory, and the losses it causes look like ordinary market noise, which delays detection. \\
S3 & A compromised agent misleads its peers & Compromised agents trade or broadcast signals that other agents, bots, and market makers act on & Harm reaches third parties through prices, and herding among correlated agents~\cite{xia2026agentictradingllmagents} amplifies it into flash-crash-like cascades~\cite{kirilenko2017flash}. \\
R1--R4 & Fail-safe behavior under faults & Agent behavior during gaps, liquidity shocks, and flash crashes & The attacker can manufacture the shock with spoofed news or thin-market trades, so a robustness failure becomes an attack path. \\
\bottomrule
\end{tabular}
\end{table*}

\begin{table*}[t]
\centering
\caption{Summary of LLM-based Financial Agents: Role, I/O, Mechanisms, and Evaluation Settings}
\label{tab:llm-agents-full}
\renewcommand{\arraystretch}{1.1}
\begin{tabular}{p{3.3cm} p{1cm} p{3cm} p{2cm} p{2.5cm} p{1.2cm}}
\toprule
\textbf{System / Paper} & \textbf{Role} & \textbf{Typical Inputs} & \textbf{Outputs} & \textbf{Mechanism} & \textbf{Code} \\
\midrule
\textsc{TradingGPT} \cite{li2023tradinggpt} & \faDollarSign &
\faNewspaper\; \faHistory\; \faLandmark\; \faFileInvoiceDollar &
\faMouse\; \faFile &
\faMemory\; \faNetworkWired\; \faUserCircle &
$\times$ \\

\textsc{FinMem} \cite{yu_finmem_2023} & \faDollarSign &
\faNewspaper\; \faHistory\; \faFileInvoiceDollar &
\faMouse\; \faFile &
\faMemory\; \faUserCircle &
\checkmark \\

\textsc{MarketSenseAI} \cite{fatouros2024can} & \faDollarSign &
\faNewspaper\; \faHistory\; \faLandmark\; \faFileInvoiceDollar &
\faMouse\; \faFile &
\faNetworkWired\; \faRavelry\; \faLink &
$\times$ \\

\textsc{FinCon} \cite{yu2025fincon} & \faDollarSign &
\faNewspaper\; \faHistory\; \faLandmark\; \faFileInvoiceDollar\; \faSmile &
\faMouse\; \faFile &
\faMemory\; \faNetworkWired &
$\times^{*}$ \\

\textsc{StockAgent} \cite{zhang_when_2024} & \faDollarSign &
\faNewspaper\; \faHistory\; \faLandmark\; \faFileInvoiceDollar &
\faMouse\; \faFile &
\faNetworkWired &
\checkmark \\

\textsc{FinAgent} \cite{zhang2024multimodal} & \faDollarSign &
\faNewspaper\; \faHistory\; \faFileInvoiceDollar\; \faUserTie &
\faMouse\; \faFile &
\faMemory &
\checkmark \\

\textsc{FinRobot} \cite{yang_finrobot_2024} & \faDollarSign &
\faNewspaper\; \faHistory\; \faLandmark\; \faFileInvoiceDollar\; \faSmile &
\faMouse\; \faFile &
\faMemory\; \faNetworkWired\; \faRavelry\; \faLink &
\checkmark \\

\textsc{Alpha-GPT 2.0} \cite{yuan_alpha-gpt_2024} & \faSearchDollar &
\faNewspaper\; \faHistory\; \faLandmark\; \faFileInvoiceDollar & \faFile &
\faUserTie &
$\times$ \\

\textsc{QuantAgent} \cite{xiong_quantagent_2025} & \faDollarSign &
\faHistory &
\faMouse\; \faFile &
\faNetworkWired &
\checkmark \\

\textsc{TradExpert} \cite{ding2024tradexpert} & \faDollarSign &
\faNewspaper\; \faHistory\; \faFileInvoiceDollar &
\faMouse\; \faFile &
\faLink &
$\times$ \\

\textsc{TradingAgents} \cite{xiao_tradingagents_2025} & \faDollarSign &
\faNewspaper\; \faHistory\; \faFileInvoiceDollar\; \faSmile &
\faMouse\; \faFile &
\faNetworkWired &
\checkmark \\

\textsc{FLAG-Trader} \cite{xiong_flag-trader_2025} & \faDollarSign &
\faNewspaper\; \faHistory &
\faMouse\; \faFile &
\faLink &
$\times$ \\

\textsc{ContestTrade} \cite{zhao_contesttrade_2025} & \faDollarSign &
\faNewspaper\; \faHistory\; \faLandmark &
\faMouse\; \faFile &
\faNetworkWired &
\checkmark \\

\textsc{TradingGroup} \cite{tian_tradinggroup_2025} & \faDollarSign &
\faNewspaper\; \faHistory\; \faFileInvoiceDollar\; \faSmile &
\faMouse\; \faFile &
\faNetworkWired &
$\times$ \\

\textsc{MarketSenseAI 2.0} \cite{fatouros2025marketsenseai} & \faDollarSign &
\faNewspaper\; \faHistory\; \faLandmark\; \faFileInvoiceDollar &
\faMouse\; \faFile &
\faNetworkWired\; \faRavelry\; \faLink &
$\times$ \\
\bottomrule
\end{tabular}

\vspace{3pt}
\raggedright
\footnotesize
\textbf{Notes:}
\faNewspaper = News,\;
\faHistory = Stock price history,\;
\faLandmark = Macro-market data,\;
\faFileInvoiceDollar = Financial statements/filings,\;
\faSmile = Market sentiment,\;
\faUserTie = Expert Opinions,\;
\faMouse = Buy/Sell/Hold decision,\;
\faFile = Summary / Rationale,\;
\faDollarSign = Trader role,\;
\faSearchDollar = Alpha Miner role,\;
\faMemory = Uses memory module,\;
\faNetworkWired = Multi-agent collaboration,\;
\faUserCircle = Multiple profile/character settings,\;
\faRavelry = RAG,\;
\faLink = Chain-of-Thought.\;
\checkmark / $\times$ = official code release linked from the paper or an author's page (checked October 2026); $\times^{*}$ = the linked repository contains no code yet.
\end{table*}
\section{Financial LLM Agent Selection and Categorization} \label{sec:taxonomy}
This section presents our approach to identifying, collecting, and categorizing financial LLM agents. We describe the paper selection criteria and dataset construction, then propose a taxonomy that classifies agents by architecture, functionality, and evaluation purpose, highlighting the traits that distinguish financial LLM agents from other domain-specific systems.

\subsection{Paper Collection and Selection Criteria}

This study focuses specifically on understanding and systematizing attacks targeting \textit{financial LLM agents}, including \textit{financial LLM trading agents} and \textit{finance-oriented LLM agents} that employ large language models for market analysis, autonomous trading, or financial decision-making. We followed a structured selection process inspired by the Preferred Reporting Items for Systematic Reviews and Meta-Analyses (PRISMA) guidelines~\cite{page2021prisma}.

\noindent\textbf{Identification.}
We systematically searched for papers published between 2023 and 2026 using keyword queries including ``financial LLM agents,'' ``LLM trading agents,'' ``LLM autonomous trading,'' and ``LLM financial decision-making.''
Our search covered publications from the \textit{USENIX Security Symposium}, the \textit{Network and Distributed System Security Symposium (NDSS)}, the \textit{ACM Conference on Computer and Communications Security (CCS)}, the \textit{IEEE Symposium on Security and Privacy (IEEE S\&P)}, the \textit{Conference on Neural Information Processing Systems (NeurIPS)}, and the \textit{International Conference on Machine Learning (ICML)}, as well as relevant preprints on arXiv. We also performed backward and forward citation snowballing on highly relevant papers to capture additional works not returned by keyword search. We targeted these venues because they publish the security and ML work that our framework builds on, and arXiv because most trading-agent systems first appear there. Snowballing added systems published elsewhere, such as FinAgent (KDD) and FLAG-Trader (ACL). The corpus therefore omits some trading agents that appear only in NLP, data-mining, or finance venues, such as HedgeAgents~\cite{li_hedgeagents_2025}, CryptoTrade~\cite{li2024cryptotrade}, and FinVision~\cite{fatemi2024finvision}. They reuse paradigms already in Table~\ref{tab:llm-agents-full} (multi-agent collaboration, reflective memory, and multimodal inputs), and the codebook applies to them unchanged. This identification phase yielded \textit{127 candidate papers}.

\noindent\textbf{Screening.}
Two reviewers independently screened all 127 papers by title and abstract, applying the following inclusion criteria: (I1)~the paper must propose, implement, or empirically evaluate a concrete financial LLM agent system; (I2)~the agent must perform market-facing tasks (e.g., trading, portfolio optimization, or alpha discovery); (I3)~the paper must be written in English and published or posted as a preprint between January 2023 and March 2026. Papers were excluded if they (E1)~only mentioned LLMs tangentially without an agent implementation, (E2)~focused exclusively on general LLM security without a financial application, or (E3)~were editorials, position papers, or surveys outside financial LLM agents, with no system contribution. Preprints passed the same inclusion, exclusion, and eligibility criteria as peer-reviewed papers, with no filter on citation count or venue. After screening, \textit{87 papers were excluded}, leaving \textit{40 papers} for full-text review.

\noindent\textbf{Eligibility and Inclusion.}
Both reviewers independently read all 40 full-text papers and assessed them against two additional eligibility criteria: (G1)~the agent architecture must be described in sufficient detail to apply our \frameworkShort{} metrics (i.e., information sources, reasoning pipeline, and output actions are identifiable); and (G2)~the system must include at least one form of autonomous or semi-autonomous decision-making rather than being purely a static prediction model. Of the 40 papers, 25 did not satisfy G2 (they are surveys, benchmarks, simulations, threat analyses, or defense frameworks with no autonomous decision-making) and are retained for background context in Section~\ref{categorization_of_agents}. The remaining \textit{15 papers} satisfy both G1 and G2, describe distinct, evaluable trading agent systems, and constitute our final evaluation set (Table~\ref{tab:llm-agents-full}).

In Section~\ref{categorization_of_agents}, we categorize these studies by their agent architecture, system design, and evaluation methodology to illustrate how the research community conceptualizes and analyzes financial LLM agents.

\begin{figure}[htbp]
\centering
\begin{tikzpicture}
\begin{axis}[
    width=0.7\linewidth,
    height=3cm,
    ybar,
    bar width=12pt,
    ylabel={\#Financial Agents},
    xlabel={Backbone LLM Model},
    symbolic x coords={GPT-4,GPT-4o,GPT-4o-mini,GPT-3.5,GPT-2,DeepSeek-V3},
    xtick=data,
    ymin=0,
    ymax=11,
    enlarge x limits=0.1,
    grid=both,
    grid style={dashed,gray!30},
    x tick label style={font=\small,rotate=45,anchor=east},
    y tick label style={font=\small},
    every node near coord/.append style={font=\footnotesize, anchor=south, color=black},
    fill=blue!45
]
\addplot coordinates {
    (GPT-4,10)
    (GPT-4o,1)
    (GPT-4o-mini,1)
    (GPT-3.5,1)
    (GPT-2,1)
    (DeepSeek-V3,1)
};
\end{axis}
\end{tikzpicture}
\caption{Distribution of backbone LLM models used across 15 financial LLM agents.}
\label{fig:llm-backbones}
\end{figure}

\subsection{Trading Agent Architectures and Ecosystem Context}
\label{categorization_of_agents}
\label{subsec:trading-architectures}

From the 40 papers retained during screening (Section~\ref{sec:taxonomy}), 25 are ecosystem studies that inform but are not themselves evaluable trading agents: surveys and SoK papers synthesize financial LLM tasks, inputs, and recurring designs~\cite{ding_large_2024,mahdavi_integrating_2025,liu_survey_2024,joshi_comprehensive_2025,batra_review_nodate,saha_large_2025,wang_sok_2025,raza_trism_2025}; benchmarks and evaluation suites probe agent-level attacks, financial metrics, and long-run robustness~\cite{zhang_agent_2025,li_investorbench_2024,chen_tasks_nodate,li_can_2025,emmanoulopoulos_trade_2025}; simulation testbeds provide controllable market environments~\cite{papadakis_stocksim_2025,gao_simulating_2024,noauthor_twinmarket_nodate}; and threat/defense analyses address domain-specific red teaming and architecture-level safeguards~\cite{cheng_uncovering_2025,shah_enforcing_2024}. While these works shape the broader landscape, our evaluation targets the remaining \textit{15 trading agent systems}, agents that autonomously transform market data into executable trading decisions. We define \textit{trading agents} as financial LLM agents that combine heterogeneous market inputs with LLM reasoning to produce trades and natural-language rationales. As detailed in Table~\ref{tab:llm-agents-full}, we organize them into six architectural paradigms.

\begin{table}[htbp!]
\centering
\caption{Reported attack success rates (ASR) for backbone LLMs of financial LLM agents: jailbreak benchmarks for the backbones that the papers report, and prompt injection for current frontier models that the released code now supports.}
\label{tab:jailbreak-asr}
\small
\setlength{\tabcolsep}{4pt}
\begin{tabular}{l p{3.6cm} r}
\toprule
\textbf{Backbone LLM} & \textbf{Attack (benchmark)} & \textbf{ASR} \\
\midrule
GPT-4 & Jailbreak (DAN~\cite{shen2024dan}) & 95\% \\
GPT-4o & Jailbreak (HarmNet~\cite{narula2025harmnet}) & 94.8\% \\
GPT-4o-mini & Jailbreak (GRAF~\cite{tang2025graf}) & 95.0\% \\
GPT-3.5-Turbo & Jailbreak (DAN~\cite{shen2024dan}) & 95\% \\
DeepSeek-V3 & Jailbreak (CNSafe-RT~\cite{ying2025deepseeksafety}) & up to 100\% \\
\midrule
GPT-5.3-Codex & Indirect injection (LivePI~\cite{zhao2026livepi}) & 27.2\% \\
Claude Opus 4.6 & Indirect injection (LivePI~\cite{zhao2026livepi}) & 10.7\% \\
Gemini 3.1 Pro & Indirect injection (LivePI~\cite{zhao2026livepi}) & 29.6\% \\
GPT-5.4 & Adaptive injection, transferred~\cite{abdelnabi2026agents} & 63.3\% \\
Gemini 3 Pro & Adaptive injection, transferred~\cite{abdelnabi2026agents} & 92.6\% \\
\bottomrule
\end{tabular}
\end{table}

\noindent\textbf{Single-Agent Memory \& Persona Systems.}
\textit{TradingGPT}~\cite{li2023tradinggpt} uses layered memory and task decomposition across cooperative roles for multi-horizon trading via GPT-4. \textit{FinMem}~\cite{yu_finmem_2023} refines this with \textit{character design} and adjustable cognitive span. Its self-evolution mechanism enables adaptive risk profiling but also creates persistent attack surfaces.

\noindent\textbf{Analyst \& Fundamental-Analysis Agents.}
\textit{MarketSenseAI}~\cite{fatouros2024can} (and \textit{MarketSenseAI 2.0}~\cite{fatouros2025marketsenseai}) applies chain-of-thought fundamental analysis over U.S. Securities and Exchange Commission (SEC) filings with minimal live API exposure. \textit{FinCon}~\cite{yu2025fincon} formalizes a manager–analyst hierarchy with feedback-driven risk control.

\noindent\textbf{Event-Driven \& Multimodal Agents.}
\textit{StockAgent}~\cite{zhang_when_2024} uses event-driven simulation with curated inputs. \textit{FinAgent}~\cite{zhang2024multimodal} targets cross-asset generality via tool-augmented, multimodal reasoning with broad API integration.

\noindent\textbf{Platform \& Alpha-Discovery Agents.}
\textit{FinRobot}~\cite{yang_finrobot_2024} provides a composable multi-LLM platform with modular design. \textit{Alpha-GPT 2.0}~\cite{yuan_alpha-gpt_2024} focuses on alpha discovery with human-in-the-loop oversight and no live execution, one of the most secure designs.

\noindent\textbf{Low-Latency \& Expert-Fusion Agents.}
\textit{QuantAgent}~\cite{xiong_quantagent_2025} integrates classical technical features with a dedicated risk-monitoring agent for HFT-like settings, achieving the strongest robustness profile. \textit{TradExpert}~\cite{ding2024tradexpert} uses a mixture-of-experts architecture fusing specialized LLM outputs.

\noindent\textbf{Multi-Agent Collaborative Systems.}
\textit{TradingAgents}~\cite{xiao_tradingagents_2025} is the most comprehensive system (over 100k GitHub stars as of October 2026), scaling the firm metaphor with analyst/researcher/risk teams and debate–synthesis, but introduces $>$240\,s decision latency. \textit{FLAG-Trader}~\cite{xiong_flag-trader_2025} blends LLM deliberation with reinforcement learning. \textit{ContestTrade}~\cite{zhao_contesttrade_2025} uses internal competition to mitigate overfitting. \textit{TradingGroup}~\cite{tian_tradinggroup_2025} probes group-scale collaboration with configurable risk parameters.

\noindent\textbf{Cross-Cutting Observations.}
Three convergences emerge across Table~\ref{tab:llm-agents-full}: (i) most systems ingest news, fundamentals, and price data to output trades with natural-language rationales; (ii) modular multi-agent structures and reflective memory modules dominate; (iii) RAG over SEC filings increasingly shapes model reasoning.

To understand what makes financial LLM agents fundamentally different, we first examined how agentic LLM systems are structured across other high-stakes domains such as law, healthcare, and education. Prior studies show that legal agents emphasize traceability and precedent grounding \cite{zhou2024lawgpt,zhu2023judgelm}, healthcare agents prioritize safety, hallucination control, and clinical validation \cite{tang2024medagents,johri2025evaluation}, and educational agents focus on pedagogical reliability and personalized instruction \cite{liedullm}. Across these domains, risks usually stem from instruction-following errors, hallucinated reasoning, or misalignment with domain constraints. But none involve real-time market microstructure or the reflexive, latency-sensitive execution loops that trading agents operate in.

\begin{figure}[t]
  \centering
  \includegraphics[width=0.4\textwidth]{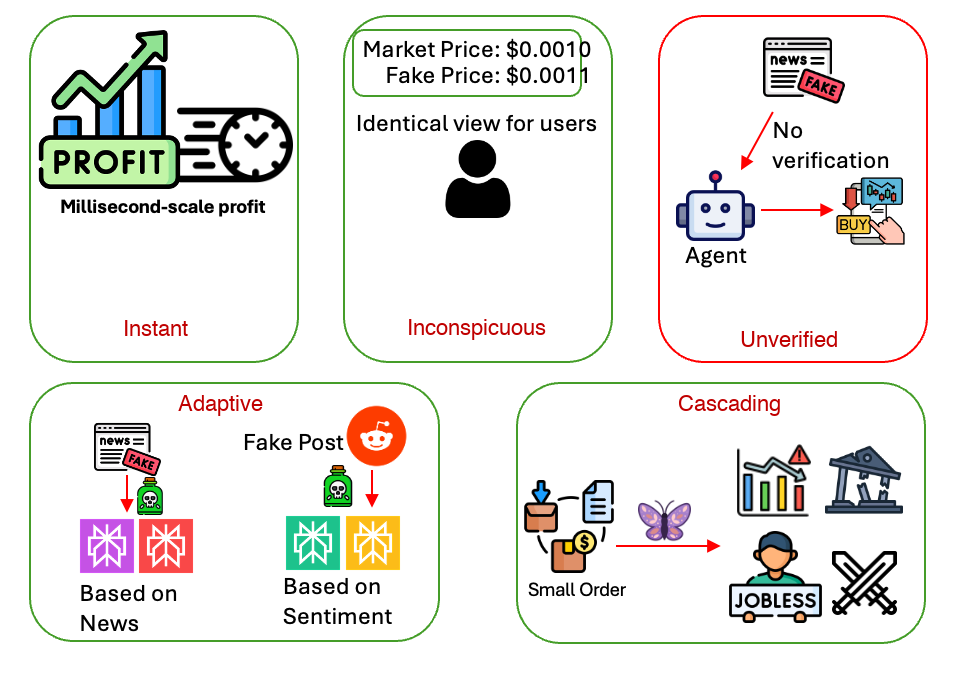}
  \caption{Distinct attack characteristics of financial LLM agents compared to other domain-specific agent systems.}
  \label{fig:attack_demos}
\end{figure}

By contrasting these domain patterns with our analysis of 15 major financial LLM trading systems, we identified a cluster of properties unique to financial agents, shaped by market speed, heterogeneous signal dependence, execution autonomy, and systemic reflexivity. As illustrated in Figure~\ref{fig:attack_demos}, financial LLM agents exhibit five defining characteristics (Appendix~\ref{appendix:distinct-characteristics} ties each one to the tables that support it):  
(i) \textit{Instant} attack windows, since markets reprice within milliseconds and an attacker can profit from a transient dislocation before a deliberating agent or a human reviewer responds;  
(ii) \textit{Inconspicuous} micro-manipulations that resemble legitimate market signals;  
(iii) \textit{Unverified} ingestion of external data due to speed-first execution pipelines;  
(iv) \textit{Adaptive} multi-source exploitation enabled by heterogeneous signal integration; and  
(v) \textit{Cascading} amplification, where small perturbations propagate system-wide through inter-agent reflexivity.
These properties combine into an attack surface that is fragile and easily amplified, and that differs fundamentally from other domain-specific agent systems.

We also analyze the backbone LLMs underlying current financial agents to understand how their foundational vulnerabilities propagate into trading environments.
As shown in Figure~\ref{fig:llm-backbones}, the papers report the \textit{GPT-4} family (GPT-4, GPT-4o, and GPT-4o-mini) as the backbone of twelve of the fifteen analyzed systems, with plain \textit{GPT-4} alone powering ten of them. For multi-model implementations, we selected the most capable model that each paper reports. Two systems use open-weight backbones (\textit{GPT-2} and \textit{DeepSeek-V3}), and the remaining thirteen depend on closed-source commercial APIs.
The released code has since moved on: four of the seven open-source repositories ship configurations for frontier models such as GPT-5.x, Claude~4.x, and Gemini~3.x, and TradingAgents now defaults to GPT-6 Sol. Because \frameworkShort{} rates controls that sit outside the model, these upgrades leave Table~\ref{tab:combined-eval} unchanged, and newer backbones do not close the attack surfaces.
As summarized in Table~\ref{tab:jailbreak-asr}, both \textit{GPT-4} and \textit{GPT-4o} exceed 94\% jailbreak ASR on standard benchmarks, \textit{GPT-4o-mini} still exhibits a 95.0\% ASR under adaptive multi-turn jailbreaking, and \textit{DeepSeek-V3} reaches up to 100\%~\cite{ying2025deepseeksafety}. For current frontier models, indirect prompt injection, the attack most relevant to agents, still succeeds in 10.7\% to 29.6\% of LivePI cases against GPT-5.3-Codex, Claude Opus~4.6, and Gemini~3.1 Pro~\cite{zhao2026livepi}, and adaptive attacks fare far better (Table~\ref{tab:jailbreak-asr})~\cite{abdelnabi2026agents}.
These are base-model or generic-agent rates, which system prompts, tool constraints, and multi-agent validation may lower, and the jailbreak rates concern harmful content rather than trading, so we treat them only as indicators of how easily a backbone's instructions can be overridden.

\begin{table}[!t]
\centering
\caption{Robustness and security evaluation of financial LLM agents under the \frameworkShort{} framework. The table summarizes robustness and security levels across nine metrics: R1 (Crash Foresight and Early Awareness), R2 (Rapid Reaction Capability under Market Shocks), R3 (Loss Containment and Stop-Loss Execution), R4 (Recovery and Post-Crash Adaptation), S1 (Attacks on Market-Relevant Information Sources), S2.1 (Prompt Injection \& Instruction Override), S2.2 (Tool-Use \& API Exploitation), S2.3 (Memory or Context Manipulation), and S3 (Agent-as-Attacker Behavior).}
\label{tab:combined-eval}

\small
\setlength{\tabcolsep}{3pt}
\begin{tabular}{l cccc|ccccc}
\toprule
\textbf{Agent System} & \textbf{R1} & \textbf{R2} & \textbf{R3} & \textbf{R4} & \textbf{S1} & \textbf{S2.1} & \textbf{S2.2} & \textbf{S2.3} & \textbf{S3} \\
\midrule
TradingGPT        & \ding{109} & \ding{109} & \ding{109} & \ding{109} & \ding{109} & \ding{109} & \ding{117} & \ding{109} & \ding{109} \\
FinMem            & \ding{117} & \ding{109} & \ding{117} & \ding{117} & \ding{109} & \ding{109} & \ding{117} & \ding{109} & \ding{109} \\
MarketSenseAI     & \ding{109} & \ding{109} & \ding{109} & \ding{109} & \ding{109} & \ding{117} & \ding{108} & \ding{108} & \ding{117} \\
FinCon            & \ding{109} & \ding{109} & \ding{109} & \ding{109} & \ding{109} & \ding{117} & \ding{117} & \ding{117} & \ding{117} \\
StockAgent        & \ding{109} & \ding{109} & \ding{109} & \ding{109} & \ding{117} & \ding{117} & \ding{117} & \ding{117} & \ding{109} \\
FinAgent          & \ding{109} & \ding{109} & \ding{109} & \ding{117} & \ding{109} & \ding{117} & \ding{117} & \ding{117} & \ding{117} \\
FinRobot          & \ding{109} & \ding{109} & \ding{109} & \ding{109} & \ding{117} & \ding{117} & \ding{117} & \ding{117} & \ding{108} \\
Alpha-GPT 2.0     & \ding{117} & \ding{109} & \ding{117} & \ding{117} & \ding{117} & \ding{108} & \ding{108} & \ding{117} & \ding{108} \\
QuantAgent        & \ding{117} & \ding{117} & \ding{117} & \ding{117} & \ding{117} & \ding{117} & \ding{117} & \ding{117} & \ding{109} \\
TradExpert        & \ding{109} & \ding{109} & \ding{109} & \ding{109} & \ding{109} & \ding{117} & \ding{117} & \ding{108} & \ding{117} \\
TradingAgents     & \ding{117} & \ding{117} & \ding{117} & \ding{117} & \ding{117} & \ding{117} & \ding{117} & \ding{117} & \ding{109} \\
FLAG-Trader       & \ding{109} & \ding{109} & \ding{109} & \ding{109} & \ding{117} & \ding{117} & \ding{117} & \ding{117} & \ding{117} \\
ContestTrade      & \ding{109} & \ding{109} & \ding{117} & \ding{117} & \ding{117} & \ding{117} & \ding{117} & \ding{117} & \ding{117} \\
TradingGroup      & \ding{117} & \ding{117} & \ding{117} & \ding{117} & \ding{117} & \ding{117} & \ding{117} & \ding{109} & \ding{109} \\
MarketSenseAI 2.0 & \ding{109} & \ding{109} & \ding{109} & \ding{109} & \ding{109} & \ding{117} & \ding{108} & \ding{108} & \ding{117} \\
\bottomrule
\end{tabular}

\vspace{4pt}
\noindent \ding{108} \textbf{(Fully Robust / Low Risk)}
\ding{117} \textbf{(Partially Robust / Medium Risk)}
\ding{109} \textbf{(Not Robust / High Risk)}

\vspace{2pt}
\noindent\footnotesize{\textit{Per-agent justifications for all partial (\ding{117}) and fully robust / low-risk (\ding{108}) scores are detailed in Appendix~\ref{appendix:metric-justifications}. Scoring criteria follow the codebook in Appendix~\ref{appendix:codebook} (Table~\ref{tab:codebook_combined}). Every security \ding{108} here reflects a surface that is absent by design, not an active defense.}}
\normalsize
\end{table}

\section{Systematization Results}
\label{sec:results}
This section reports the ratings that the \frameworkShort{} framework of Section~\ref{framework} assigns to the 15 schemes. We focus on two aspects: the robustness of financial LLM agents under market shocks and their security against diverse adversarial attacks. A \ding{109} means that the paper does not evidence the control, not that a running system lacks it.

\begin{figure}[t]
  \centering
  \includegraphics[width=\columnwidth]{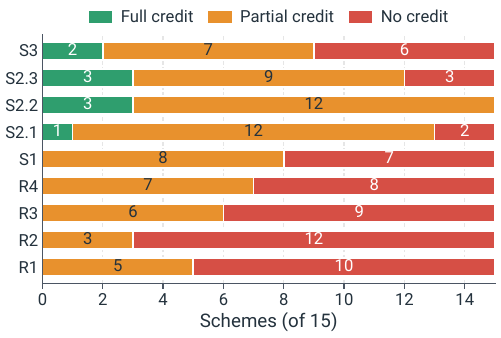}
  \caption{Distribution of the paper-level ratings of Table~\ref{tab:combined-eval} across the 15 schemes. No scheme earns full credit on any robustness metric or on S1, and every security full credit comes from a surface that the design removes.}
  \label{fig:rating-distribution}
\end{figure}

Figure~\ref{fig:rating-distribution} condenses Table~\ref{tab:combined-eval} into one view. The robustness metrics are the weakest: \vtwoNoneRobustCells{} of the \vtwoRobustCells{} robustness ratings give no credit, and none gives full credit. Security looks better only because partial credit is common, and full credit appears in \vtwoFullSecurityCells{} of \vtwoSecurityCells{} security ratings, each for a surface that the design leaves out.

\subsection{Findings on the Robustness of Financial LLM Agents}

Table~\ref{tab:combined-eval} summarizes the robustness ratings of fifteen financial LLM-based trading systems across four resilience dimensions (R1–R4). Overall, robustness remains weak: no scheme earns full credit on any robustness metric, 12 of the 15 receive no credit on at least one, and those that earn partial credit do so through ad-hoc or conceptual mechanisms without complete automation, verification, or closed-loop control.

\noindent\textbf{Analysis of Crash Foresight and Early Awareness.}
Five agents earn partial R1 scores, all through qualitative or reactive mechanisms with no structured quantitative foresight.
\textit{FinMem} adjusts trading aggressiveness after realized drawdowns, which is a purely reactive mechanism.
\textit{Alpha-GPT 2.0} uses offline human-in-the-loop oversight during research cycles, with no continuous anomaly surveillance in live trading.
\textit{QuantAgent} includes a risk monitoring agent but lacks concrete metrics (e.g., VIX tracking, correlation dislocation).
\textit{TradingAgents} and \textit{TradingGroup} assess the direction of the market regime from multi-source sentiment, a high-level awareness with no quantitative foresight.
None implement automated anomaly-detection pipelines comparable to traditional quantitative systems (e.g., VIX-based alerts, entropy divergence, principal component analysis (PCA)-based volatility clustering), revealing a consistent bias toward qualitative interpretation over measurable early warning.

\noindent\textbf{Analysis of Rapid Reaction Capability.}
Only \textit{QuantAgent}, \textit{TradingAgents}, and \textit{TradingGroup} partially address rapid reaction.
\textit{QuantAgent} can trigger alerts upon abnormal volatility but is bottlenecked by multi-agent inference latency.
\textit{TradingAgents} includes a risk management team that can override risky behavior, yet the team acts through message-passing and debate, with no direct execution path.
\textit{TradingGroup} offers configurable stop-loss and take-profit thresholds but lacks autonomous shock detection or circuit-breaking logic.
All systems prioritize analytical deliberation over reflexive action. None implement fast-path interrupt mechanisms, safe-mode pipelines, or hardware-level latency reduction.

As summarized in Tables~\ref{tab:latency-distribution} and~\ref{tab:latency-comparison}, all 15 systems exhibit LLM inference delay and I/O latency, while 10 add multi-agent communication overhead. The most widely used framework, \textit{TradingAgents}, introduces over four minutes of total decision latency, orders of magnitude slower than algorithmic or high-frequency trading requirements.
These latency gaps translate into critical vulnerabilities: \textit{(1) price staleness}, where agents act on outdated information; \textit{(2) stop-loss failures}, where delayed risk-control activation leads to uncontrolled drawdowns; and \textit{(3) slippage and execution drift}, where I/O and API delays cause execution prices to deviate from targets.

\begin{table}[htbp]
\centering
\caption{Occurrence of major latency sources across 15 financial LLM trading agents.}
\label{tab:latency-distribution}
\begin{tabular}{l c}
\toprule
\textbf{Latency Source} & \textbf{Count} \\
\midrule
LLM Inference Delay & 15/15 \\
Agent Communication Overhead & 10/15 \\
I/O \& API Latency & 15/15 \\
\bottomrule
\end{tabular}
\end{table}

\begin{table}[htbp]
\centering
\caption{Latency comparison between traditional trading styles and representative LLM trading agents.}
\label{tab:latency-comparison}
\begin{tabular}{p{4cm} p{2.2cm} p{0.8cm}}
\toprule
\textbf{Trading Type/Agent} & \textbf{Typical Latency} & \textbf{Src.} \\
\midrule
High-Frequency Trading (HFT) & $<$0.0001 s \cite{Moallemi2009,Hasbrouck2013} & [R] \\
Algorithmic Trading & $<$0.1\text{s}\cite{LuxAlgoLatency,CoinAPI_CryptoLatency} & [R] \\
Intraday Trading & $\sim$1 s \cite{LuxAlgoLatency} & [R] \\
Daily / Swing Trading & $\sim$10 s \cite{LuxAlgoLatency} & [R] \\
\textbf{TradingAgents} & $\mathbf{>240}$ \text{s} & [M] \\
\bottomrule
\end{tabular}

\vspace{2pt}
\noindent\footnotesize{[R] = Reported in cited literature; [M] = Measured: median wall-clock latency over five end-to-end runs of the open-source TradingAgents release in its fastest configuration (smallest supported backbone and shortest chain-of-thought), and therefore a lower bound. See Section~\ref{subsec:trading-architectures} for architectural details.}\normalsize
\end{table}

\noindent\textbf{Analysis of Loss Containment and Stop-Loss Execution.}
Six agents earn partial R3 scores, but none formalize containment into deterministic enforcement.
\textit{FinMem} adapts risk modes after performance declines yet specifies no drawdown thresholds or liquidation triggers.
\textit{Alpha-GPT 2.0} and \textit{QuantAgent} acknowledge risk management roles but lack algorithmic stop-loss execution.
\textit{TradingAgents} achieves smaller drawdowns empirically through its risk management team, but without transparent liquidation logic.
\textit{ContestTrade} downranks poorly performing agents only \emph{after} losses materialize, and \textit{TradingGroup} offers configurable stop-loss settings without dynamic volatility-adjusted calibration.
Across all systems, containment remains discretionary: none implement rule-based position limits, volatility-linked exposure scaling, or forced liquidation protocols, showing that current LLM agents treat \textit{risk control} as a concept but rarely enforce it as a hard limit.

\noindent\textbf{Analysis of Recovery and Post-Crash Adaptation.}
R4 shows the most partial scores, with seven agents displaying forms of adaptive learning or reflection, yet all rely on \textit{continuous learning}, and none has \textit{crash-triggered recovery}.
\textit{FinMem} updates memory representations over time but lacks event-triggered retraining to distinguish normal drift from tail events.
\textit{FinAgent} and \textit{Alpha-GPT 2.0} enable post-hoc behavioral adjustments (via self-reflection and human guidance, respectively) but provide no automated crash-recovery pipeline.
\textit{QuantAgent}’s two-loop optimization gradually integrates outcomes yet cannot accelerate recalibration after extreme losses, and \textit{TradingAgents} shares this limitation.
\textit{ContestTrade} promotes surviving agents via competition, but this selection process fails under systemic crashes that degrade all agents simultaneously.
\textit{TradingGroup} synthesizes data for continuous model updates, but recovery remains passive, with no explicit trigger.
None implement structured fault analysis, regime reclassification, or volatility-dependent retraining triggers.

\subsection{Findings on the Security of Financial LLM Agents}

Table~\ref{tab:combined-eval} summarizes the comparative security risk levels of fifteen financial LLM-based trading systems under the \frameworkShort{} framework. Full-credit (\ding{108}) security ratings are rare: none on S1, one on S2.1, three each on S2.2 and S2.3, and two on S3, and each of them goes to a scheme that removes the surface by design. None exhibit end-to-end validation, isolation, or post-compromise containment, underscoring a systematic neglect of proactive security design in current financial LLM ecosystems.

\noindent\textbf{Analysis of Attacks on Market-Relevant Information Sources.}
Trading agents that continuously consume external market data or sentiment feeds are especially exposed to data-layer manipulation. \textit{TradingGPT}, \textit{FinMem}, \textit{MarketSenseAI}, \textit{MarketSenseAI 2.0}, \textit{FinCon}, \textit{FinAgent}, and \textit{TradExpert} receive \ding{109} on S1: they ingest news, social sentiment, macroeconomic indicators, and fundamental documents from heterogeneous APIs and RAG pipelines, yet their papers evidence none of the four S1 controls (provenance tracking, cross-source corroboration, structured aggregation, or curation). Although many of these systems deliberately diversify inputs and stage multi-round debates to avoid dependence on a single feed, such mechanisms primarily filter noise and disagreement rather than targeted misinformation. A poisoned but plausible source can still be amplified into trading theses and written back into memory. 

The remaining eight schemes receive \ding{117}. Six of them, \textit{TradingAgents}, \textit{TradingGroup}, \textit{FinRobot}, \textit{QuantAgent}, \textit{FLAG-Trader}, and \textit{ContestTrade}, earn it through structured aggregation such as voting, ranking, debate, or policy learning, which mitigates but does not eliminate the impact of corrupted feeds. \textit{Alpha-GPT 2.0} and \textit{StockAgent} earn it through curated or simulated inputs. None of the 15 papers records source provenance or requires cross-source corroboration, the two indicators that a full S1 rating needs, so the data layer remains the primary attack surface.

\noindent\textbf{Analysis of Attacks on Agent Logic and Prompts.}
Prompt-layer and logic-layer threats are pervasive but often under-addressed. In S2.1 (Prompt-Injection and Instruction Override), we rate \textit{TradingGPT} and \textit{FinMem} as \textit{high risk}: both expose natural-language coordination channels that accept unverified inputs, enabling adversaries to override instructions or seed malicious reasoning. By contrast, \textit{Alpha-GPT 2.0} (an alpha-focused, idea-generation design) is \textit{low risk} in S2.1 because it operates on curated inputs rather than open prompt channels. \textit{StockAgent} is \textit{medium risk}, since its simulated inputs reduce but do not eliminate exposure to prompt-layer manipulation.
In S2.2 (Tool-Use and API Exploitation), \textit{MarketSenseAI}, \textit{MarketSenseAI 2.0}, and \textit{Alpha-GPT 2.0} are assessed as \textit{low risk} because their pipelines emphasize static document retrieval and analysis rather than live API-driven tool invocation, reducing the attack surface for real-time API spoofing or tool misuse. Systems that do invoke external tools remain more exposed, since injected prompts can escalate into unauthorized API actions.

In S2.3 (Memory or Context Manipulation), we identify \textit{TradingGPT}, \textit{FinMem}, and \textit{TradingGroup} as the \textit{high risk} group: their persistent memory and self-updating pipelines mean injected content can be stored, reinforced, and replayed across future decision cycles. Conversely, \textit{MarketSenseAI}, \textit{MarketSenseAI 2.0}, and \textit{TradExpert} are \textit{low risk} for memory attacks because they lack long-term, self-evolving memory layers. The most dangerous failures arise when prompt, tool, and memory weaknesses interact and no provenance, isolation, or sanitization control stands between them.

\noindent\textbf{Analysis of Agent-as-Attacker Behavior.}
Once compromised, agents that carry order-placement authority pose the greatest threat. In \textit{S3 (Agent-as-Attacker)}, we identify \textit{TradingGPT}, \textit{FinMem}, \textit{StockAgent}, \textit{QuantAgent}, \textit{TradingAgents}, and \textit{TradingGroup} as \textit{high risk}. Five interact with live markets through execution APIs or automated pipelines. \textit{StockAgent} instead trades inside a multi-agent simulation, but through a shared order book with no isolation, so a compromised instance can still steer peer agents and downstream artifacts trained on that market. Once the internal logic or prompt layer of any of these agents is compromised, an attacker could issue real trades, manipulate orders, or coordinate malicious strategies across multiple sub-agents, amplifying both velocity and market impact. 

Agents such as \textit{FinRobot} and \textit{Alpha-GPT 2.0} are comparatively \textit{low risk}, as their design intentionally excludes any live execution or external trading connection. They operate strictly as analytical or simulation frameworks. The remaining agents, which can interface with market APIs but are not inherently autonomous, fall into a \textit{medium risk} tier: their exposure depends on whether external tools or execution endpoints are activated at runtime.  

\noindent\textbf{Insights on Robustness.}
As summarized in Table~\ref{tab:combined-eval}, none of the evaluated agents achieved full robustness. Analytical sophistication does not translate into resilience: ten schemes receive no credit on R1 and twelve none on R2, and none has a crash-triggered recovery protocol. Partial ratings rise from five on R1 and three on R2 to six on R3 and seven on R4, indicating limited attention to adaptation but a persistent absence of proactive mechanisms such as volatility-aware controls or circuit-breaker logic.

\noindent\textbf{Insights on Security.}
High-risk ratings fall from seven on S1 to six on S3, but this apparent improvement is misleading: no system actively defends against the pervasive attack vectors (prompt injection, data poisoning, tool misuse). Securing these agents requires defense-in-depth: authenticated data sources, sandboxed reasoning, restricted tool privileges, and multi-layer anomaly detection. Across the schemes, greater agent autonomy expands the attack surface, while strict security constraints reduce autonomy. Appendix~\ref{appendix:investment-schemes} compares these agents with passive-income schemes.

\subsection{Code-Level Validation of Open-Source Schemes}
\label{subsec:code-audit}
Table~\ref{tab:combined-eval} rates each scheme from its published design, so we also asked whether the released implementations behave as the papers describe. Of the \auditSchemesTotal{} schemes, \auditSchemes{} have an official code release. FinCon's repository is a placeholder, and the remaining \auditNoCode{} schemes have none. We audited the \auditSchemes{} releases statically at pinned commits with the same codebook. Claude Sonnet~5.5, given our paper-level ratings to locate the described mechanisms, labeled every indicator with file-and-line evidence in \auditRuns{} independent runs per repository, deterministic pattern matching checked the code paths that decide the ratings (order submission, code execution, persistence, and risk limits), and a separate reading of the cited source lines ruled on every cross-run disagreement (Appendix~\ref{appendix:code-audit}). The runs agreed on \auditUnanimous{} of \auditRatings{} ratings (\auditUnanimousPct{}) and on \auditIndAgree{} of \auditIndTotal{} indicator judgments (\auditIndAgreePct{}). The pattern scan, which ran independently of the LLM, found no order-submission or broker call in any release and located the host-side \texttt{exec} calls in ContestTrade that weakness (i) below describes. Table~\ref{tab:code-audit} lists the resulting code-level ratings.

\begin{table}[t]
\centering
\caption{Code-level ratings of the seven schemes with an official code release, derived from three independent audit runs per repository (pinned commits in Table~\ref{tab:audit-commits}). $\uparrow$ ($\downarrow$) marks a rating that is better (worse) than the paper-level rating in Table~\ref{tab:combined-eval}.}
\label{tab:code-audit}
\small
\setlength{\tabcolsep}{3pt}
\begin{tabular}{l cccc|ccccc}
\toprule
\textbf{Scheme} & \textbf{R1} & \textbf{R2} & \textbf{R3} & \textbf{R4} & \textbf{S1} & \textbf{S2.1} & \textbf{S2.2} & \textbf{S2.3} & \textbf{S3} \\
\midrule
FinMem & \ding{117} & \ding{109} & \ding{117} & \ding{117} & \ding{117}$^{\uparrow}$ & \ding{117}$^{\uparrow}$ & \ding{108}$^{\uparrow}$ & \ding{117}$^{\uparrow}$ & \ding{108}$^{\uparrow}$ \\
StockAgent & \ding{109} & \ding{109} & \ding{117}$^{\uparrow}$ & \ding{109} & \ding{117} & \ding{117} & \ding{108}$^{\uparrow}$ & \ding{117} & \ding{109} \\
FinAgent & \ding{117}$^{\uparrow}$ & \ding{109} & \ding{117}$^{\uparrow}$ & \ding{117} & \ding{117}$^{\uparrow}$ & \ding{117} & \ding{108}$^{\uparrow}$ & \ding{117} & \ding{108}$^{\uparrow}$ \\
FinRobot & \ding{117}$^{\uparrow}$ & \ding{109} & \ding{109} & \ding{117}$^{\uparrow}$ & \ding{117} & \ding{117} & \ding{117} & \ding{117} & \ding{108} \\
QuantAgent & \ding{117} & \ding{109}$^{\downarrow}$ & \ding{109}$^{\downarrow}$ & \ding{109}$^{\downarrow}$ & \ding{117} & \ding{117} & \ding{108}$^{\uparrow}$ & \ding{108}$^{\uparrow}$ & \ding{108}$^{\uparrow}$ \\
TradingAgents & \ding{117} & \ding{117} & \ding{117} & \ding{117} & \ding{117} & \ding{117} & \ding{108}$^{\uparrow}$ & \ding{117} & \ding{108}$^{\uparrow}$ \\
ContestTrade & \ding{117}$^{\uparrow}$ & \ding{109} & \ding{117} & \ding{117} & \ding{117} & \ding{117} & \ding{117} & \ding{117} & \ding{108}$^{\uparrow}$ \\
\bottomrule
\end{tabular}
\end{table}

\begin{figure}[t]
  \centering
  \includegraphics[width=\columnwidth]{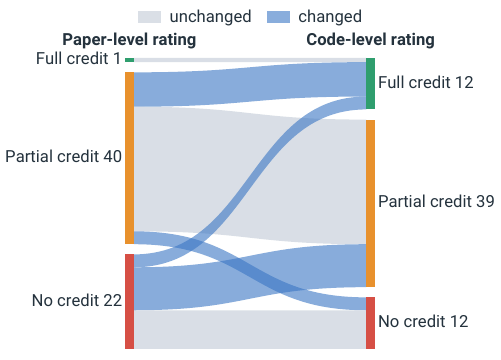}
  \caption{Movement of the \auditRatings{} ratings of the \auditSchemes{} open-source schemes from the paper (left) to the released code (right). Blue flows mark ratings that change: \auditCodeBetter{} rise and \auditCodeWorse{} fall. Of the rises, \auditBetterNoExec{} come from releases that contain no order path, and all falls belong to QuantAgent.}
  \label{fig:paper-vs-code}
\end{figure}

\noindent\textbf{Paper and code diverge in both directions.} Code-level and paper-level ratings agree on \auditMatch{} of \auditRatings{} ratings (\auditMatchPct{}), and Figure~\ref{fig:paper-vs-code} traces where the other \the\numexpr\auditCodeBetter+\auditCodeWorse\relax{} move. Of these, \vtwoNoneToPartial{} rise from no credit to partial credit, \vtwoPartialToFull{} from partial to full credit, \vtwoNoneToFull{} from no credit to full credit, and \vtwoPartialToNone{} fall from partial to no credit. Of the \auditCodeBetter{} ratings where the code scores better, \auditBetterNoExec{} fall on S2.2 and S3 and share one cause: no release contains an order-execution path, so \auditSThreeFull{} of the \auditSchemes{} releases satisfy the no-execution indicator of S3 (all but StockAgent, whose agents share an order book). These research prototypes emit trading decisions or run backtests, and the execution surface that Table~\ref{tab:combined-eval} rates returns once a deployer wires their output to a brokerage account. The opposite gap is more worrying. All \auditCodeWorse{} code-worse ratings belong to QuantAgent, whose release contains neither the risk agent nor the stop-loss that its paper describes. ContestTrade's contest mechanism and FinMem's risk-averse mode are likewise inactive in the released pipelines. Their R3 ratings in Table~\ref{tab:code-audit} survive only because this dormant code counts as a partial indicator (both drop to \ding{109} under the strict convention), so Table~\ref{tab:combined-eval} overstates what these releases do.

\noindent\textbf{The code exposes weaknesses that the papers do not show.} None of the \auditSchemes{} releases enforces a stop-loss, position cap, or drawdown limit outside the LLM. The only stop-loss values in the code are a field that the model fills into its report (TradingAgents) and a strategy that the pipeline never calls (FinAgent). Beyond these missing controls, the audit found four concrete weaknesses, each tied to a surface of \S\ref{framework}.
\textit{(i) Model-written code runs on the host.} ContestTrade's default stock-selection tool extracts Python from the model's reply and runs it with \texttt{exec} in the host process, and FinRobot's default assistants execute model-written code locally with neither a sandbox nor human approval. Because news and social-media text share the context with these tools, an injected headline can turn into code execution on the operator's machine, which chains S1 into the worst S2.2 outcome.
\textit{(ii) Secrets reach the model or the logs.} If an SEC filing download fails, a FinRobot data tool returns the exception text, which contains the request URL with the API token, to the model, and FinAgent prints its market-data API key to standard output.
\textit{(iii) Memory is stored without validation.} FinMem, FinAgent, FinRobot, and TradingAgents write memory or retrieval stores to disk and reuse them without checking what is written. FinMem and FinAgent load these stores with \texttt{pickle}, so a tampered memory file executes code when it is loaded. The codebook does not score deserialization, so Table~\ref{tab:code-audit} omits this risk, and FinMem's higher S2.3 rating there reflects memory decay and feedback weighting, not write validation. TradingAgents appends model-written text to a decision log whose entry delimiter the model can reproduce, so injected content can forge a past decision and its lesson for later runs (S2.3).
\textit{(iv) Peers are trusted.} In StockAgent, every agent's free-text forum post reaches all other agents' next-day prompts unfiltered, and the market price is simply the last matched price, so one agent can broadcast false signals and move the price that every other agent reads (S3).
Paper-level and code-level assessments therefore answer different questions, and a deployer needs both: the paper states what the design intends, and the code shows what the deployer actually inherits.

\section{A Flash-Crash Experiment}
\label{sec:market-experiment}
Section~\ref{sec:results} shows which controls the 15 designs lack. No published design halts trading without an LLM call (R2) or checks both the provenance of its news and a second source (S1), and no code release enforces a stop-loss or position limit outside the model (R3). Ratings cannot show the cost of these gaps, because the harm appears once many agents trade the same stock.

We therefore measure how three LLMs trade one position after a fake headline and after four real hoaxes (\S\ref{subsec:sim-probe}), and we let populations of minimal agents that decide as these models do trade in a simulated market (\S\ref{subsec:sim-market}). We then record when the agents cause or amplify a flash crash (\S\ref{subsec:sim-crash}), switch on the three missing controls and the trading pause of U.S. exchanges (\S\ref{subsec:sim-controls}), and vary the market, the agents, and the policy (\S\ref{subsec:sim-robust}). The agents keep the two properties that the ratings and the code audit found across designs, an LLM that makes each trading decision and a decision cycle as long as one round of LLM reasoning, and we drop the rest of each design on purpose.

\subsection{Trading Decisions of Three LLMs}
\label{subsec:sim-probe}
To measure decisions, we gave gpt-5.6-luna, deepseek-v4-flash, and gemini-3.5-flash 108 market snapshots each, three times per snapshot at temperature 0.7 (\probeCalls{} calls per model). Each snapshot gives the stock, the last price, the one-minute prices and the change over the last five minutes, and the position, which is 300 shares with their average entry price and unrealized profit and loss. It also states a 600-share position limit, a hard cap on how many shares the account may hold, and contains either no headline or one unsourced report that the company is under SEC investigation over accounting irregularities.

The system prompt asks for one order in JSON, BUY, SELL, or HOLD with a share count. The grid crosses nine five-minute price changes from $-6\%$ to $+6\%$, the two headline states, two position states (just bought, or 3\% under water), and three stocks (MSFT, JPM, and XOM). We chose these states because an agent passes through them in a crash: a quiet market, a falling price, a headline, and a position that has already lost money.

\begin{figure}[t]
\centering
\includegraphics[width=\columnwidth]{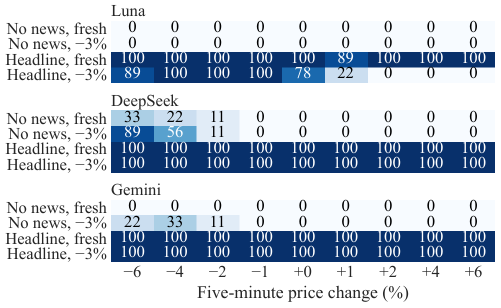}
\caption{Share of sell answers (\%) of the three LLMs by five-minute price change, headline, and position (fresh or 3\% under water), nine answers per entry.}
\label{fig:policy-heatmap}
\end{figure}

As Figure~\ref{fig:policy-heatmap} shows, all three models sell on the headline, and they sell the whole position. Luna sells in \lunaSellNews{} of headline cases, DeepSeek in \dsSellNews{}, and Gemini in \gemSellNews{}, and each sell order covers all or nearly all 300 shares. On price changes alone, however, the models differ. Luna does not trade without a headline (\lunaTradeNoNews{} of such cases), whereas DeepSeek sells after large falls once its position is under water (\dsSellDropUnder{} of cases at $-6\%$). Gemini buys dips on a fresh position (in up to \gemBuyDipMax{} of cases), yet it also sells after falls once under water (up to \gemSellDropUnderMax{}). DeepSeek and Gemini buy \dsBuySize{} and \gemBuySize{} shares on average. These differences matter in the market, because a model that sells on falling prices can amplify a crash that it did not start, whereas a model that ignores prices can start a crash but cannot amplify one.

The sales follow the headline itself, whatever its wording and the prompt. A second probe gave Luna and DeepSeek a guidance cut or a CEO resignation in place of the SEC report, and a trader prompt in the style of multi-role designs, in which analysts summarize the snapshot and a risk manager urges capital protection, at five-minute changes of $-2\%$, 0, and $+2\%$ (\robProbeCases{} cases per headline and prompt). DeepSeek sold in at least \robSellMinDs{} and Luna in at least \robSellMinLuna{} of the \robProbeCases{} cases for each headline and prompt. Without a headline, the trader prompt sold in \robTraderNoNewsSellMin{} to \robTraderNoNewsSellMax{} of \robProbeCases{} cases. Most sales therefore come from the headline itself. A multi-role design that ends in one LLM trader therefore inherits the response that we measure here with a single prompt.

We also replayed four hoaxes that moved real markets, each under the source it appeared under: the fake Emulex press release of 2000, which cost \$2.2 billion of market value within 16 minutes~\cite{sec2000litigation}, the hacked AP tweet of 2013~\cite{karppi2015hackcrash}, a fake report of an explosion at the Pentagon from 2023~\cite{bloomberg2023fake}, and a fake Eli Lilly tweet from 2022~\cite{semafor2022fake}. Each hoax received nine answers per market state (\histCalls{} calls in total). All three models sell the whole position on the press release, in \histSellAllEmuLuna{} of decisions. DeepSeek and Gemini also sell on the AP tweet (\histSellApDs{} and \histSellApGem{} of decisions) and on the Pentagon report (\histSellPentDs{} and \histSellPentGem{}). Luna chooses Hold in \histLunaHolds{} of its \histLunaTwo{} decisions on these two, and \histLunaHoldsConfirm{} of these replies say that the item needs confirmation. On the Lilly tweet, Gemini sells in \histSellLillyGem{} of decisions, DeepSeek in \histSellLillyDs{}, and Luna in \histSellLillyLuna{}.

The replays stay close to the originals. The Emulex release follows the SEC's description~\cite{sec2000litigation} with the probe's stock in place of Emulex, the Pentagon report follows news coverage, and the Lilly tweet is verbatim on LLY. The AP tweet is verbatim except that ``Barack Obama'' becomes ``the President''. With the original name, Gemini sells in \histSellApVerbGem{} of decisions, and \histGemVerbKnown{} of its other \histGemVerbOther{} replies name the 2013 hack or call the item a hoax. Gemini thus appears to remember the 2013 event instead of judging the claim, since the same tweet with one name changed makes it sell in \histSellApGem{} of decisions. A model can recognize a hoax that is already in its training data, but an attacker writes a new hoax, as the SEC-investigation headline of our grid shows.

\subsection{Simulated Market and Shocks}
\label{subsec:sim-market}
The simulated market trades one asset for one hour in one-second steps. The asset has a constant fundamental value of \$100, which makes any price below \$100 a mispricing that we can measure. Each second, the price changes by $\lambda_t$ times the net order flow plus Gaussian noise ($\sigma = \$0.01$). Value traders, which buy below the fundamental and sell above it, demand 80 shares per second per dollar of mispricing. The price impact $\lambda_t$, the price change per share of net selling, starts at \$0.0004 and rises up to fivefold with the realized volatility of the last 60 seconds, because liquidity thinned as volatility rose in 2010~\cite{kirilenko2017flash}. This setup follows agent-based market models that pair fundamental traders with other trader types~\cite{brock1998heterogeneous}, which also reproduce the leading explanation of the 2010 Flash Crash~\cite{paddrik2012agent}.

Each agent manages 300 shares for one retail investor and decides once per 240 seconds on a staggered clock, the decision latency that we measured for TradingAgents (Table~\ref{tab:latency-comparison}). At each decision, it sells or buys with the probabilities measured in \S\ref{subsec:sim-probe}, interpolated linearly over the five-minute price change and its unrealized profit and loss between the just-bought and the 3\%-under-water states. An agent that sells sells its whole position in one order, as the models did. We measure each investor's profit and loss relative to holding the 300 shares. We run 0 to 800 agents with \crashSeeds{} paired seeds per setting, so that two settings with the same seed see the same noise and the same shock, and we report the mean deepest drawdown, the largest fall of the price below \$100 within the hour.

Three shocks start at minute ten. An attacker's fake headline, the SEC-investigation report of the probe, reaches each agent's feed with probability 14/15, the share of rated designs that take news as input, and alone lowers the price by \ddFakeBase{}. A sell program with no attacker sells 120{,}000 shares over ten minutes and alone lowers the price by \ddSellBase{}. A crash that starts outside the agents, a larger sell program of 220{,}000 shares over five minutes like the one in 2010, alone takes the price \ddOutBase{} down and back within minutes. We call a fall of at least 5\% below the fundamental within the hour a flash crash and a fall of at least 10\% a severe one, since the 2010 crash fell about 9\% and recovered within the hour~\cite{cftcsec2010flash}. For scale, selling 120{,}000 shares, about \$12 million, over ten minutes lowers our price by \ddSellBase{}, roughly as in a thinly traded stock, and the agent counts below describe a market of that depth.

The three controls follow the missing ratings. The fast path of R2 freezes new orders for ten minutes once the price falls by more than three calm 60-second standard deviations within 60 seconds, and it does so without waiting for any LLM call. The hard limits of R3 cap each position at 300 shares and stop any agent whose loss exceeds 3\% of a full position from trading, using broker prices in place of any text the LLM receives. Source corroboration (S1) lets an agent act on a headline only if a second source confirms it, which the fake headline lacks in our setup. We also model Limit Up-Limit Down (LULD), the pause that U.S. exchanges apply to large stocks~\cite{sec2012luld}. Under LULD, the price cannot cross a 5\% band around its mean over the previous five minutes, and if it remains at the band for 15 seconds, trading pauses for five minutes, during which the sell programs stop and the agents' orders wait in sequence.

\subsection{Causing and Amplifying Crashes}
\label{subsec:sim-crash}

\begin{figure*}[t]
\centering
\includegraphics[width=\textwidth]{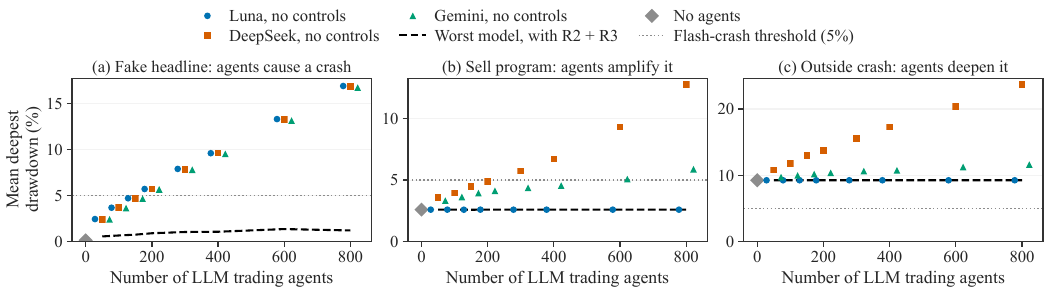}
\caption{LLM agents cause and amplify flash crashes, and the fast path and hard limits (R2 + R3) stop both. Mean deepest drawdown over \crashSeeds{} seeds as the number of agents grows, for (a) an attacker's fake headline, (b) a sell program with no attacker, and (c) a 2010-sized crash that starts outside the agents. All three models appear in each panel, shifted slightly along the x axis. Because Luna does not trade on price changes alone, it matches the agent-free level in (b) and (c).}
\label{fig:crash-threshold}
\end{figure*}

As Figure~\ref{fig:crash-threshold}a shows, the fake headline barely affects a market without agents, yet it sets off a flash crash in \crashFakeHundred{} of runs with 100 DeepSeek agents, \crashFakeOneFifty{} with 150, and \crashFakeTwoHundred{} with 200. The other two models come within \fakeCrashGap{} points of these rates, and from 300 agents on, all runs crash with each model. Between 100 and 300 agents the crash thus goes from rare to certain, although these populations hold only a quarter to three quarters as many shares as the sell program that the market absorbs. They need fewer shares because each agent sells its whole position in one order and the orders of all agents arrive within one 240-second decision cycle, faster than the value traders can buy them. With 400 agents the investors lose \pnlFakeFourHundred{} of their holdings, and with 800 agents \pnlFakeEightHundred{}.

\begin{figure}[t]
\centering
\includegraphics[width=\columnwidth]{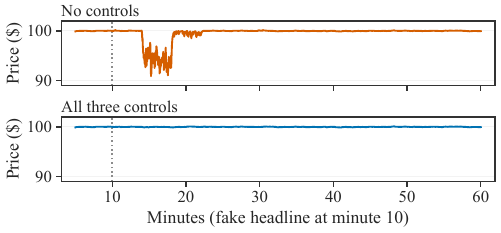}
\caption{Price path of one run with 400 agents following the measured DeepSeek policy after a fake headline at minute 10, without controls and with all three controls.}
\label{fig:crash-path}
\end{figure}

Figure~\ref{fig:crash-path} follows one run with 400 DeepSeek agents. The price remains at \$100 until the orders arrive four minutes after the headline, falls to about \$91, and recovers within minutes as the value traders buy the mispricing, the same shape as the 2010 crash. In such runs, the agents sell \soldLowFakeFourHundred{} of their investors' shares at least 5\% below value, and this loss falls on the investors who own the agents. An attacker who sold the stock short (sold borrowed shares) before posting the headline profits from these sales. In this S3 path of \frameworkShort{}, the attacker compromises no agent, and the agents' own orders execute the attack. The real hoaxes do the same. After the Emulex press release, 200 agents of each model crash the market in \histCrashEmuLunaTwoHundred{} of runs, and 300 DeepSeek or Gemini agents crash it in all runs after the AP tweet or the Pentagon report. Luna's agents choose Hold on these two, and the same tweet therefore has no effect on a market of Luna agents.

Without any attacker, agents whose model sells when prices fall change a shock that the market would absorb into a crash (Figure~\ref{fig:crash-threshold}b). DeepSeek agents sell into the program once under water and change its \ddSellBase{} fall into a flash crash in \crashSellTwoHundred{} of runs with 200 agents and \crashSellThreeHundred{} with 300, and into a severe one in \severeSellEightHundred{} of runs with 800. Gemini agents buy part of the dip, yet they still cause a flash crash in \crashSellGemSixHundred{} of runs with 600 agents and \crashSellGemEightHundred{} with 800. Luna agents do not trade on price changes alone and match the agent-free level. This failure needs neither an attacker nor a compromised agent. It is the robustness failure that R2 and R3 rate, and it follows from a design that lets the model's reading of the price decide each sale.

\begin{figure}[t]
\centering
\includegraphics[width=\columnwidth]{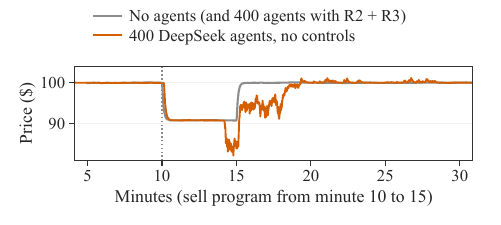}
\caption{One outside-crash run. The agents' sales arrive near the end of the sell program and push the price from about \$91 to \$82. With the fast path and hard limits, the price path remains within \$0.2 of the agent-free path.}
\label{fig:outside-path}
\end{figure}

A crash that starts outside the agents also gets deeper (Figures~\ref{fig:crash-threshold}c and~\ref{fig:outside-path}). Its \ddOutBase{} fall grows to \ddOutDsTwoHundred{} with 200 DeepSeek agents, \ddOutDsFourHundred{} with 400, and \ddOutDsEightHundred{} with 800, and to \ddOutGemEightHundred{} with 800 Gemini agents. In this crash, 400 DeepSeek agents sell \soldLowDsFourHundred{} of their investors' shares at least 5\% below value, and their investors end \holdOutDsFourHundred{} behind a buy-and-hold investor.

Because the policy in each market state comes from nine answers, we also resampled the nine answers with replacement in each of the 36 market states, \resPolicies{} times. Between the 5th and 95th percentile of these policies, 400 DeepSeek agents deepen the outside crash to \resDsOutDdLo{} to \resDsOutDdHi{}. The sell program becomes a crash in \resDsSellLo{} to \resDsSellHi{} of runs with 300 DeepSeek agents and in \resGemSellLo{} to \resGemSellHi{} with 800 Gemini agents, which give few sell answers on price alone and therefore depend more on sampling. The headline crash depends less on sampling, since DeepSeek and Gemini sell on the headline in all answers and Luna in \lunaSellNews{}.

\subsection{Effect of the Controls}
\label{subsec:sim-controls}

\begin{table*}[t]
\centering
\caption{The framework's controls keep the agents from causing or amplifying a flash crash. Mean deepest drawdown (\%) with 400 agents per measured policy (DS: DeepSeek), 50 paired seeds per entry. Bold marks a crash the agents cause or amplify: at least 5\% of runs fall 5\% below the fundamental, or the agents add at least one point to the outside crash.}
\label{tab:crash-controls}
\small
\setlength{\tabcolsep}{6pt}
\begin{tabular}{l rrr rrr rrr}
\toprule
& \multicolumn{3}{c}{\textbf{Fake headline}} & \multicolumn{3}{c}{\textbf{Sell program}} & \multicolumn{3}{c}{\textbf{Outside crash}} \\
\cmidrule(lr){2-4}\cmidrule(lr){5-7}\cmidrule(lr){8-10}
\textbf{Controls} & Luna & DS & Gemini & Luna & DS & Gemini & Luna & DS & Gemini \\
\midrule
No agents & 0.1 & 0.1 & 0.1 & 2.6 & 2.6 & 2.6 & 9.2 & 9.2 & 9.2 \\
\midrule
No controls & \textbf{9.6} & \textbf{9.6} & \textbf{9.5} & 2.6 & \textbf{6.7} & \textbf{4.6} & 9.2 & \textbf{17.3} & \textbf{10.8} \\
Fast path (R2) & 1.1 & 1.1 & 1.1 & 2.6 & 2.6 & 2.6 & 9.2 & 9.2 & 9.2 \\
Hard limits (R3) & \textbf{3.6} & \textbf{3.6} & \textbf{3.6} & 2.6 & 3.3 & 3.3 & 9.2 & 9.2 & 9.2 \\
Corroboration (S1) & 0.1 & 0.1 & 0.1 & 2.6 & \textbf{6.7} & \textbf{4.6} & 9.2 & \textbf{17.3} & \textbf{10.8} \\
R2 + R3 & 1.1 & 1.1 & 1.1 & 2.6 & 2.6 & 2.6 & 9.2 & 9.2 & 9.2 \\
R2 + R3 + S1 & 0.1 & 0.1 & 0.1 & 2.6 & 2.6 & 2.6 & 9.2 & 9.2 & 9.2 \\
\bottomrule
\end{tabular}
\end{table*}

Table~\ref{tab:crash-controls} switches the controls on one at a time with 400 agents, and each control stops the failure that its metric names. Source corroboration (S1) removes the agents' effect after the fake headline, because no agent acts on it, but it does not change the two price shocks, because there the agents act on prices. Corroboration can also fail in part, for instance if a second outlet copies the first. If it misses a random quarter of 800 agents, the headline still crashes the market in \leakQuarterEightMin{} to \leakQuarterEightMax{} of runs, and if it misses half of 400 agents, in \leakHalfFourMin{}. With the fast path and hard limits, no run crashes even when corroboration misses half of the agents. The replayed hoaxes show a further limit of source checks. They arrived through a hacked news account, fake verified accounts, and a wire-service press release, and a check of the source label alone would therefore likely pass all four.

Hard limits (R3) stop the agents from amplifying either price shock, because an agent that has lost 3\% of a full position stops trading before its sales can deepen the fall. Yet \crashFakeLimits{} of fake-headline runs with 400 agents still crash, because the agents sell on the headline before any of them is under water. The limits halt an agent instead of selling its position, and this choice matters. A stop-loss, an order that sells once the price falls to a trigger level, changes \revStopCollapse{} of \revStopRuns{} shocks into a collapse of more than half the price at the same 3\% threshold, because the first sales push other agents past their thresholds, and their sales push the next agents further. Stop-loss sales run on broker prices and bypass the fast path, and even with the fast path the outside crash collapses in \revStopFastOutCollapse{} of \revStopFastOutRuns{} runs. In a population of agents, a hard limit should therefore stop trading, as a kill switch does, instead of selling.

The fast path (R2) works against all three shocks, since the first sharp fall freezes the orders still in flight. With the fast path alone, the mean deepest drawdown after the headline falls from about 9.6\% to 1.1\% with each model, and both price shocks return to their agent-free levels (Table~\ref{tab:crash-controls}). With all three controls, each setting matches its agent-free drawdown. The fast path also works at a looser threshold. At five standard deviations in place of three, the fast path and hard limits keep each shock's drawdown within \fiveSigmaGapMax{} points of the agent-free level. These controls do not require the agents to be diverse or slow, because they act on prices and positions, which the broker observes, and not on the content that the LLM reads.

The fast path has a cost in hours without a shock, because the agents' own trades can move the price by three calm standard deviations within a minute. In a calm hour, it halts 400 DeepSeek agents \calmDsHaltsThree{} times on average at three standard deviations and \calmDsHaltsFive{} times at five. Gemini agents move the price more, because they buy dips, and the fast path halts 400 of them \calmGemHaltsThree{} times per calm hour (\calmGemHaltsFive{} at five standard deviations), and with both controls their trading falls from \calmGemVolNone{} to \calmGemVolThree{} of their holdings per hour. The 300-share cap equals the starting position and blocks buys. With a 600-share cap, 400 Gemini agents trade \calmGemVolCapSix{} of their holdings per calm hour, as many as without controls, and the drawdown under the hard limits changes by at most \capSixDiffMax{} points. Luna agents do not trade in a calm hour. A deployer can therefore trade false halts against protection by moving one threshold.

The controls also protect investors after true news. If the value really falls \revTrueDrop{}, 400 agents without controls push the price \revTrueOverMin{} to \revTrueOverMax{} below the new value and their investors end \revTrueBehindMin{} to \revTrueBehindMax{} points behind holding, while with the fast path and hard limits the price ends within \revTrueCtlOverMax{} of the new value. A halt thus does not trap investors in a falling stock. Instead, the agents' own sales after true news overshoot the new value, and the investors who sell near the bottom lose more than those who hold. Faster models do not help either. If 400 DeepSeek agents decide every 60 seconds instead of every 240, they sell into the fall sooner and deepen the outside crash to \sensOutSixty{} instead of \sensOutTwoForty{}, whereas with the fast path and hard limits it reaches \sensOutSixtyCtl{}. The part of a trading agent that must be fast is therefore the rule-based path outside the LLM.

Existing exchange pauses miss these crashes. Table~\ref{tab:crash-robustness} adds LULD to the market, and the headline with 200 or 400 agents starts a pause in \luldFakePausedRuns{} of \luldFakeRuns{} runs across the three models. Because the agents' orders arrive over four minutes, the five-minute average that defines the band falls along with the price. With 400 DeepSeek agents the price still falls \luldFakeDd{}, and their investors lose \luldFakePnl{} of their holdings. In the outside crash, trading pauses \luldOutBasePauses{} times per run, yet 400 DeepSeek agents deepen the fall from \luldOutBase{} to \luldOutDs{}, and the fast path and hard limits return it to \luldOutDsCtl{}. The four replayed hoaxes give the same result, since LULD pauses trading in \histLuldPausedRuns{} of their \histLuldRuns{} runs with 200 or 400 agents, while the fast path and hard limits cap the mean deepest drawdown at \histCtlDdMax{}. Because LULD responds to price alone, it cannot replace controls inside the trading stack.

\subsection{Robustness of the Results}
\label{subsec:sim-robust}

\begin{table}[t]
\centering
\caption{Robustness checks (DeepSeek policy, 30 paired seeds; 50 for mixed population, personas, and LULD): crashed headline runs (\%), mean deepest drawdown (\%) of price shocks, and the largest gap (points) between 400 agents under R2+R3 and none.}
\label{tab:crash-robustness}
\footnotesize
\setlength{\tabcolsep}{1.9pt}
\begin{tabular}{l rr rr rr r}
\toprule
& \multicolumn{2}{c}{\textbf{Headline}} & \multicolumn{2}{c}{\textbf{Sell program}} & \multicolumn{2}{c}{\textbf{Outside crash}} & \textbf{R2+R3} \\
\cmidrule(lr){2-3}\cmidrule(lr){4-5}\cmidrule(lr){6-7}
\textbf{Setting} & 200 & 400 & None & 400 & None & 400 & gap \\
\midrule
Our setting & 93 & 100 & 2.6 & 6.6 & 9.2 & 17.3 & 0.9 \\
Decisions every 60\,s & 100 & 100 & 2.6 & 24.2 & 9.2 & 35.9 & 1.1 \\
Decisions every 120\,s & 100 & 100 & 2.6 & 13.6 & 9.2 & 24.6 & 0.9 \\
Value traders $\times$0.5 & 100 & 100 & 5.1 & 18.0 & 18.4 & 31.6 & 1.0 \\
Value traders $\times$2 & 0 & 100 & 1.3 & 2.2 & 4.6 & 8.6 & 1.0 \\
Price impact $\times$0.5 & 13 & 100 & 2.6 & 5.2 & 9.3 & 15.7 & 0.5 \\
Price impact $\times$2 & 100 & 100 & 2.6 & 8.7 & 9.2 & 19.4 & 1.1 \\
No liquidity thinning & 0 & 100 & 2.6 & 4.3 & 9.2 & 13.4 & 0.4 \\
Noise $\times$3 & 93 & 100 & 2.8 & 6.1 & 9.4 & 17.2 & 1.1 \\
Headline reaches half & 0 & 100 & 2.6 & 6.6 & 9.2 & 17.3 & 0.8 \\
Positions vary & 100 & 100 & 2.6 & 7.7 & 9.2 & 18.1 & 0.9 \\
Cycles 60--600\,s & 0 & 13 & 2.6 & 5.4 & 9.2 & 13.3 & 0.9 \\
Three models mixed & 92 & 100 & 2.6 & 4.9 & 9.2 & 12.8 & 0.9 \\
Five investor personas & 4 & 100 & 2.6 & 7.9 & 9.2 & 14.1 & 0.8 \\
Exchange LULD pause & 92 & 100 & 2.6 & 6.7 & 9.2 & 12.9 & 0.9 \\
\bottomrule
\end{tabular}
\end{table}

The crash does not depend on our lookup table of probe answers. In a live check, gpt-5.6-luna decided for each agent in a separate call during the ten minutes after the headline, with the probe's prompt and that second's simulated prices, and the agents followed Luna's measured policy otherwise (\liveCalls{} calls). With 200 agents the market crashed in \liveTwoHundred{} runs, as with the measured policy (\liveTwoHundredMeasured{}), and with 400 agents in \liveFourHundred{} runs, again as with the measured policy (\liveFourHundredMeasured{}). The live model received prices that its own agents had already moved, and this check therefore also covers the feedback between the agents' sales and their next decisions, which the lookup table models through interpolation.

Table~\ref{tab:crash-robustness} changes one setting at a time with the DeepSeek policy: decisions every 60 or 120 seconds, half or double the value traders, half or double the price impact, no liquidity thinning, three times the noise, a headline that reaches half of the agents, positions that vary, cycles that vary, three models mixed, five investor personas, and LULD. With 400 agents, the headline crashes the market in all runs of each setting but one.

Faster decisions and fewer value traders deepen both price shocks, because the market then absorbs less of the agents' selling before their next decisions. Agents deepen the outside crash by at least \robOutAddMin{} points in each of the 15 settings, and the fast path and hard limits keep each drawdown within \robGapMax{} points of the agent-free level. With double depth or no thinning, however, 400 DeepSeek agents deepen the sell program to only \robDeepSell{} and \robNoThinSell{}. Amplifying a small shock therefore needs a thin market, whereas the headline crash does not.

The one exception matters for real agents, which do not share a clock. If each agent trades on its own cycle, drawn uniformly between 60 and 600 seconds, the orders spread out, and 400 agents crash the market in \cycFourHundred{} of runs, although 600 still crash it in \cycSixHundred{}. Positions that vary, drawn lognormally with a mean of 300 shares (10th to 90th percentile about 110 to 560) and with orders that scale with them, still give a crash after the headline in all runs with 200 and 400 agents. Mixed populations, in which agents cycle through the three models, crash after the headline in 92\% of runs with 200 agents and in all runs with 400. They deepen the outside crash less than DeepSeek alone, because one third of them follow Luna, which does not sell on price.

Splitting sales raises the number of agents that a crash needs, yet more agents still crash the market. If each agent splits its sale into two equal orders over one cycle, 200 agents crash the market in \splitTwoHundredTwoMin{} to \splitTwoHundredTwoMax{} of headline runs, and with four orders in \splitTwoHundredFour{}, against \splitTwoHundredOneMin{} to \splitTwoHundredOneMax{} with one order. With 400 or more agents, all runs crash with either split. Split orders also deepen the outside crash, to \splitOutDsTwo{} and \splitOutDsFour{} with 400 DeepSeek agents, and with the fast path and hard limits no split run crashes. An agent that splits its orders, as an execution algorithm would, thus raises the threshold but keeps the mechanism.

Five investor personas in place of one prompt make the agents less alike but do not remove the crash. A system-prompt sentence makes the investor a buy-and-hold saver, a momentum trader, a value investor, a conservative retiree, or a day trader, and agent $i$ follows the policy measured for persona $i \bmod 5$ (\personaProbeCalls{} calls per model). Sell answers after the headline fall from \personaSellSharedDs{} to \personaSellMixDs{} for DeepSeek and from \personaSellSharedLuna{} to \personaSellMixLuna{} for Luna, because the buy-and-hold and value investors hold. Then 200 DeepSeek agents crash the market in \personaFakeTwoHundredDs{} of headline runs and 400 in all runs. Personas also add a failure. Momentum traders sell on falling prices, so 400 Luna agents deepen the outside crash from \personaOutBase{} to \personaOutLuna{} and change the sell program into a crash in \personaSellCrashLuna{} of runs, against \personaSellCrashSharedLuna{} with one prompt.

The number of agents that a crash needs grows with the depth of the market. In markets $k = 2$, 5, and 10 times deeper, with value-trader demand multiplied by $k$ and price impact divided by $k$, $200k$ agents crash the market in \revDepthLoMin{} to \revDepthLoMax{} of runs, because a larger population spreads its orders more evenly over the cycle. By contrast, $300k$ agents crash it in \revDepthHiMin{} to \revDepthHiMax{} of runs (\revDepthFive{} at $k = 5$, which takes 1{,}500 agents). The mechanism, whole-position sales within one decision cycle, does not change with depth. For scale, 1{,}500 agents are about 1\% of the agentic trading accounts that Robinhood reported in September 2026~\cite{godbole2026robinhood}, if they traded one stock on one model's decisions.

\section{Recommendations for Secure and Robust Financial LLM Agent Design}
\label{sec:recommendations}

Building on Table~\ref{tab:combined-eval}, we give concise design recommendations, each tied to a specific failure pattern observed across the 15 surveyed agents and mapped to the robustness (R1--R4) and security (S1--S3) metrics of \frameworkShort{}. Several of these controls are already mandatory for conventional algorithmic trading. SEC Rule 15c3-5 requires broker-dealers with market access to block orders that exceed pre-set credit or capital limits~\cite{sec2010rule15c35}, and the EU's RTS~6 requires algorithmic traders to apply pre-trade controls on order entry and to keep a kill functionality that can cancel any or all unexecuted orders~\cite{eu2017rts6}. No scheme earns full credit on R2 or R3 (Table~\ref{tab:combined-eval}), so none describes such limits or interrupts outside the LLM, which puts these designs below the bar already set for non-LLM algorithms.

Section~\ref{sec:market-experiment} measures what this gap costs. With 400 simulated agents, the fast path of R2 and the hard limits of R3 removed the crashes that agents caused and amplified with all three models (Table~\ref{tab:crash-controls}), so we treat these two controls as the first to add. The experiment also shapes both. A breached loss limit has to halt the agent instead of selling its position (R3 below), and the fast path costs a few false halts per calm hour that a deployer sets through one threshold. For LLM agents, the new problem is that the inputs to these controls can carry injected content.

\noindent\textbf{R1: Crash Foresight and Early Awareness.}
No agent in Table~\ref{tab:combined-eval} receives a Full rating on R1: \textit{FinMem} reacts only to realized drawdowns, while \textit{TradingAgents}/\textit{TradingGroup} infer regimes from sentiment rather than microstructure. This recommendation primarily targets R1 and additionally strengthens S1 via cross-source verification.
History motivates the baseline: the 2013 AP Twitter hack erased \$136\,B in minutes because automated systems acted on a single unverified headline~\cite{karppi2015hackcrash}, and the 2010 Flash Crash was preceded by liquidity and toxic-flow signals no surveyed agent monitors~\cite{kirilenko2017flash, cftcsec2010flash}. Agents should therefore source-rank and cross-check every market-moving input, downweighting any single-source headline until an independent high-reputation feed confirms it, and maintain a library of quantitative precursors that fire triggers independently of LLM reasoning: VIX, equity--credit correlation breakdowns, order-book flow toxicity~\cite{easley2012flow}, and exchange-traded fund (ETF)--underlying decoupling. Inputs arriving in thin-liquidity windows (pre-market, after-hours) warrant stricter confirmation.

\noindent\textbf{R2: Rapid Reaction Capability under Shocks.}
Table~\ref{tab:combined-eval} shows zero Full ratings on R2, and Table~\ref{tab:latency-comparison} measures \textit{TradingAgents} at over four minutes of decision latency, orders of magnitude slower than the circuit breakers instituted after the 2010 Flash Crash~\cite{kirilenko2017flash, cftcsec2010flash}. This recommendation targets R2 and cross-covers S2.2 (by constraining which tools run during shock states) and S3 (by giving a supervisor override authority).
Trading stacks should separate a \textit{thinking path} (LLM reasoning, multi-agent debate) from a \textit{fast path} (predefined, latency-bounded safety actions). On shock detection (price gap $>3\sigma$ in $\leq 1$ minute, spread blow-up, or volatility regime jump), the agent must not launch a new reasoning round. Instead, it freezes new orders, cancels non-essential resting orders, and switches to capital-preservation mode via a rules engine that does not call the LLM. In multi-agent settings the risk agent outranks analyst/trader agents during shocks, and reactions should be hysteresis-based to resist transient spikes.

\noindent\textbf{R3: Loss Containment and Stop-Loss Execution.}
Agents \textit{consider} risk but rarely \emph{enforce} it: \textit{FinMem} and \textit{Alpha-GPT 2.0} describe risk modes without algorithmic stops, \textit{ContestTrade} downranks losers only after damage, and \textit{TradingGroup}'s stops lack volatility-aware calibration. This recommendation targets R3, cross-covers S2.1 (risk invariants live outside the LLM) and S3 (hard caps bound compromised-agent damage).
Designs should hard-code per-instrument, per-sector, and per-day loss limits that LLM output cannot override, with stop-losses evaluated on broker/exchange data rather than LLM-parsed feeds to remain intact under data poisoning~\cite{markowitz1952portfolio, mishkin2018}. Containment must be multi-layered (position-level stop, strategy-level halt, portfolio-level circuit breaker) and volatility-aware (wider stops with smaller sizes in high-vol regimes). Contest-style agents should feed stop-triggered events back as negative evidence. Once many agents trade the same stock, a breached limit should halt the agent, as a kill switch does, instead of selling its position. In our simulated market, a stop-loss that sells at a 3\% loss turned \revStopCollapse{} of \revStopRuns{} shocks into a collapse of more than half the price, because each wave of sales pushed further agents past their thresholds (\S\ref{subsec:sim-controls}).

\noindent\textbf{R4: Recovery and Post-Crash Adaptation.}
No surveyed agent triggers event-gated recovery: \textit{FinMem} updates memory continuously, \textit{QuantAgent}'s two-loop optimization cannot accelerate after tail events, and \textit{TradingGroup}'s recovery is passive. This recommendation targets R4, reinforces S1 via clean-data replay and S2.3 via quarantine memory.
Crisis economics consistently shows that liquidity spirals demand \textit{explicit} post-event rebalancing rather than incremental drift~\cite{brunnermeier2009deciphering, reinhart2009, minsky1986stabilizing}. A recovery pipeline should freeze trading and snapshot model/memory state, run a post-mortem against historical stress scenarios, downgrade aggressiveness (leverage, confidence thresholds, confirmation requirements), and replay the day with clean data to identify corrupted signals. Memory agents should route uncertain experiences to a \textit{quarantine memory} revalidated before promotion, and no long-lived instruction should be accepted from a single event.

\noindent\textbf{S1: Attacks on Market-Relevant Information Sources.}
Seven agents that ingest RAG feeds or social sentiment with no S1 control (e.g., \textit{FinMem}, \textit{FinCon}, \textit{MarketSenseAI}, \textit{MarketSenseAI 2.0}) receive a high-risk S1 rating, and the rest earn only partial credit because none both tracks provenance and requires corroboration. Real incidents~\cite{karppi2015hackcrash} and indirect prompt injection~\cite{greshake2023indirect} show a single poisoned source is sufficient to coerce market-moving actions. This recommendation targets S1 and reinforces R1 and R4 by ensuring both foresight and post-mortem replay run on verified inputs.
Agents should tag every inbound item with provenance (domain, timestamp, authentication), require cross-source majority or diversity before acting, prioritize structured official feeds over unstructured social sources, maintain a denylist of easily spoofed channels, and surface single-source anomalies to the operator rather than silently trading on them. Provenance has to cover the content and not only the channel. The four hoaxes that we replayed arrived through a hacked news account, fake verified accounts, and a wire-service press release (\S\ref{subsec:sim-probe}), so a check of the source label alone would likely pass all four.

\noindent\textbf{S2: Prompt/Logic, Tool/API, and Memory/Context Attacks.}
Security research already shows that LLM-level alignment cannot be the primary defense for autonomous trading. Jailbreaks bypass safety training~\cite{wei2023jailbroken}, persist across model generations~\cite{shen2024dan}, and transfer to closed models~\cite{zou2023universal}. Direct~\cite{perez2022ignore, liu2024formalizing} and indirect~\cite{greshake2023indirect} prompt injection override instructions, tool-calling agents follow injected content~\cite{debenedetti2025agentdojo, zhang_agent_2025}, backdoors~\cite{yang2024watchout} and memory poisoning~\cite{chen2024agentpoison} persist inside agents, training corpora can be poisoned cheaply~\cite{carlini2024poisoningweb}, fine-tuning data can be extracted~\cite{carlini2021extracting}, and tool misuse cascades~\cite{ruan2024toolemu}. Because financial agents expose all three of the affected surfaces (natural-language prompts, external tool invocation, and persistent memory), trading invariants must be enforced \emph{outside} the LLM and treated as security boundaries in the systems-security sense rather than as soft policies the model is asked to follow. This recommendation covers S2.1--S2.3 and reinforces R2 (fast-path safety cannot be stalled by an injected prompt) and R3 (position caps must sit beyond the prompt surface).

Concretely, three hardening layers must interlock so that the system degrades gracefully under partial compromise. At the logic layer (S2.1), immutable invariants (\textit{never exceed 2\% of portfolio per new position}, \textit{never trade restricted symbols}, \textit{no market orders after hours}) are enforced by an external policy engine, and every inbound text passes through a sanitizer that strips tool-invocation phrases, role reassignments, and hidden-channel payloads documented in published injection taxonomies~\cite{greshake2023indirect, liu2024formalizing}. At the tool layer (S2.2), tool bindings expose fixed schemas, per-agent capability scopes separating analyst-tier read permissions from execution-tier write permissions~\cite{ruan2024toolemu, debenedetti2025agentdojo}, rate limits, and two-signal approval for high-notional or high-urgency orders, mirroring the dual-control discipline long used in financial operations. At the memory layer (S2.3), only trusted and market-consistent observations are promoted to long-term memory through explicit write-gates with provenance checks, while suspicious experiences remain in a revalidated short-term store, directly countering the retrieval-poisoning and backdoor-implantation attacks documented in~\cite{chen2024agentpoison, yang2024watchout}. These three layers together move the agent away from relying on model alignment alone and toward defense-in-depth, matching the assurance level that non-LLM trading infrastructure already provides outside the model.

\noindent\textbf{S3: Agent-as-Attacker Containment.}
Because financial LLM agents act in the real market, a compromised instance becomes an adversarial participant, the dynamic behind AI-amplification concerns raised by regulators~\cite{kirilenko2017flash, imf2024ai_markets, sec2025ai_task_force, lin2016}. In Table~\ref{tab:combined-eval}, six agents receive a high-risk S3 rating. This recommendation targets S3 and cross-covers R2 (supervisor override in shocks) and R3 (per-agent notional caps).
Systems should assume a subset of agents will be compromised, and provide behavioral anomaly detection (sudden order-frequency, asset-class, or liquidity shifts), single-agent isolation without whole-system shutdown, and coordination guards so that multiple agents cannot place correlated large trades without risk-agent approval. Inter-agent messages are untrusted inputs, mediated by a policy engine. Finally, a human \textit{hard stop} must always override agent behavior, so that operators can revoke the agent's trading authority at any time.

\section{Beyond Trading: Lessons for High-Stakes Agents}
\label{sec:beyond}

The controls that the 15 schemes lack are missing for reasons that have little to do with finance. Each scheme moves a decision that markets used to wrap in hard limits into a language model, reads inputs that outsiders can write, keeps state that nobody validates, and passes its output to other decision-makers. As a result, any agent that acts with real authority on such inputs inherits the same failure pattern. This section lifts \frameworkShort{} out of trading. We first restate its metrics as domain-independent properties (\S\ref{subsec:general-model}), then rate seven high-stakes domains on what makes their agents dangerous (\S\ref{subsec:domain-profiles}), and finally work through healthcare and law in detail (\S\ref{subsec:healthcare}, \S\ref{subsec:law}) before turning to the remaining domains (\S\ref{subsec:other-domains}).

\subsection{A Domain-Independent Threat Model}
\label{subsec:general-model}

Every scheme in Table~\ref{tab:llm-agents-full} has the same anatomy, and so do agents outside finance. An agent reads from an \textit{information plane} (feeds, documents, records, web pages), decides in a \textit{control plane} (prompts, tools, memory), acts on an \textit{actuation plane} (orders, prescriptions, filings, commits, payments), and writes into a \textit{shared environment} that other agents and people read. Figure~\ref{fig:general-model} places the \frameworkShort{} metrics on this anatomy. S1 protects the information plane, S2.1--S2.3 protect the three parts of the control plane, and S3 guards the edge from the agent's actions back into the inputs of others. R1--R4 describe how the whole loop behaves when the environment itself turns hostile, which the NIST Cybersecurity Framework calls detection, response, containment, and recovery~\cite{nist2024csf}. The anatomy also separates two failures that the trading schemes tend to conflate. A compromised agent fails inside its control plane, whereas a non-robust agent fails because the environment changes faster than its control plane can respond. Because both failures reach the world only through the actuation boundary, \frameworkShort{} places its strictest requirements there (R2, R3, and S2.2).

\begin{figure*}[t]
\centering
\begin{tikzpicture}[
  font=\footnotesize,
  plane/.style={draw=#1, line width=0.8pt, rounded corners=4pt, fill=#1!7},
  card/.style={draw=#1!70, rounded corners=3pt, fill=white, line width=0.6pt, align=left, inner sep=3pt},
  badge/.style={fill=#1, text=white, rounded corners=2pt, font=\footnotesize\bfseries, inner sep=1.4pt},
  ttl/.style={font=\small\bfseries, text=#1},
  flow/.style={-{Stealth[length=2.6mm]}, line width=1.1pt, draw=#1},
  phase/.style={draw=fsGreen, fill=fsGreenL, rounded corners=3pt, line width=0.6pt, minimum height=0.95cm, align=center, text width=3.65cm, inner sep=2pt}
]
\fill[plane=fsBlue] (0,1.75) rectangle (3.45,5.65);
\node[ttl=fsBlue, anchor=north west] at (0.12,5.58) {Information plane};
\node[badge=fsOrange, anchor=north east] at (3.37,5.58) {S1};
\node[anchor=north west, align=left, text width=3.2cm] at (0.12,5.08) {%
\faNewspaper\ news, filings, feeds\\[2pt]
\faFileMedical\ records, scans, notes\\[2pt]
\faGavel\ briefs, productions\\[2pt]
\faCode\ issues, docs, tool specs\\[2pt]
\faGlobe\ web pages, social posts};

\fill[plane=fsPurple] (4.2,1.75) rectangle (8.95,5.65);
\node[ttl=fsPurple, anchor=north west] at (4.32,5.58) {Agent (control plane)};
\node[anchor=north east, text=fsPurple] at (8.85,5.6) {\Large\faRobot};
\node[card=fsPurple, text width=4.25cm, anchor=north west] (ins) at (4.35,4.95) {\textbf{Instructions and context}\hfill\tikz[baseline=-0.6ex]\node[badge=fsOrange]{S2.1};\\ untrusted text can act as a command};
\node[card=fsPurple, text width=4.25cm, anchor=north west] (tools) at (4.35,3.85) {\textbf{Tools}\hfill\tikz[baseline=-0.6ex]\node[badge=fsOrange]{S2.2};\\ order entry, e-filing, shell, payments};
\node[card=fsPurple, text width=4.25cm, anchor=north west] (mem) at (4.35,2.75) {\textbf{Memory}\hfill\tikz[baseline=-0.6ex]\node[badge=fsOrange]{S2.3};\\ reflections, notebooks, clause banks};

\fill[draw=fsGreen, fill=fsGreenL, line width=0.8pt, rounded corners=3pt] (9.3,1.75) rectangle (10.3,5.65);
\node[text=fsGreen, align=center] at (9.8,5.05) {\Large\faIcon{shield-alt}};
\node[text=fsGreen, align=center, font=\footnotesize\bfseries] at (9.8,3.7) {\rotatebox{90}{hard limits}};
\node[badge=fsGreen] at (9.8,2.2) {R2/R3};

\fill[plane=fsRed] (10.65,1.75) rectangle (13.5,5.65);
\node[ttl=fsRed, anchor=north west] at (10.77,5.58) {Actuation};
\node[anchor=north west, align=left, text width=2.65cm] at (10.77,5.08) {%
\faDollarSign\ orders\\[2pt]
\faPills\ prescriptions\\[2pt]
\faFileSignature\ filings\\[2pt]
\faCodeBranch\ commits, packages\\[2pt]
\faCoins\ transfers};

\fill[plane=fsGray] (14.25,1.75) rectangle (17.65,5.65);
\node[ttl=fsGray, anchor=north west] at (14.37,5.58) {Shared environment};
\node[anchor=north west, align=left, text width=3.15cm] at (14.37,5.08) {%
\faChartLine\ market prices\\[2pt]
\faHospital\ health record\\[2pt]
\faBalanceScale\ case law\\[2pt]
\faDatabase\ repos, ledgers\\[2pt]
\faUsers\ other agents read it};

\draw[flow=fsBlue] (3.45,3.7) -- node[above, font=\footnotesize, text=fsBlue]{reads} (4.2,3.7);
\draw[flow=fsPurple] (8.95,3.7) -- (9.3,3.7);
\draw[flow=fsGreen] (10.3,3.7) -- (10.65,3.7);
\draw[flow=fsRed] (13.5,3.7) -- node[above, font=\footnotesize, text=fsRed]{acts} (14.25,3.7);
\draw[flow=fsOrange, rounded corners=6pt] (15.95,5.65) -- (15.95,6.45) -- (2.5,6.45) -- (2.5,5.65);
\node[anchor=south, inner sep=2pt, text=fsOrange, font=\footnotesize\bfseries] at (9.15,6.5) {S3: what one agent writes, other agents read as input};
\node[text=fsRed, font=\Large] (adv) at (0.55,6.45) {\faUserSecret};
\node[text=fsRed, anchor=west, font=\footnotesize] at (0.85,6.45) {adversary};
\draw[flow=fsRed, dashed] (0.55,6.15) -- (0.55,5.65);

\node[phase, anchor=west] (r1) at (0,0.55) {\textbf{R1 Detect}\\early warning of abnormal states};
\node[phase, anchor=west] (r2) at (4.63,0.55) {\textbf{R2 Respond}\\interrupt without model inference};
\node[phase, anchor=west] (r3) at (9.27,0.55) {\textbf{R3 Contain}\\hard limits outside the model};
\node[phase, anchor=west] (r4) at (13.9,0.55) {\textbf{R4 Recover}\\replay, quarantine, recalibrate};
\draw[flow=fsGreen] (r1.east) -- (r2.west);
\draw[flow=fsGreen] (r2.east) -- (r3.west);
\draw[flow=fsGreen] (r3.east) -- (r4.west);
\end{tikzpicture}
\caption{\frameworkShort{} as a domain-independent threat model. An agent reads an information plane that adversaries can write (S1), decides with instructions, tools, and memory (S2.1--S2.3), acts through limits that should sit outside the model (R2/R3), and writes into a shared environment that other agents read (S3). The resilience band (R1--R4) describes the whole loop under a hostile environment. Each plane lists one example per domain: trading, healthcare, law, software, and the open web.}
\label{fig:general-model}
\end{figure*}

Table~\ref{tab:farsight-general} restates each metric as a property that any high-stakes agent needs and instantiates it in four domains. The property column is the useful abstraction. Crash foresight becomes early warning of abnormal environment states, rapid reaction becomes an interrupt path that does not wait for model inference, and loss containment becomes a hard limit enforced outside the model. Although none of these properties mentions prices, each has an established counterpart in at least one other profession: hospitals block out-of-range doses at order entry~\cite{kaushal2003effects}, courts hold the signing attorney responsible for every citation~\cite{sdny2023mata}, and software teams scope credentials by least privilege. The trading schemes show what happens when a design moves the decision into a model and forgets that the old limit lived outside it.

\begin{table*}[t]
\centering
\caption{\frameworkShort{} metrics restated as domain-independent properties and instantiated in four domains. The trading column summarizes this paper; the other columns show where the same property sits in healthcare, law, and software engineering.}
\label{tab:farsight-general}
\footnotesize
\renewcommand{\arraystretch}{1.18}
\setlength{\tabcolsep}{3.5pt}
\begin{tabular}{>{\raggedright\arraybackslash}p{1.25cm} >{\raggedright\arraybackslash}p{2.7cm} >{\raggedright\arraybackslash}p{3.0cm} >{\raggedright\arraybackslash}p{3.05cm} >{\raggedright\arraybackslash}p{3.0cm} >{\raggedright\arraybackslash}p{3.05cm}}
\toprule
\rowcolor{fsGrayL}
\textbf{Metric} & \textbf{Domain-independent property} & \textbf{Trading (this paper)} & \textbf{Healthcare} & \textbf{Law} & \textbf{Software engineering} \\
\midrule
\multicolumn{6}{l}{\cellcolor{fsGreenL}\textbf{Resilience when the environment turns hostile}} \\
R1 & Early warning of abnormal environment states & Volatility spikes, correlation breaks, toxic order flow & Deterioration scores, abnormal order patterns & Contradictory filings, sudden changes in opposing productions & Unusual dependency, permission, or CI changes \\
R2 & An interrupt path that does not wait for model inference & Order freeze and cancel outside the LLM & Hard stop at order entry, pharmacist hold & Citation check before anything is filed or sent & Session kill switch, token revocation \\
R3 & Hard limits enforced outside the model & Position caps, stop-loss, daily loss limit & Dose limits in order entry and infusion pumps & Limits on what a drafting agent may file or send & Least-privilege scopes, sandbox quotas, rate limits \\
R4 & Post-incident replay and recalibration & Clean-data replay, memory quarantine & Incident reporting, root-cause analysis & Corrected filings, notice to the court & Rollback and audit of agent commits \\
\midrule
\multicolumn{6}{l}{\cellcolor{fsOrangeL}\textbf{Security of the agent and of the actors it affects}} \\
S1 & Integrity and provenance of inputs & Forged headlines, spoofed filings, coordinated posts & Poisoned literature and training data~\cite{alber2025medical}, patient-supplied images & Documents produced by the opposing party, fabricated precedents & Malicious issues, pull requests, and tool descriptions~\cite{hou2026context} \\
S2.1 & Separation of instructions from data & Injected text overrides trading rules & Hidden prompts in medical images~\cite{clusmann2025prompt} & Hidden prompts in submitted documents~\cite{lin2026hidden} & Instructions planted in READMEs and web pages \\
S2.2 & Least privilege for actuating tools & Order submission, model-written code on the host (\S\ref{subsec:code-audit}) & Order entry, prescriptions, record writes & E-filing, email to counterparties & Shell, package publishing, cloud credentials~\cite{halloran2026mcp} \\
S2.3 & Integrity of persistent state & Poisoned reflections and decision logs & Notebooks that persist across patients~\cite{schmidgall2024agentclinic} & Clause libraries and precedent banks & Agent memory files and vector stores \\
S3 & Containment of harm to third parties & Orders move the prices that others read & Entries in a shared record that later readers trust & Fabricated citations copied into later filings & Malicious commits and packages pulled by others \\
\bottomrule
\end{tabular}
\end{table*}

The five traits of \S\ref{categorization_of_agents} generalize in the same way. \textit{Instant} becomes the gap between how fast harm arrives and how fast a reviewer can act, and \textit{Inconspicuous} becomes how well harm hides in normal variation. \textit{Unverified} becomes how much of the input outsiders write, \textit{Adaptive} becomes how many input channels an attacker can choose from, and \textit{Cascading} becomes how strongly the agent's outputs couple to other actors. Together with irreversibility and reliance on in-model safety, these generalized traits are the axes on which \S\ref{subsec:domain-profiles} compares domains.

The anatomy also explains why generic agent benchmarks miss these failures. Benchmarks such as AgentDojo~\cite{debenedetti2025agentdojo}, InjecAgent~\cite{zhan2024injecagent}, and ASB~\cite{zhang2025agent} score whether one agent completes or refuses one task. Because they score single tasks, they exercise the information and control planes but stop at the actuation boundary, and none of them asks what the action does to the next agent. Robustness under a hostile environment and propagation through the shared environment, the two halves of \frameworkShort{} that are specific to trading, are exactly the halves that transfer to other high-stakes domains.

\subsection{Risk Profiles of Seven High-Stakes Domains}
\label{subsec:domain-profiles}

\begin{figure*}[t]
  \centering
  \includegraphics[width=\textwidth]{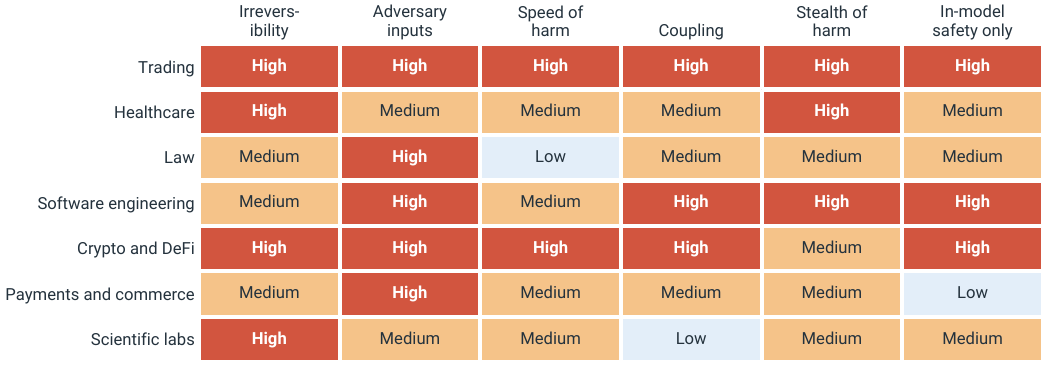}
  \caption{Risk profiles of LLM agents in seven high-stakes domains, rated on the six properties that make trading agents dangerous. High means the property raises the cost of a compromised agent in that domain. Ratings are our assessment; Appendix~\ref{appendix:domain-profiles} gives the evidence for each cell.}
  \label{fig:domain-profiles}
\end{figure*}

Trading agents combine six properties that raise the cost of a compromise: actions that cannot be undone, inputs that adversaries write, harm that arrives faster than review, coupling to other actors, harm that hides in normal variation, and safety that rests on the model alone. Figure~\ref{fig:domain-profiles} rates seven domains on the same six properties. The ratings are our assessment from the cited literature and deployments, and Appendix~\ref{appendix:domain-profiles} gives the reason for every cell.

Figure~\ref{fig:domain-profiles} shows that trading is the only domain rated high on all six properties, which is why it makes a sharp test case: every weakness it exposes is amplified. Crypto agents come closest, because on-chain transfers are final, blocks confirm within seconds, and front-running bots react within one block~\cite{daian2020flash}. Crypto is also a domain where context manipulation of a widely used agent framework has been shown to trigger unauthorized asset transfers~\cite{patlan2025real}. Software engineering shares the two properties that matter most for propagation. Outsiders write the issues, pull requests, and tool descriptions that coding agents read~\cite{hou2026context,halloran2026mcp}, and other developers and their agents pull the code that coding agents write.

The figure also separates domains by where their last line of defense sits. Payments wrap every payer, human or agent, in transaction limits, authentication, and fraud scoring, so an agent that is talked into a bad payment still meets a control it cannot argue with. Healthcare and law place a trained human at the gate, the pharmacist or the signing attorney, which works only while that human checks what the agent produced. Trading, crypto, and software agents in their current designs often have neither, so their safety rests on the model following its instructions. This property, more than the domain itself, decides how far a compromise can go.

Read column by column, the profiles also say which control each domain needs first. Where harm is fast, as in trading and crypto, only an interlock that does not wait for inference can respond in time (R2). Where adversaries author the inputs, as in law and software, provenance and the separation of instructions from data come first (S1, S2.1). Where coupling is high, controls belong on the channel itself, such as fan-out limits and circuit breakers (S3). Where harm is stealthy, as in healthcare, monitoring against ground truth and replay after an incident matter most (R1, R4).

\begin{insightbox}[Takeaway]
The trading-specific parts of \frameworkShort{}, robustness under a hostile environment and propagation through shared state, are the parts that transfer. A domain is as safe as the controls that sit between its agents and their irreversible actions, and the ratings show that few agent designs place those controls outside the model.
\end{insightbox}

\subsection{Healthcare Agents}
\label{subsec:healthcare}

Clinical agents have moved from question answering to dialogue and action. AMIE, a conversational diagnostic agent, outperformed 20 primary-care physicians on 30 of 32 axes rated by specialists in a randomized, double-blind study of text consultations over 159 case scenarios~\cite{tu2025conversational}. Polaris runs patient-facing voice calls with a primary agent and specialist support agents that take over nursing tasks to improve safety~\cite{mukherjee2024polaris}, and ambient scribes already draft clinical notes at health-system scale~\cite{tierney2024ambient}. Research testbeds go further. AgentClinic gives agents tools and a notebook that persists across patients~\cite{schmidgall2024agentclinic}, and Agent Hospital lets doctor agents evolve by treating tens of thousands of simulated patients~\cite{li2024agenthospital}. These designs reproduce the multi-role, memory-heavy architectures of Table~\ref{tab:llm-agents-full}, with a care team in place of an analyst desk.

\noindent\textbf{Information plane (S1).} Poisoning the knowledge that clinical models learn from is cheap. Replacing 0.001\% of the training tokens of a web corpus with medical misinformation produced models that spread more medical errors while matching clean models on standard medical benchmarks~\cite{alber2025medical}, and editing about 1.1\% of a model's weights implants a targeted false fact without hurting other tasks~\cite{han2023medical}. As in trading, a corrupted agent therefore looks healthy on the metrics that its builders track.

\noindent\textbf{Control plane (S2).} Patients and outside providers write much of what a clinical agent reads. Clusmann et al. hid instructions in medical images that human readers could not see and found all four tested vision-language models susceptible across 594 attacks~\cite{clusmann2025prompt}. If the agent can also place orders, such an injection becomes an S2.2 action, and if it keeps a cross-patient notebook~\cite{schmidgall2024agentclinic} or evolves from past cases~\cite{li2024agenthospital}, the injection gains the persistence that S2.3 describes. The adversary need not be a criminal outsider. Finlayson et al. point out that reimbursement rules already reward manipulating the inputs of medical models~\cite{finlayson2019adversarial}, which gives healthcare the profit motive that trading has.

\noindent\textbf{Shared environment (S3).} The electronic health record plays the role of the order book. An agent's note or order enters a record that every later clinician, scribe, and decision-support tool reads, and copied-forward text carries an error from visit to visit. A compromised scribe does not need to harm the current patient to cause harm. It only needs to write a plausible entry that the next reader trusts.

\noindent\textbf{Resilience (R1--R4).} Medicine already has the out-of-model controls that the trading schemes lack. Computerized order entry with dose checks reduces medication errors~\cite{kaushal2003effects}, and pharmacist verification adds a second, independent review. The lesson from trading is to keep these controls in the path of the agent, because a control re-implemented inside a prompt can be argued away. However, the weak point is the human gate. Automation bias makes reviewers accept wrong advice from decision support~\cite{goddard2012automationbias}, so a reviewer who approves a hundred correct agent drafts per day is a poor detector of the one poisoned draft.

\subsection{Legal Agents}
\label{subsec:law}

Legal practice adopted retrieval-augmented research tools and drafting assistants quickly, and the evidence on their reliability is sobering. On specific, verifiable questions about federal court cases, general-purpose models hallucinated between 58\% (ChatGPT~4) and 88\% (Llama~2) of the time~\cite{dahl2024legal}. Commercial legal research tools that advertised hallucination-free citations still hallucinated on 17\% to 33\% of queries in a preregistered evaluation~\cite{magesh2024hallucination}. Courts have sanctioned lawyers who filed briefs citing cases that a chatbot invented, and the sanctioning court stressed that existing rules give attorneys a gatekeeping duty to ensure the accuracy of their filings~\cite{sdny2023mata}.

\noindent\textbf{Information plane (S1).} Law differs from trading in one respect that makes it harder. In litigation the adversary writes part of the agent's input by design. The opposing party produces the documents that a review agent reads in discovery, drafts the contract that a negotiation agent marks up, and files the briefs that a research agent summarizes. Whereas a trading attacker must plant a headline, a litigant hands the planted content over through a channel the agent is obliged to read. Hidden instructions already appear in documents that AI systems review. In July 2025, 18 arXiv manuscripts carried concealed prompts that asked AI reviewers for positive reviews~\cite{lin2026hidden}.

\noindent\textbf{Control and actuation (S2).} A drafting agent with e-filing or email tools can send privileged material to the wrong party, file on the wrong docket, or miss a deadline, and a firm's clause library or precedent bank is a long-lived memory that every later draft reads (S2.3). Legal harm is slower than market harm, but much of it is final in practice, because a disclosed privileged document cannot be unread and a missed deadline can end a claim.

\noindent\textbf{Shared environment (S3).} Citations are the order book of law. A fabricated or mischaracterized precedent that enters one filing can be quoted in the next, and model-drafted clauses become templates that other agents retrieve. As with herding in markets~\cite{xia2026agentictradingllmagents}, the agents that copy the error are not compromised. They behave correctly given what they read.

\noindent\textbf{Resilience (R1--R4).} The legal profession has a hard stop that trading agents lack, since a filing needs a signature and the signer answers for it. The hard stop works only if the signer verifies, and the hallucination rates above show how much verification it takes. The \frameworkShort{} recommendations translate directly: verify every citation against a primary source before anything leaves the firm (the R2 interrupt), restrict which tools a drafting agent may call (S2.2), and treat documents from the other side as untrusted input that cannot issue instructions (S1, S2.1).

\subsection{Other High-Stakes Domains}
\label{subsec:other-domains}

Table~\ref{tab:domains} puts the domains side by side and adds general multi-agent platforms. For each, it lists what the agent changes in the world, who writes its inputs, how its effects reach others, the strongest published evidence of attacks, and the control that already exists outside the model.

\begin{table*}[t]
\centering
\caption{High-stakes agent domains compared on the questions that \frameworkShort{} asks of trading agents: what the agent changes in the world, who writes the inputs it reads, how its effects reach other actors, the strongest published evidence of attacks, and the control that already exists outside the model.}
\label{tab:domains}
\footnotesize
\renewcommand{\arraystretch}{1.18}
\setlength{\tabcolsep}{3.5pt}
\begin{tabular}{>{\raggedright\arraybackslash}p{1.55cm} >{\raggedright\arraybackslash}p{2.35cm} >{\raggedright\arraybackslash}p{2.6cm} >{\raggedright\arraybackslash}p{2.75cm} >{\raggedright\arraybackslash}p{3.6cm} >{\raggedright\arraybackslash}p{2.95cm}}
\toprule
\rowcolor{fsGrayL}
\textbf{Domain} & \textbf{What the agent changes} & \textbf{Who writes its inputs} & \textbf{How effects reach others} & \textbf{Published attack evidence} & \textbf{Control outside the model} \\
\midrule
Trading (this paper) & Orders and positions & News outlets, social media, data vendors & Prices that every participant reads & Adversarial headlines and misinformation mislead trading agents~\cite{rizvani2026adversarial,liu2026autoredtrader}; poisoned posts spread inside agent stacks~\cite{sua2026contagion} & Pre-trade limits and kill switches~\cite{sec2010rule15c35,eu2017rts6}; no scheme describes them outside the LLM \\
Healthcare & Orders, notes, patient messages & Patients, outside providers, published literature & Shared health record, copied-forward notes & 0.001\% poisoned training tokens raise medical errors~\cite{alber2025medical}; hidden prompts in images~\cite{clusmann2025prompt} & Order-entry dose checks~\cite{kaushal2003effects}, pharmacist verification \\
Law & Filings, contracts, client emails & Opposing parties, courts, counterparties & Citations and templates reused in later work & 17--33\% hallucination in legal research tools~\cite{magesh2024hallucination}; hidden prompts in submitted documents~\cite{lin2026hidden} & Attorney signature and duty to verify~\cite{sdny2023mata} \\
Software engineering & Code, packages, infrastructure & Issue authors, dependency maintainers, tool servers & Repositories and registries that other agents pull & MCP tools coerce credential theft and remote access~\cite{halloran2026mcp}; risky code execution by coding agents~\cite{guo2024redcode} & Code review, CI, sandboxes, least-privilege tokens \\
Multi-agent platforms & Messages to other agents, delegated tool calls & Other agents, web content, files & Agent-to-agent messages and shared memory & Arbitrary code execution in 58--90\% of trials~\cite{triedman2025multi}; self-replicating prompts~\cite{lee2024promptinfection,cohen2024here} & Message tagging~\cite{lee2024promptinfection}, topology-aware filtering~\cite{wang2025gsafeguard} \\
Crypto and DeFi & Signed on-chain transactions & Social posts, stored agent context, on-chain data & Composable contracts and a shared mempool & Memory injection triggers unauthorized transfers~\cite{patlan2025real}; front-running within one block~\cite{daian2020flash} & Few; immutable contracts leave the signing key as the last control \\
Scientific labs & Physical experiments and syntheses & Literature, protocols, tool outputs & Published results and shared protocols & Autonomous synthesis planning and execution~\cite{boiko2023autonomous,mbran2024augmenting}; safety risks of AI scientists~\cite{tang2025risks} & Lab safety protocols, human sign-off on hazardous steps \\
\bottomrule
\end{tabular}
\end{table*}

Web and computer-use agents read the most open information plane of all. Adversarial pop-ups that human users would ignore made vision-language agents click them in 86\% of cases on average and cut task success by 47\%~\cite{zhang2024attacking}. In the WASP benchmark, prompt injections in realistic web pages partially succeeded in up to 86\% of cases, although agents often failed to complete the attacker's full goal~\cite{evtimov2025wasp}. That gap reflects limited capability, and it will close as agents become more capable. Embodied agents add physical actuation, and RoboPAIR elicited harmful physical actions from LLM-controlled robots, including a ground robot planned by GPT-4o~\cite{robey2024jailbreaking}.

Three patterns run through Table~\ref{tab:domains}. First, code execution is the most common route from text to action. ContestTrade and FinRobot execute model-written code on the host (\S\ref{subsec:code-audit}), multi-agent frameworks can be steered into executing arbitrary malicious code in 58\% to 90\% of trials with a GPT-4o backbone~\cite{triedman2025multi}, and MCP servers expose tools that coerce models into credential theft and remote access~\cite{halloran2026mcp}. In all three cases, a sandbox outside the model is the single control that would have stopped the attack. Second, physical and financial actuation share the property that matters most, finality: a synthesis that Coscientist runs~\cite{boiko2023autonomous}, a transfer that a Web3 agent signs~\cite{patlan2025real}, and an order that a trading agent places cannot be recalled by a later model call. Third, the domains with the most mature controls, payments and healthcare, built them for human operators decades before agents arrived. Agents that call those controls inherit decades of engineering, and agents that replace them with prompts inherit none of it.

\section{Cascades, Contagion, and Agent Worms}
\label{sec:cascades}

S3 is the surface that turns one compromised agent into a system failure, and it is the surface that the literature on agent security has studied least. Markets have a century of experience with cascades, computer security has three decades of experience with worms, and LLM agents now combine both. This section draws the three literatures together. We trace how market failures already propagated through automated participants (\S\ref{subsec:market-cascades}), classify the channels through which agents affect one another (\S\ref{subsec:channels}), compare agent worms with classic worms and market cascades (\S\ref{subsec:worms}), and use a branching-process model to show when one compromised agent can start an outbreak (\S\ref{subsec:branching}).

\begin{figure*}[t]
\centering
\begin{tikzpicture}[
  font=\footnotesize,
  ev/.style={draw=#1, fill=#1!8, rounded corners=3pt, line width=0.6pt, align=center, text width=2.6cm, inner sep=2.5pt, minimum height=1.25cm},
  yr/.style={fill=#1, text=white, font=\footnotesize\bfseries, rounded corners=2pt, inner sep=1.5pt},
  stem/.style={draw=#1, line width=0.7pt}
]
\draw[line width=1.6pt, draw=fsGray!60, -{Stealth[length=3mm]}] (0.0,0) -- (17.2,0);
\foreach \x/\y/\c/\t/\s in {
  1.40/1988/fsPurple/{Morris worm\\exploits bugs, spreads in hours}/1,
  3.43/1991/fsPurple/{Epidemic model\\of computer viruses}/-1,
  5.46/2003/fsPurple/{Slammer infects\\most hosts in minutes}/1,
  7.49/2010/fsRed/{Flash Crash\\automated traders amplify a shock}/-1,
  9.52/2012/fsRed/{Knight Capital\\dormant code, \$460M lost}/1,
  11.55/2013/fsRed/{AP Twitter hack\\one false tweet, \$136B}/-1,
  13.58/2023/fsOrange/{Indirect prompt\\injection in LLM apps}/1,
  15.61/2024/fsOrange/{Morris~II, Agent Smith\\self-replicating prompts}/-1}
{
  \draw[stem=\c] (\x,0) -- (\x,{0.55*\s});
  \fill[\c] (\x,0) circle (2.2pt);
  \node[yr=\c] at (\x,{0.0-0.32*\s}) {\y};
  \node[ev=\c, anchor={\ifnum\s=1 south\else north\fi}] at (\x,{0.55*\s}) {\t};
}
\node[anchor=west, text=fsPurple, font=\footnotesize\bfseries] at (0.1,-2.55) {\rule{8pt}{8pt}~classic worms};
\node[anchor=west, text=fsRed, font=\footnotesize\bfseries] at (3.6,-2.55) {\rule{8pt}{8pt}~market cascades};
\node[anchor=west, text=fsOrange, font=\footnotesize\bfseries] at (7.4,-2.55) {\rule{8pt}{8pt}~LLM agent attacks};
\end{tikzpicture}
\caption{Three lineages of propagating failure converge in LLM agents: classic worms~\cite{spafford1989internet,kephart1991directed,moore2003inside}, market cascades~\cite{kirilenko2017flash,sec2013matter,karppi2015hackcrash}, and attacks on LLM agents~\cite{greshake2023indirect,cohen2024here,gu2024agentsmith}. Since 2025, MCP exploits~\cite{halloran2026mcp}, code execution through multi-agent frameworks~\cite{triedman2025multi}, and public warnings from regulators~\cite{boe2026breeden,donachie2026lawmakers} have followed.}
\label{fig:timeline}
\end{figure*}

\subsection{From Flash Crashes to Agent Cascades}
\label{subsec:market-cascades}

Figure~\ref{fig:timeline} places the events that shaped this section on one time line. Automated markets have already failed in the three ways that populations of agents can fail. On May 6, 2010, a large sell program met high-frequency traders that passed inventory back and forth, and nearly one trillion dollars of market value disappeared within minutes before prices recovered~\cite{kirilenko2017flash,cftcsec2010flash}. In April 2013, a single false tweet from a hijacked news account erased \$136 billion in minutes, because automated systems traded on the headline before anyone checked it~\cite{karppi2015hackcrash}. On August 1, 2012, Knight Capital's router sent more than 4 million orders in 45 minutes while trying to fill 212 customer orders and lost more than \$460 million~\cite{sec2013matter}. The SEC found that the firm's controls did not stop the erroneous orders before they reached the market, which made Knight the first enforcement action under the market access rule, Rule 15c3-5~\cite{sec2010rule15c35}.

The three events correspond to the three surfaces of \frameworkShort{}. The AP hack is an S1 attack, in which one unverified source moved every automated reader at once. The Flash Crash is an S3 cascade, in which participants that behaved correctly given their inputs amplified a shock that none of them caused alone. Knight Capital is a robustness failure with no attacker. Its order flood started when code for a function that the firm had stopped using years earlier was reactivated on one server, and nothing outside the trading logic stopped it~\cite{sec2013matter}. Our audit found the agent-era version of the last pattern. ContestTrade ships a contest mechanism that its pipeline never calls, FinMem's risk-averse branch is commented out, and no release enforces a stop-loss outside the LLM (\S\ref{subsec:code-audit}). Code that the paper describes but the pipeline does not run is the kind of dormant path that turned a deployment error into a \$460 million loss.

Economics explains why such cascades are self-reinforcing. In an informational cascade, each participant rationally sets aside its own signal and follows the actions of earlier participants, so a few early mistakes can lock in a wrong consensus~\cite{bikhchandani1992theory}. If losses force sales, the sales cause further losses, as in the liquidity spirals that turned the 2007--2008 subprime losses into a market-wide crisis~\cite{brunnermeier2008market,brunnermeier2009deciphering}. LLM trading agents inherit and amplify herding~\cite{xia2026agentictradingllmagents}, and LLM pricing agents reach supracompetitive prices without being told to collude~\cite{fish2024algorithmic}, echoing earlier findings for reinforcement-learning pricing agents~\cite{calvano2020artificial} and trading agents~\cite{dou2023powered}. Regulators have drawn the same conclusion. The Financial Stability Board warns that common models and data sources can produce correlated behavior across institutions~\cite{fsb2024financial}, and the Bank of England has said publicly that autonomous agents could trigger a market meltdown~\cite{boe2026breeden}. Our simulated market in Section~\ref{sec:market-experiment} reproduces two of the three patterns with agents that decide as measured LLMs do. Agents that sell on one fake headline start a crash as the automated readers of the AP tweet did, and agents whose model sells when prices fall deepen a 2010-sized crash from \ddOutBase{} to \ddOutDsFourHundred{} with 400 agents, with no attacker and no compromise.

\subsection{Propagation Channels between Agents}
\label{subsec:channels}

A compromised agent can reach other agents through six channels, which Table~\ref{tab:channels} lists with a trading instance from our data, evidence from general agents, and the control that would interrupt the channel. Two of the channels need no compromised receiver at all.

\begin{table*}[t]
\centering
\caption{Channels through which one agent affects others. ``Needs a compromised receiver'' marks whether harm spreads only through receivers that the payload compromises (\cmark) or also through receivers that behave correctly on what they read (\xmark). The trading instances come from our ratings and code audit.}
\label{tab:channels}
\footnotesize
\renewcommand{\arraystretch}{1.18}
\setlength{\tabcolsep}{3.5pt}
\begin{tabular}{>{\raggedright\arraybackslash}p{1.6cm} >{\raggedright\arraybackslash}p{2.75cm} >{\raggedright\arraybackslash}p{3.05cm} >{\raggedright\arraybackslash}p{3.65cm} >{\centering\arraybackslash}p{1.35cm} >{\raggedright\arraybackslash}p{3.05cm}}
\toprule
\rowcolor{fsGrayL}
\textbf{Channel} & \textbf{Mechanism} & \textbf{Trading instance} & \textbf{Evidence beyond trading} & \textbf{Needs a compromised receiver} & \textbf{Control on the channel} \\
\midrule
Shared inputs & One poisoned item reaches many agents in parallel & \vtwoNewsSchemes{} of 15 schemes read news; the 2013 AP hack~\cite{karppi2015hackcrash} & Five passages per question poison RAG answers 90\% of the time~\cite{zou2025poisonedrag}; web injection~\cite{liao2025eia} & \xmark & Provenance and cross-source corroboration (S1) \\
Shared models & One weakness works on every agent with the same backbone & \vtwoGptFourFamily{} of 15 schemes run the GPT-4 family & Transferable adversarial prompts~\cite{zou2023universal}; correlated errors across 350+ LLMs~\cite{kim2025correlated} & \xmark & Backbone diversity; limits outside the model (R3) \\
Agent-to-agent messages & Each receiver follows the payload and re-emits it & \vtwoMultiAgentSchemes{} of 15 schemes pass free text between roles & Prompt Infection~\cite{lee2024promptinfection}; communication attacks~\cite{he2025red}; code execution in 58--90\% of trials~\cite{triedman2025multi} & \cmark & Treat peer messages as data (S3-E3); message tagging; bounded fan-out \\
Shared environment & Agents read state that other agents write & Prices; StockAgent's price is the last matched trade (\S\ref{subsec:code-audit}) & Training on model-written text degrades models~\cite{shumailov2024collapse}; shared records and case law & \xmark & Circuit breakers; provenance on records \\
Shared memory and tools & Long-lived stores are read across sessions and agents & \auditPersistStore{} releases write stores without validation, \auditPickle{} load them with \texttt{pickle} & MCP tool descriptions carry instructions~\cite{hou2026context}; memory injection moves crypto assets~\cite{patlan2025real} & \cmark & Write gates and quarantine (S2.3); signed stores; safe loading \\
Human relays & People copy agent output to where other agents read it & Written rationales, which all 15 schemes emit (Table~\ref{tab:llm-agents-full}) & Fabricated citations reached court filings~\cite{sdny2023mata} & \xmark & Duty to verify filings~\cite{sdny2023mata}; labels for generated text~\cite{kirchenbauer2023watermark} \\
\bottomrule
\end{tabular}
\end{table*}

\begin{figure}[t]
  \centering
  \includegraphics[width=\columnwidth]{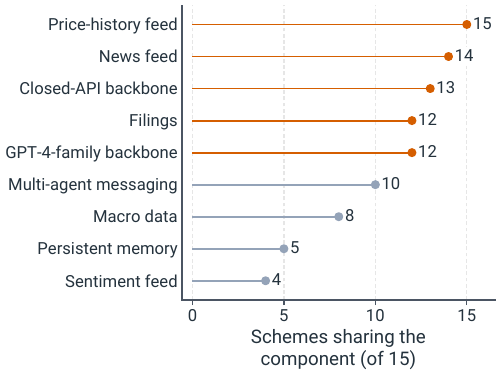}
  \caption{Components shared by the 15 schemes. A poisoned price or news feed, or a weakness in the GPT-4 family, reaches most designs at once. Input counts come from Table~\ref{tab:llm-agents-full} and backbone counts from Figure~\ref{fig:llm-backbones}; red marks components shared by at least 12 schemes.}
  \label{fig:common-mode}
\end{figure}

\noindent\textbf{Shared inputs.} If many agents read one poisoned item, the item reaches all of them at once. Figure~\ref{fig:common-mode} shows how concentrated the inputs of the 15 schemes are: \vtwoPriceSchemes{} read price history, \vtwoNewsSchemes{} read news, and \vtwoFilingSchemes{} read company filings. For example, a fabricated headline in a widely syndicated feed would reach almost every design at once, which is the AP hack scaled to agents. Retrieval systems widen the channel further. With five crafted passages per target question, an attacker makes a RAG system return the chosen answer 90\% of the time, even in a knowledge base of millions of texts~\cite{zou2025poisonedrag}.

\noindent\textbf{Shared models.} The same backbone fails in the same way. \vtwoGptFourFamily{} of the 15 schemes run on the GPT-4 family, so a jailbreak or injection that works on that family works on most of the field. Monoculture has measurable costs even without an attacker. If many decision-makers share one algorithm, overall decision quality can fall below what independent algorithms would achieve~\cite{kleinberg2021algorithmic}, deployed models fail on the same individuals~\cite{toups2023ecosystem}, and sharing models and training data homogenizes outcomes~\cite{bommasani2022picking}. Across more than 350 LLMs, two models that both err agree on the wrong answer 60\% of the time on one leaderboard, and larger, more accurate models have more correlated errors even across providers~\cite{kim2025correlated}. Diversity across vendors therefore helps less than it appears to. In finance specifically, Gensler and Bailey warned that deep learning could promote herding through uniform models and shared data~\cite{gensler2020deep}, and Danielsson et al. argue that AI can destabilize the financial system by amplifying procyclicality and creating new tail risks~\cite{danielsson2022artificial}.

\noindent\textbf{Agent-to-agent messages.} Multi-agent designs pass free text between roles, and \vtwoMultiAgentSchemes{} of the 15 schemes use multi-agent messaging. Prompt Infection shows that a payload can make each receiving agent re-emit it to the next~\cite{lee2024promptinfection}, communication attacks that intercept messages between agents mislead whole systems~\cite{he2025red}, and manipulated knowledge spreads through agent communities without any explicit prompt manipulation and survives in their retrieval stores~\cite{ju2026flooding}. Topology also matters, because highly connected networks spread attacks further~\cite{yu2024netsafe} and hierarchical structures lose less performance to faulty agents than flat ones~\cite{huang2025resilience}. Control-flow hijacking is the worst case. Adversarial web content steers multi-agent frameworks into executing arbitrary code in 58\% to 90\% of trials with a GPT-4o backbone, even when no single agent is susceptible to prompt injection on its own~\cite{triedman2025multi}.

\noindent\textbf{Shared environment.} Agents read the world that other agents write. Prices are the canonical case, and StockAgent's code makes the channel explicit. Every agent reads a market price that is simply the last matched price, so one agent's order moves what every other agent sees (\S\ref{subsec:code-audit}). Outside markets the shared environment is the health record, the case law, the package registry, and the open web. If models are trained on text that earlier models wrote, they degrade over generations~\cite{shumailov2024collapse}, which is the same feedback loop at the scale of a training corpus.

\noindent\textbf{Shared memory and tools.} Retrieval stores, memory files, and tool servers are long-lived state that several agents or sessions read. Of the seven audited releases, \auditPersistStore{} write such stores without validation and \auditPickle{} load them with \texttt{pickle} (\S\ref{subsec:code-audit}). In the wider agent world, MCP servers distribute tool descriptions that can carry instructions~\cite{hou2026context,halloran2026mcp}, and memory injection into a Web3 agent framework produced unauthorized asset transfers that prompt-injection defenses did not stop~\cite{patlan2025real}.

\noindent\textbf{Human relays.} People copy agent output into places that other agents read: a model-drafted brief into a court filing, a summary into a research report, a scribe's note into the record. Each copy launders the output's origin. The fabricated citations that reached court filings~\cite{sdny2023mata} traveled this way.

\subsection{Anatomy of an Agent Worm}
\label{subsec:worms}

A worm replicates without human help. The 1988 Morris worm exploited a buffer overflow, a sendmail debugging feature, and weak passwords, and it spread across the early Internet within hours~\cite{spafford1989internet}, Code Red reached hundreds of thousands of hosts in less than a day~\cite{zou2002code}, and Slammer infected most vulnerable hosts within about ten minutes~\cite{moore2003inside}. Kephart and White modeled virus spread as an epidemic on a directed graph and showed that infection persists only above a threshold set by contact and cure rates~\cite{kephart1991directed}. In scale-free networks, that threshold falls toward zero~\cite{pastorsatorras2001epidemic}.

LLM agents have produced the first worms of their own. Morris~II plants an adversarial self-replicating prompt in an email that a RAG-based assistant retrieves, so each assistant that processes the email leaks data and forwards the prompt to the next~\cite{cohen2024here}. Agent Smith feeds one adversarial image into the memory of one agent in a simulated population of up to one million multimodal agents, and pairwise chats spread the jailbreak to almost all of them exponentially fast~\cite{gu2024agentsmith}. Prompt Infection makes agents inside one multi-agent system pass a payload to each other like a virus~\cite{lee2024promptinfection}. Table~\ref{tab:worms} compares these worms with classic worms and with the market cascades of \S\ref{subsec:market-cascades}.

\begin{table*}[t]
\centering
\caption{Classic computer worms, LLM agent worms, and market cascades compared. The market column has no replication step: other participants spread the effect by acting correctly on corrupted prices or news.}
\label{tab:worms}
\footnotesize
\renewcommand{\arraystretch}{1.18}
\setlength{\tabcolsep}{4pt}
\begin{tabular}{>{\raggedright\arraybackslash}p{2.2cm} >{\raggedright\arraybackslash}p{4.45cm} >{\raggedright\arraybackslash}p{4.9cm} >{\raggedright\arraybackslash}p{4.65cm}}
\toprule
\rowcolor{fsGrayL}
 & \textbf{Classic worm} & \textbf{LLM agent worm} & \textbf{Market cascade} \\
\midrule
Exploited weakness & Memory-safety bugs and weak authentication~\cite{spafford1989internet} & Instruction following on untrusted text~\cite{cohen2024here,lee2024promptinfection} & Correct reactions to corrupted prices or news~\cite{kirilenko2017flash,karppi2015hackcrash} \\
Replication & Copies its code to the next host & The model re-emits the payload in its output~\cite{cohen2024here} & None; each participant acts for its own reasons \\
Propagation vector & Network services and email & RAG stores, email, agent messages, shared memory~\cite{cohen2024here,gu2024agentsmith} & Prices, order books, and news feeds \\
Target discovery & Random scanning of the address space~\cite{zou2002code,moore2003inside} & Retrieval, contact lists, and broadcast channels & Every participant reads the same price \\
Payload & Arbitrary code & Data exfiltration, tool calls, trades, misinformation & Correlated selling and withdrawn liquidity \\
Observed speed & Most vulnerable hosts within about ten minutes~\cite{moore2003inside} & Exponential in the number of chat rounds~\cite{gu2024agentsmith} & Minutes~\cite{kirilenko2017flash}; 45 minutes for Knight Capital~\cite{sec2013matter} \\
Patching & Fixing the bug immunizes the host & No complete fix; adaptive attacks bypass published defenses~\cite{nasr2025attacker} & Nothing to patch; the reaction is correct given the input \\
Containment & Firewalls, patching, rate limits & Message validation, tagging~\cite{lee2024promptinfection}, propagation guardrails~\cite{cohen2024here} & Circuit breakers, kill switches, pre-trade limits~\cite{sec2010rule15c35,eu2017rts6} \\
\bottomrule
\end{tabular}
\end{table*}

Two differences stand out. First, an agent worm exploits instruction following, the very feature that makes agents useful, so no patch removes the vulnerability. Instead, defenses lower the per-hop success rate but leave it above zero, and adaptive attacks bypassed 12 published defenses with success rates above 90\% for most of them~\cite{nasr2025attacker}. Second, the market column shows a propagation mode that neither classic nor agent worms need: \textit{propagation without compromise}. A compromised trading agent's sell orders lower a price, and every other agent that reads the price sells for its own, correct reasons. The effect spreads as belief rather than as code. The same mode appears in law, where a fabricated citation spreads through lawyers who cite it in good faith, and in medicine, where an erroneous entry spreads through clinicians who trust the record. Defenses that detect and isolate compromised agents (S3-E1 and S3-E2) cannot stop this mode, because no receiving agent is compromised. Only controls on the channel itself, such as circuit breakers on prices, provenance on records, and corroboration of citations, can.

\begin{insightbox}[Takeaway]
Populations of agents propagate harm in two ways. In infection, a payload makes each receiver re-emit it. In propagation without compromise, correct agents amplify what a compromised agent wrote into a shared environment. Agent-level defenses address only the first. Markets already show that the second needs controls on the channel, such as circuit breakers.
\end{insightbox}

\subsection{Outbreak Threshold for One Compromised Agent}
\label{subsec:branching}

The worm literature measures spread with a reproduction number, the expected number of new infections that one infected node causes. For agents, a simple version is $R = k \cdot p$, where $k$ is the fan-out of a channel (how many agents read each message, post, or price) and $p$ is the probability that a receiving agent both follows the payload and passes it on. We model spread from one seed as a Galton--Watson branching process with $\mathrm{Binomial}(k, p)$ offspring. The chance that the seed starts an outbreak that never dies out is $1 - q$, where $q$ is the smallest solution in $[0,1]$ of $q = (1 - p + p q)^k$. An outbreak is possible only when $R > 1$. The model assumes independent hops and a tree-like topology, so it describes the early phase of spread, and Agent Smith observed the same exponential regime in simulation~\cite{gu2024agentsmith}.

\begin{figure*}[t]
  \centering
  \includegraphics[width=0.98\textwidth]{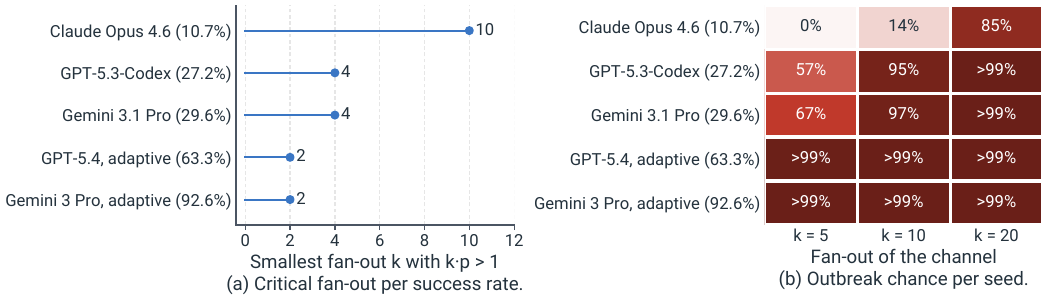}
  \caption{When one compromised agent can start an outbreak. (a) Smallest fan-out $k$ at which a channel turns supercritical ($k \cdot p > 1$) for the per-hop success rates that Table~\ref{tab:jailbreak-asr} reports. (b) Probability that one seed starts an outbreak that never dies out, from a Galton--Watson branching process with $\mathrm{Binomial}(k,p)$ offspring. A broadcast channel such as StockAgent's forum, where every agent reads every post, has a fan-out equal to the number of agents minus one.}
  \label{fig:branching}
\end{figure*}

We take $p$ from the injection success rates that Table~\ref{tab:jailbreak-asr} reports for current frontier models. Because a payload must also be passed on, these rates are upper reference points for $p$. Figure~\ref{fig:branching} shows the result. With the most resistant model in LivePI, Claude Opus~4.6 at 10.7\%~\cite{zhao2026livepi}, a channel needs a fan-out of \vtwoOpusCriticalFanout{} before one seed is expected to cause more than one new infection, and at a fan-out of 20 the chance of an outbreak reaches \vtwoOpusOutbreakTwenty{}. With Gemini~3.1 Pro at 29.6\%, a fan-out of \vtwoGeminiCriticalFanout{} suffices, and a fan-out of 10 gives \vtwoGeminiOutbreakTen{}. Under the transferred adaptive attacks~\cite{abdelnabi2026agents}, two readers per message are enough.

The model turns three qualitative concerns into design rules. First, hardening the model and limiting fan-out trade off one for one, because only their product matters. For example, a tenfold drop in per-hop success buys nothing if the channel's fan-out grows tenfold, and a broadcast forum like StockAgent's, where each post reaches every other agent (\S\ref{subsec:code-audit}), pushes even the most resistant model in Table~\ref{tab:jailbreak-asr} past the threshold once the population exceeds ten agents. Second, the defenses that work are those that reduce $k$ or $p$ for whole channels: message validation that treats peer output as data (S3-E3), provenance that lets a receiver discount unverified content, and rate limits that cap how many agents one message can reach. Third, monoculture removes the variation in $p$ that would otherwise slow spread. Because \vtwoGptFourFamily{} of 15 designs share one model family, a payload tuned for that family meets the same $p$ at almost every hop.

\begin{insightbox}[Takeaway]
Outbreak risk is the product of per-hop success and fan-out, so model hardening alone cannot make a broadcast channel safe. Even 10.7\% per-hop success, the best frontier rate in LivePI, makes any channel with ten or more readers supercritical.
\end{insightbox}

\section{Open Problems and Research Agenda}
\label{sec:open}

The ratings, the code audit, and the two preceding sections leave a set of problems that no single paper can close. We state 14 of them. Each starts from a gap that our data exposes, names what is known, and proposes a first step that a team could take within a year. Figure~\ref{fig:agenda} places the problems on the planes of Figure~\ref{fig:general-model} and on a time horizon, and Table~\ref{tab:open-problems} summarizes them. Existing agendas for multi-agent security~\cite{csdw2025open,hammond2025multi} and surveys of agent threats~\cite{deng2025agents} argue from first principles and examples. The agenda below starts from measured gaps, namely the controls that 15 published designs and 7 code releases leave out.

\begin{figure*}[t]
\centering
\begin{tikzpicture}[
  font=\footnotesize,
  colhead/.style={fill=#1, text=white, font=\footnotesize\bfseries, rounded corners=3pt, text width=3.05cm, minimum width=3.2cm, minimum height=0.8cm, align=center, inner sep=1pt},
  rowhead/.style={fill=fsGrayL, text=fsGray, font=\footnotesize\bfseries, rounded corners=3pt, minimum width=1.4cm, minimum height=1.8cm, align=center, inner sep=1pt},
  cell/.style={fill=#1!5, draw=#1!35, rounded corners=3pt, line width=0.4pt},
  chip/.style={draw=#1, fill=white, line width=0.7pt, rounded corners=6pt, text width=2.85cm, align=left, inner sep=2.6pt, minimum height=0.66cm},
  num/.style={fill=#1, text=white, circle, inner sep=0.2pt, minimum size=0.4cm, font=\footnotesize\bfseries}
]
\foreach \i/\c/\t in {0/fsBlue/{Inputs (S1)}, 1/fsPurple/{Agent (S2)}, 2/fsGreen/{Actions (R2, R3)}, 3/fsOrange/{Propagation (S3)}, 4/fsGray/{Oversight and governance}}{
  \node[colhead=\c] at ({1.55 + \i*3.25 + 1.6}, 6.15) {\t};
}
\foreach \y/\t in {4.75/{Near\\term}, 2.75/{Mid\\term}, 0.75/{Long\\term}}{
  \node[rowhead] at (0.7, \y) {\t};
}
\foreach \i/\c in {0/fsBlue, 1/fsPurple, 2/fsGreen, 3/fsOrange, 4/fsGray}{
  \foreach \y in {4.75, 2.75, 0.75}{
    \fill[cell=\c] ({1.55 + \i*3.25}, \y-0.9) rectangle ({1.55 + \i*3.25 + 3.2}, \y+0.9);
  }
}
\def\chipat#1#2#3#4#5{%
  \node[num=#3] (n) at ({1.55 + #1*3.25 + 0.36}, #2) {#4};
  \node[anchor=west, text width=2.2cm, align=left, inner sep=1pt, font=\footnotesize] (t) at ([xshift=1.5pt]n.east) {#5};
  \node[draw=#3, line width=0.7pt, rounded corners=6pt, fit=(n)(t), inner sep=2.4pt] {};}
\chipat{1}{4.75}{fsPurple}{3}{Auditing agent code at scale}
\chipat{2}{4.75}{fsGreen}{6}{Interlocks outside the model}
\chipat{3}{4.75}{fsOrange}{4}{Harm metrics for third parties}
\chipat{4}{5.2}{fsGray}{2}{Assessing deployed agents}
\chipat{4}{4.3}{fsGray}{10}{Codebooks for other domains}
\chipat{0}{2.75}{fsBlue}{5}{Provenance with correlated sources}
\chipat{1}{2.75}{fsPurple}{7}{Memory integrity and unlearning}
\chipat{3}{3.2}{fsOrange}{1}{Spillover testbeds}
\chipat{3}{2.3}{fsOrange}{8}{Breaking replication}
\chipat{4}{3.2}{fsGray}{12}{Identity and delegation}
\chipat{4}{2.3}{fsGray}{13}{Regulation and incident reports}
\chipat{0}{0.75}{fsBlue}{14}{Agents reading agents}
\chipat{1}{0.75}{fsPurple}{9}{Diversity against monoculture}
\chipat{2}{0.75}{fsGreen}{11}{Oversight at machine speed}
\end{tikzpicture}
\caption{Research agenda. The 14 open problems of \S\ref{sec:open} placed on the planes of the domain-independent model (Figure~\ref{fig:general-model}) and on a time horizon. Near-term problems need engineering and shared benchmarks, mid-term problems need new defenses or testbeds, and long-term problems need advances in science and governance.}
\label{fig:agenda}
\end{figure*}

\begin{table*}[t]
\centering
\caption{Open problems at a glance: the gap in our data that motivates each problem, the surfaces and domains it touches, and a first step.}
\label{tab:open-problems}
\footnotesize
\renewcommand{\arraystretch}{1.16}
\setlength{\tabcolsep}{3.5pt}
\begin{tabular}{>{\raggedleft\arraybackslash}p{0.3cm} >{\raggedright\arraybackslash}p{3.05cm} >{\raggedright\arraybackslash}p{4.3cm} >{\raggedright\arraybackslash}p{1.25cm} >{\raggedright\arraybackslash}p{1.95cm} >{\raggedright\arraybackslash}p{5.55cm}}
\toprule
\rowcolor{fsGrayL}
\textbf{\#} & \textbf{Open problem} & \textbf{Gap in our data} & \textbf{Surfaces} & \textbf{Domains} & \textbf{First step} \\
\midrule
1 & Agent-to-market spillover & 6 schemes have no S3 control; no study measures spillover & S3 & Trading & Shared-order-book simulator with a controlled fraction of compromised agents \\
2 & Assessing deployed agents & Code differs from paper in \auditCodeBetter{} + \auditCodeWorse{} of \auditRatings{} ratings & All & All & Disclosure template for out-of-model controls \\
3 & Auditing code at scale & LLM audit runs agree on \auditUnanimousPct{} of ratings & All & All & Expert-labeled benchmark of agent repositories \\
4 & Harm metrics for third parties & Benchmarks score only the attacked agent & S3 & All & Report price impact, downstream reads, reproduction number \\
5 & Provenance with correlated sources & No scheme combines provenance and corroboration & S1 & All & Count sources by upstream independence \\
6 & Interlocks outside the model & No interrupt without inference; no enforced loss limit & R2, R3 & All & Latency-bounded, verified policy engine based on Rule 15c3-5 and RTS~6 \\
7 & Memory integrity and unlearning & \auditPersistStore{} releases store memory unvalidated, \auditPickle{} with \texttt{pickle} & S2.3, R4 & All & Tamper-evident memory with quarantine and rollback of affected decisions \\
8 & Breaking replication & Fan-out 10 suffices even at 10.7\% per-hop success & S3, S2.1 & Multi-agent & Defense with a provable spread bound under adaptive attack \\
9 & Diversity against monoculture & \vtwoGptFourFamily{} of 15 schemes share one model family & S2, S3 & All & Joint attack success across backbone pairs \\
10 & Codebooks for other domains & Controls missing for design reasons, not finance & All & Healthcare, law & Rate published clinical and legal agents with adapted indicators \\
11 & Oversight at machine speed & Human gates fail under automation bias & R2, R3 & Healthcare, law & Decide which actions get a human gate and how long it needs \\
12 & Identity and delegation & Cascades cross firms without attribution & S3 & All & Scoped, revocable delegation for agent protocols \\
13 & Regulation and incident reports & Mandatory controls exist but are not mapped to agents & All & All & Incident taxonomy indexed by surface and phase \\
14 & Agents reading agents & Inputs increasingly written by agents & S1, S3 & All & Provenance and friction in the fastest channels \\
\bottomrule
\end{tabular}
\end{table*}

\subsection{Measuring the Problem}
\label{subsec:op-measure}

\noindent\textbf{1. Empirical evidence for agent-to-market spillover.} Six schemes receive no credit on S3, and we found no study that measures how far a compromised LLM agent moves a market in which other LLM agents trade. Although LLM market simulators exist~\cite{papadakis_stocksim_2025,gao_simulating_2024,noauthor_twinmarket_nodate} and attack studies measure harm inside one agent stack~\cite{sua2026contagion,na2026poisoning}, none combines a shared order book, a population of heterogeneous agents, and a controlled fraction of compromised ones. A first step is a simulator with an exchange-grade matching engine in which the fraction of compromised agents, the backbone mix, and the channel fan-out of \S\ref{subsec:branching} are experimental variables, and price impact, herding, and recovery time are outcomes. The same testbed would let the community design and test circuit breakers for agents, whose reaction times differ from those of human traders.

\noindent\textbf{2. Assessing deployed agents.} Our ratings describe published designs, and the audit shows that released code diverges from them in \auditCodeBetter{} ratings upward and \auditCodeWorse{} downward (\S\ref{subsec:code-audit}). Products add a third layer that neither papers nor repositories reveal. The AI Agent Index found that developers of deployed agents publish ample information about capabilities but little about safety and risk management~\cite{casper2025agent}. A first step is a disclosure template, the agent analogue of a model card, that asks deployers which \frameworkShort{} controls they implement outside the model, how fast their interrupt path is, and which inputs they treat as untrusted.

\noindent\textbf{3. Auditing agent code at scale.} Our three LLM audit runs agreed on \auditUnanimousPct{} of ratings and \auditIndAgreePct{} of indicator judgments, which is useful but leaves room for systematic error. Because the auditor saw our paper-level ratings, its labels may also be anchored to them (Appendix~\ref{appendix:code-audit}). The next questions are whether blind audits converge to the same ratings, how to calibrate an LLM auditor against human experts, and how to combine LLM judgment with static analysis that cannot be talked out of a finding. A shared benchmark of agent repositories with expert labels for each codebook indicator would let the community measure auditors the way it measures models.

\noindent\textbf{4. Harm metrics for third parties.} Agent security benchmarks report attack success rate on the attacked agent~\cite{debenedetti2025agentdojo,zhang2025agent,zhan2024injecagent}. The harms that make trading high-stakes fall on others: counterparties who trade at manipulated prices, patients who inherit a corrupted record, courts that receive fabricated citations. Yet the field lacks metrics for these externalities, such as the price impact per compromised dollar, the number of downstream decisions that read a poisoned record, or the reproduction number of \S\ref{subsec:branching} measured on real channels. Benchmarks that report such metrics would make the cost of a vulnerability comparable across domains.

\subsection{Building Defenses}
\label{subsec:op-defend}

\noindent\textbf{5. Provenance and corroboration when sources are correlated.} No scheme combines source provenance with cross-source corroboration (Table~\ref{tab:combined-eval}), and both are harder than they look. For example, corroboration assumes independent sources, but syndication, shared wire services, and coordinated posting make sources dependent, so an attacker who controls one hub controls the majority. Provenance standards for media, signed data feeds, and watermarks for generated text~\cite{kirchenbauer2023watermark} each cover part of the problem. A first step is a corroboration rule that counts sources by their upstream independence, evaluated against an adversary who can create sources cheaply.

\noindent\textbf{6. Interlocks outside the model with bounded latency.} No paper describes an interrupt that does not wait for LLM inference, and no release enforces a loss limit outside the LLM. For general-purpose agents, system-level designs that separate control flow from untrusted data~\cite{debenedetti2026defeating,beurerkellner2025design}, enforce privileges on tool calls~\cite{shi2025progent,wu2025isolategpt}, or track information flow~\cite{costa2025securing,wu2024system} show how to place a policy engine around an agent. The missing piece is a trading-grade interlock whose worst-case latency is bounded and verified, whose rules a regulator can read, and whose inputs cannot carry injected content. Pre-trade controls under Rule 15c3-5 and RTS~6~\cite{sec2010rule15c35,eu2017rts6} are the specification to start from.

\noindent\textbf{7. Memory integrity, forensics, and unlearning.} Of the audited releases, \auditPersistStore{} store memory without validation and \auditPickle{} load it with \texttt{pickle}, and TradingAgents' decision log can be forged by model output that reproduces its delimiter (\S\ref{subsec:code-audit}). Memory injection succeeds more often than prompt injection against Web3 agents, and prompt-injection defenses do little once the stored context is corrupted~\cite{patlan2025real}. Agents need tamper-evident memory that records who wrote each entry, a quarantine for unverified experience, and a way to undo the decisions that a poisoned entry influenced. Forensic replay, which the recommendations for R4 ask for, needs this record to exist first.

\noindent\textbf{8. Breaking replication and bounding fan-out.} The branching model shows that outbreak risk is the product of per-hop success and fan-out (\S\ref{subsec:branching}). Morris~II's guardrail detects self-replicating prompts with a true-positive rate of 1.0 and a false-positive rate of 0.015 in its evaluation~\cite{cohen2024here}, LLM tagging reduces infection inside multi-agent systems~\cite{lee2024promptinfection}, and topology-guided detection on the utterance graph recovers over 40\% of the performance lost to prompt injection~\cite{wang2025gsafeguard}. Agent Smith's authors derive a condition under which a defense provably restrains infectious jailbreaks but leave its practical design open~\cite{gu2024agentsmith}. The open problem is a defense that meets such a condition under adaptive attack~\cite{nasr2025attacker}, together with protocol-level fan-out limits for agent-to-agent standards.

\noindent\textbf{9. Diversity against monoculture.} In our corpus, \vtwoGptFourFamily{} of the 15 schemes share one model family, and errors are correlated even across providers~\cite{kim2025correlated}. N-version programming tolerated software faults by running independently developed versions and voting~\cite{avizienis1985version}. The same idea protects agents only if different backbones fail independently under attack. A first step is to measure the joint success rate of the same payloads against pairs of backbones and to ask whether a committee of diverse models raises the cost of a market-moving attack enough to justify its latency.

\subsection{Generalization and Governance}
\label{subsec:op-general}

\noindent\textbf{10. Codebooks for other high-stakes domains.} Table~\ref{tab:farsight-general} suggests which \frameworkShort{} indicators transfer unchanged (provenance, out-of-model limits, memory gates, peer-message validation) and which need domain content (what counts as an early warning in a hospital, or as an interrupt in a law firm). A first step is to apply a healthcare codebook to published clinical agents such as AgentClinic, Agent Hospital, and Polaris~\cite{schmidgall2024agentclinic,li2024agenthospital,mukherjee2024polaris}, and a legal codebook to drafting and research agents, and to compare their rating distributions with Figure~\ref{fig:rating-distribution}. If the same controls are missing there, the cause lies in how agents are designed, and finance is only where it shows first.

\noindent\textbf{11. Human oversight at machine speed.} Every domain in Figure~\ref{fig:domain-profiles} that relies on a human gate rates medium on reliance on in-model safety, because the human is the control. Human review fails in predictable ways. Automation bias leads reviewers to accept wrong advice~\cite{goddard2012automationbias}, and automation leaves operators least practiced exactly when they must take over~\cite{bainbridge1983ironies}. In trading, a human cannot review a decision that must happen within a second. The open problem is to decide which actions deserve a human gate (irreversible, large, or unfamiliar ones), how much time the gate needs, and how to keep reviewers alert when nearly every draft they see is correct.

\noindent\textbf{12. Identity, delegation, and accountability.} If a cascade crosses firms, attribution requires knowing which agent acted for whom. Proposals for agent identifiers, activity logs, and real-time monitoring~\cite{chan2024visibility,chan2025infrastructure}, and for authenticated delegation built on OAuth~2.0 and OpenID Connect~\cite{south2025authenticated}, give agents verifiable principals. Agent-to-agent protocols and MCP need the same properties for messages and tools~\cite{hou2026context}. Open questions include how to scope a delegation so that an injected instruction cannot widen it, and how to revoke an agent's authority across every service it touches within the time a market cascade takes.

\noindent\textbf{13. Mapping regulation to agent controls and reporting incidents.} The controls that the schemes lack are already mandatory for algorithmic trading~\cite{sec2010rule15c35,eu2017rts6}, and regulators now ask about correlated trading by AI agents~\cite{donachie2026lawmakers,boe2026breeden}. Healthcare and law have their own anchors in order-entry checks and attorney verification duties~\cite{kaushal2003effects,sdny2023mata}. Yet we know of no public mapping that translates these duties into agent controls that an auditor can check. Aviation collects near misses and the AI Incident Database collects AI failures~\cite{mcgregor2021preventing}, but agent failures lack a shared taxonomy that names where in the agent they occurred. The \frameworkShort{} indicators offer a starting vocabulary for such a mapping, and an incident taxonomy indexed by surface (S1--S3) and phase (R1--R4) would let reports from different domains be compared.

\noindent\textbf{14. Agents that read what agents write.} Prices, records, case law, and the web increasingly contain text and signals that agents produced. If models train on such text, they degrade over generations~\cite{shumailov2024collapse}, and when agents read each other's outputs, they can lock into the informational cascades that economists describe for people~\cite{bikhchandani1992theory}. The long-term question is how an information environment stays reliable when most of its writers and readers are agents. Provenance for generated content, diversity of sources, and deliberate friction in the fastest channels are candidate answers, and none has been tested at the scale at which agents now operate.

\begin{figure*}[t]
\centering
\begin{tikzpicture}[
  font=\footnotesize,
  ticket/.style={draw=#1, line width=0.8pt, rounded corners=4pt, fill=white},
  head/.style={fill=#1, rounded corners=3pt},
  field/.style={draw=#1, fill=#1!9, rounded corners=2.5pt, line width=0.6pt},
  cost/.style={-{Stealth[length=2.3mm]}, line width=0.8pt, draw=#1}
]
\draw[ticket=fsGray] (0,1.0) rectangle (4.4,5.2);
\fill[head=fsGray!18] (0.08,4.62) rectangle (4.32,5.12);
\node[anchor=west, font=\footnotesize\bfseries] at (0.2,4.87) {\faFileInvoiceDollar\ \ Stock order};
\draw[field=fsBlue] (0.15,3.85) rectangle (4.25,4.4);
\node[anchor=west, font=\footnotesize\bfseries, text=fsBlue] at (0.25,4.125) {Direction};
\node[anchor=east] at (4.15,4.125) {buy or sell};
\draw[field=fsGray] (0.15,3.17) rectangle (4.25,3.72);
\node[anchor=west, font=\footnotesize\bfseries, text=fsGray] at (0.25,3.445) {Size};
\node[anchor=east] at (4.15,3.445) {shares};
\node[anchor=north west, text width=3.95cm, align=left] at (0.2,2.95) {Worst case: the wrong direction. A loss without bound needs an explicit short, the decision that risk checks already watch.};
\fill[fsBlue] (0.15,0.55) rectangle (0.4,0.8);
\node[anchor=west] at (0.45,0.675) {checked by direction-based tests};
\fill[fsRed] (0.15,0.12) rectangle (0.4,0.37);
\node[anchor=west] at (0.45,0.245) {corruptible with the right direction};

\draw[ticket=fsGray] (5.0,0.3) rectangle (10.6,5.2);
\fill[head=fsGray!18] (5.08,4.62) rectangle (10.52,5.12);
\node[anchor=west, font=\footnotesize\bfseries] at (5.2,4.87) {\faFileContract\ \ Options order};
\foreach \y/\c/\a/\b in {
  3.85/fsBlue/Direction/{call or put, long or short},
  3.17/fsRed/Strike/{sets the breakeven},
  2.49/fsRed/Expiry/{sets the deadline},
  1.81/fsRed/{Implied volatility}/{sets the price paid},
  1.13/fsRed/Legs/{set the maximum loss},
  0.45/fsGray/Size/{contracts}}
{
  \draw[field=\c] (5.15,\y) rectangle (10.45,\y+0.55);
  \node[anchor=west, font=\footnotesize\bfseries, text=\c] at (5.25,\y+0.275) {\a};
  \node[anchor=east] at (10.35,\y+0.275) {\b};
}

\node[anchor=west, font=\footnotesize\bfseries] at (11.25,4.87) {With the direction still correct};
\draw[cost=fsGreen] (10.6,4.125) -- (11.25,4.125);
\node[anchor=west, text=fsGreen] at (11.3,4.125) {\faCheckCircle\ a direction check passes};
\foreach \y/\t in {
  3.445/{the move never reaches the breakeven},
  2.765/{the catalyst arrives after expiry},
  2.085/{volatility collapses after the event},
  1.405/{a \$400 cap becomes \$9{,}900, or none}}
{
  \draw[cost=fsRed] (10.6,\y) -- (11.25,\y);
  \node[anchor=west, text=fsRed] at (11.3,\y) {\t};
}
\draw[cost=fsPurple] (10.6,0.725) -- (11.25,0.725);
\node[anchor=west, text=fsPurple] at (11.3,0.725) {\faUserSecret\ a counterparty fills the visible order};
\end{tikzpicture}
\caption{An options order carries five decisions where a stock order carries one. An attacker can corrupt four of them (red) without touching the direction the agent chose, so a check that compares the order's direction with the agent's own analysis passes while the strike, expiry, volatility, or legs carry the loss. In a thin contract, the order itself can reveal the agent to a counterparty who trades against it.}
\label{fig:option-order}
\end{figure*}

\noindent\textbf{Actions for each community.} The problems above take years, but four groups can act on our findings today. \textit{Agent researchers} can report \frameworkShort{} ratings for new designs, release code that matches the paper, and evaluate attacks by their effect on other agents, not only on the target. \textit{Builders and deployers} can move every limit that matters out of the prompt: corroborate sources before acting, keep an interrupt path that does not call the model, cap positions and fan-out, validate memory writes, and never execute model-written code on the host. \textit{Researchers in other high-stakes domains} can apply the codebook of Appendix~\ref{appendix:codebook} with the substitutions of Table~\ref{tab:farsight-general} and test whether their agents repeat the pattern we found. \textit{Regulators and standards bodies} can ask agent deployers the questions that Rule 15c3-5 already asks of algorithmic traders, and require incident reports that name the surface and the phase in which a failure occurred.

\section{Discussion}
\label{sec:discussion}

The 15 schemes trade stocks, and the first agents that brokers opened to retail customers did too. This section asks what changes when the same agents trade options (\S\ref{subsec:options}), then discusses the deployment questions that our ratings raise, the limitations of this work, and its potential harms (\S\ref{subsec:deployment}).

\subsection{Options Agents: More Decisions, Lower Floors}
\label{subsec:options}

Options are the next market that agents enter, and they raise the stakes of every surface in \frameworkShort{}. Robinhood announced in July 2026 that every US customer can now let an agent trade options~\cite{robinhood2026agenticoptions}, and tastytrade publishes an MCP server through which an agent with write access can trade options on stocks, indexes, and futures, multi-leg orders included~\cite{tastytrade2026tastytrade}. The market that these agents enter is large and increasingly short-dated. In July 2026, contracts that expire on the day they trade accounted for a record 66.2\% of all S\&P~500 index-options volume~\cite{cboe2026cboe}, and by mid-2026 their volume was up 46.2\% year to date, at more than 20 million contracts a day~\cite{cboe2026state}.

\noindent\textbf{One decision becomes five.} A stock order gives an attacker one decision to corrupt, the direction, plus a size that risk limits already bound. An options order carries five decisions: the direction, the strike that sets the breakeven, the expiry that sets the deadline, the implied volatility that sets the price paid, and the legs that set the maximum loss. Figure~\ref{fig:option-order} contrasts the two orders. The four added fields share one property that makes them dangerous. An attacker can move each of them while the agent's view of the market stays correct, so a check that compares the order's direction with the agent's own analysis passes. Attack studies on equities agents measure whether a decision flips or a profit falls~\cite{rizvani2026adversarial,yan2025tradetrap}, and neither measure registers an attack that leaves the decision intact.

\noindent\textbf{Losing with the right view.} Each added field turns a correct view into a loss in its own way. A call bought at a strike far above the expected move expires worthless even if the stock rises, and an expiry chosen before the catalyst ends the position before the news arrives. Volatility is subtler. Before a scheduled event, option prices embed a high implied volatility that usually collapses once the event passes, so an agent persuaded to buy just before an earnings release pays for a move that the market had already priced in. Figure~\ref{fig:payoff}(b) shows such a case. A call bought at 80\% implied volatility loses money once implied volatility drops to 25\%, unless the stock gains more than 8.7\%. None of these losses requires a false headline. Urgent framing (``two days left'') or an inflated volatility statistic is enough, and both are hard to catch, because the Greeks and implied volatility come out of a pricing model~\cite{black1973pricing} and cannot be checked against a price chart.

\begin{figure}[t]
\centering
\begin{tikzpicture}
\begin{groupplot}[
  group style={group size=1 by 2, vertical sep=1.45cm},
  scale only axis, width=6.55cm, height=2.1cm,
  tick label style={font=\footnotesize}, label style={font=\footnotesize},
  title style={font=\footnotesize\bfseries, yshift=-4pt},
  axis lines=left,
  every axis plot/.append style={line width=1.0pt},
  xlabel style={font=\footnotesize, yshift=3pt}, ylabel style={font=\footnotesize, yshift=-4pt},
]
\nextgroupplot[title={(a) One leg dropped}, xlabel={Stock price at expiry (\$)},
  ylabel={P\&L per contract (\$)}, xmin=0, xmax=112, ymin=-10.6, ymax=2.0,
  xtick={0,50,100},
  ytick={-10,-5,0}, yticklabels={$-$10k,$-$5k,0}]
\addplot[name path=NK, color=okverm] coordinates {(0,-9.9) (100,0.1) (112,0.1)};
\addplot[name path=SP, color=okblue, line width=1.3pt] coordinates {(0,-0.4) (95,-0.4) (100,0.1) (112,0.1)};
\addplot[fill=okverm!16, draw=none, forget plot] fill between[of=SP and NK, soft clip={domain=0:95}];
\node[font=\footnotesize, text=okblue, anchor=south west, align=left, inner sep=1pt] at (axis cs:3,-0.1) {spread: max loss \$400};
\node[font=\footnotesize, text=okverm, anchor=south east, align=right, inner sep=1pt] at (axis cs:111,-10.3) {long put dropped:\\ max loss \$9{,}900\\ (25$\times$)};
\nextgroupplot[title={(b) Volatility collapse}, xlabel={Stock price after the event (\$)},
  ylabel={P\&L per share (\$)}, xmin=90, xmax=120, ymin=-10, ymax=10,
  xtick={90,100,110,120}, ytick={-10,-5,0,5,10}]
\addplot[name path=CR, color=okverm] coordinates {(90.00,-8.932) (90.50,-8.893) (91.00,-8.848) (91.50,-8.796) (92.00,-8.738) (92.50,-8.671) (93.00,-8.595) (93.50,-8.511) (94.00,-8.416) (94.50,-8.311) (95.00,-8.194) (95.50,-8.066) (96.00,-7.926) (96.50,-7.773) (97.00,-7.607) (97.50,-7.427) (98.00,-7.234) (98.50,-7.026) (99.00,-6.805) (99.50,-6.569) (100.00,-6.319) (100.50,-6.055) (101.00,-5.777) (101.50,-5.485) (102.00,-5.180) (102.50,-4.861) (103.00,-4.530) (103.50,-4.186) (104.00,-3.830) (104.50,-3.463) (105.00,-3.085) (105.50,-2.697) (106.00,-2.299) (106.50,-1.891) (107.00,-1.475) (107.50,-1.051) (108.00,-0.620) (108.50,-0.181) (109.00,0.264) (109.50,0.714) (110.00,1.171) (110.50,1.632) (111.00,2.097) (111.50,2.567) (112.00,3.040) (112.50,3.517) (113.00,3.996) (113.50,4.478) (114.00,4.962) (114.50,5.449) (115.00,5.937) (115.50,6.427) (116.00,6.918) (116.50,7.411) (117.00,7.905) (117.50,8.399) (118.00,8.894) (118.50,9.390) (119.00,9.887) (119.50,10.384) (120.00,10.882)};
\addplot[color=okblue, dashed] coordinates {(90.00,-4.692) (90.50,-4.509) (91.00,-4.322) (91.50,-4.129) (92.00,-3.932) (92.50,-3.731) (93.00,-3.524) (93.50,-3.313) (94.00,-3.098) (94.50,-2.878) (95.00,-2.653) (95.50,-2.423) (96.00,-2.189) (96.50,-1.951) (97.00,-1.707) (97.50,-1.459) (98.00,-1.207) (98.50,-0.950) (99.00,-0.689) (99.50,-0.423) (100.00,-0.153) (100.50,0.122) (101.00,0.401) (101.50,0.684) (102.00,0.972) (102.50,1.264) (103.00,1.560) (103.50,1.860) (104.00,2.164) (104.50,2.473) (105.00,2.785) (105.50,3.102) (106.00,3.422) (106.50,3.746) (107.00,4.075) (107.50,4.407) (108.00,4.742) (108.50,5.082) (109.00,5.425) (109.50,5.771) (110.00,6.121) (110.50,6.475) (111.00,6.832) (111.50,7.193) (112.00,7.556) (112.50,7.923) (113.00,8.294) (113.50,8.667) (114.00,9.044) (114.50,9.423) (115.00,9.806) (115.50,10.191) (116.00,10.580) (116.50,10.971) (117.00,11.365) (117.50,11.762) (118.00,12.161) (118.50,12.563) (119.00,12.968) (119.50,13.375) (120.00,13.784)};
\addplot[name path=ZERO, draw=none, forget plot] coordinates {(90,0) (120,0)};
\addplot[fill=okorange!45, draw=none, forget plot] fill between[of=CR and ZERO, soft clip={domain=100:108.70}];
\draw[black!45, line width=0.4pt] (axis cs:90,0) -- (axis cs:120,0);
\draw[dashed, black!45, line width=0.5pt] (axis cs:100,-6.32) -- (axis cs:100,10);
\node[font=\footnotesize, text=okblue, anchor=south east, inner sep=1pt] at (axis cs:107.3,5.6) {IV held};
\node[font=\footnotesize, text=okverm, anchor=north west, inner sep=1pt] at (axis cs:114.2,3.4) {IV falls};
\node[font=\footnotesize, text=black!85, anchor=north west, inner sep=0.5pt] at (axis cs:100.6,-7.2) {stock up, losing};
\draw[black!60, line width=0.4pt, -{Stealth[length=3pt]}] (axis cs:104.2,-7.1) -- (axis cs:104.2,-2.6);
\draw[black!60, line width=0.4pt, -{Stealth[length=3pt]}] (axis cs:112.6,-2.3) -- (axis cs:108.9,-0.3);
\node[font=\footnotesize, text=black!75, anchor=north east, align=right, inner sep=1pt] at (axis cs:119.6,-2.4) {breakeven\\ +8.7\%};
\end{groupplot}
\end{tikzpicture}
\caption{Losing money while the direction stays correct. (a)~A bull put spread (short 100 put, long 95 put, \$1.00 net credit) can lose at most \$400 per contract. Without its long put, the same view can lose \$9{,}900, and the shading marks the loss that the missing leg would have bounded. (b)~A 30-day at-the-money call bought at 80\% implied volatility before a scheduled event loses money once implied volatility drops to 25\%, unless the stock gains more than 8.7\%. The shaded band marks a correct directional view that still loses. Black--Scholes values with illustrative parameters~\cite{black1973pricing,hull1997options}.}
\label{fig:payoff}
\end{figure}

\noindent\textbf{One missing leg removes the floor.} The worst case needs no change of direction at all. Traders cap the risk of a written option by buying a further option of the same expiry, and the bound then comes from the pair, not from either leg~\cite{hull1997options}. Remove the protective leg and the same view becomes an uncovered position. Figure~\ref{fig:payoff}(a) uses a bull put spread, short the 100 put and long the 95 put, which collects a \$1.00 credit and can lose at most \$400 per contract. Without the long put, the position can lose \$9{,}900, about 25 times as much, and an uncovered call has no bound at all, because a stock price has none~\cite{sec2026investorbulletin}. On a stock, an unbounded position requires an explicit short, which is the decision that risk checks already watch. On an option, it requires only that the agent be talked out of a leg that, in most months, looks like a cost that never pays. The pattern has a precedent with human traders. In 2018, after natural gas jumped 18\%, a fund that had sold uncovered options on it informed about 290 clients that their money was gone and that they might owe their clearing broker more~\cite{cnbc2018risky}.

\noindent\textbf{The counterparty can be the attacker.} Options add a channel that pays the attacker directly. Liquidity is split across many strikes and expiries, so a single contract trades thinly, and retail option flow is concentrated: nearly 90\% of payment for order flow goes to three wholesalers~\cite{bryzgalova2023retail}, which specialize in executing against retail option orders~\cite{ernst2026payment}. In a thin contract, an agent that trades by fixed rules can be the only visible interest, and its displayed orders reveal its entry conditions, strikes, and timing. An adversary who learns these rules can publish an input that triggers the agent's next order and then fill that order as its counterparty. Each trade is small and ordinary, the victim's loss is the attacker's profit, and the attack survives a ban on uncovered writing because it needs only bought premium.

\noindent\textbf{Option flow reaches the stock market.} Options also widen the S3 surface of Section~\ref{sec:cascades}. Dealers who sell options hedge them in the underlying, and when they are short gamma, the hedging trades in the direction of the price move. That hedging demand has been linked to intraday momentum across futures markets~\cite{baltussen2021hedging}. Correlated option orders from many agents that share one input can therefore move the stock through the dealers who take the other side, a shared-environment channel that needs no compromised receiver.

\begin{table*}[t]
\centering
\caption{Stock orders and options orders compared from the attacker's side. Each row names a property of the order or its market, and the last column states what the property gives an attacker who can write text that the agent reads.}
\label{tab:options-vs-stock}
\footnotesize
\renewcommand{\arraystretch}{1.18}
\setlength{\tabcolsep}{4pt}
\begin{tabular}{>{\raggedright\arraybackslash}p{2.6cm} >{\raggedright\arraybackslash}p{3.7cm} >{\raggedright\arraybackslash}p{4.6cm} >{\raggedright\arraybackslash}p{5.3cm}}
\toprule
\rowcolor{fsGrayL}
\textbf{Property} & \textbf{Stock order} & \textbf{Options order} & \textbf{What it gives the attacker} \\
\midrule
Decisions in the order & Direction, plus size & Direction, strike, expiry, implied volatility, and legs, plus size & Four fields to move while the direction stays correct \\
Worst case & Capital committed; no bound only for an explicit short~\cite{sec2026investorbulletin} & No bound for an uncovered call; one dropped leg removes the cap of a spread~\cite{hull1997options} & A choice of risk class, not just of direction \\
Loss with a correct view & Rare, since a correct view gains & Common: a strike the move never reaches, a late move, or a volatility collapse & Harm that a direction check reports as correct \\
Numbers a reviewer can check & Price and volume, visible on a chart & Greeks and implied volatility, derived from a pricing model~\cite{black1973pricing} & Misstated values that few reviewers can catch \\
Liquidity & Deep books and tight spreads & Thin books per contract and wide spreads & A displaced strike or expiry pays the spread \\
Visibility of the agent's orders & Anonymous in a deep book & Possibly the only displayed interest in a contract; flow concentrated at three wholesalers~\cite{bryzgalova2023retail} & Profiling of a fixed-rule agent, then trading against it \\
Lifecycle events & None & Exercise, assignment, and pin risk, compressed into hours for contracts that expire the same day~\cite{cboe2026state} & Unplanned positions from events the agent mishandles \\
Effect on other markets & Orders move the stock & Dealer hedging carries option flow into the stock~\cite{baltussen2021hedging} & A second route from agent orders to prices (S3) \\
\bottomrule
\end{tabular}
\end{table*}

\noindent\textbf{Controls for options agents.} Table~\ref{tab:options-vs-stock} summarizes the differences, and each row points to a control that works without detecting manipulated text. The loss limit that R3 asks for must be the maximum loss of the whole position at expiry, computed from all legs before entry and enforced outside the model, because a notional or premium limit does not see a dropped leg. Agent orders should default to defined-risk structures, and every leg, strike, and expiry should be checked against the plan the agent stated before it read new inputs. Greeks and implied volatility should be recomputed from the option chain, not taken from text. A ban on uncovered writing helps, but it is a platform rule that a product change can lift, and it leaves the strike, expiry, volatility, and counterparty channels open. Finally, agents that trade thin contracts should randomize timing and size or route through mechanisms that hide them, so that their rules are harder to learn.

\begin{insightbox}[Takeaway]
On options, an attacker does not need to flip a decision. Four of the five fields of an options order can be corrupted while the direction stays correct, a single dropped leg turns a bounded loss into an unbounded one, and the attacker can profit as the counterparty. Every \frameworkShort{} surface carries over, and each becomes more costly.
\end{insightbox}

\subsection{Deployment, Practice, and Scope}
\label{subsec:deployment}

\noindent \textbf{Flash crashes versus normal pullbacks.}
Volatility is a normal feature of markets, so our robustness metrics target adversarial flash crashes, not ordinary price moves. The two differ in explainability. A normal pullback follows identifiable fundamentals or news, such as an earnings miss or a rate decision, and is interpretable and, in principle, recoverable. An adversarial flash crash is deliberately manufactured, decoupled from fundamentals, and leaves participants unable to identify a cause. This lack of a clear cause is what makes adversarial flash crashes dangerous, because an unexplained drop can invite panic selling, stop-loss cascades, and quant-bot liquidation, which feed back into trading agents and amplify the shock.

\noindent \textbf{Applicability to deployed systems.}
Our evaluation focuses on academic schemes, which raises the question of whether production financial agents share the same vulnerabilities. Proprietary systems, such as those deployed by quantitative hedge funds, are not publicly accessible for security auditing, but three observations suggest that the risks we identify apply to them as well.
First, production systems such as RockAlpha~\cite{rockalpha2025}, Composer~\cite{composer2025}, and Nof1.ai~\cite{nof1ai2025} employ architecturally similar patterns (LLM-driven signal interpretation, API-based execution, and multi-source data ingestion) that expose the same attack surfaces documented in our S1--S3 analysis.
Second, the U.S. SEC's establishment of a dedicated AI Task Force~\cite{sec2025ai_task_force} reflects regulatory recognition that LLM-based trading agents pose systemic risks that merit oversight.
Third, the architectural weaknesses we identify (unverified data ingestion, lack of prompt isolation, absence of circuit breakers) follow from the speed-first ethos of financial systems, which production systems share. Production systems operating under the same constraints are unlikely to have solved these problems without explicit security engineering.

\noindent \textbf{Practicality and latency.}
A natural concern is whether our recommended defenses introduce unacceptable latency. The fast path that we recommend for R2 runs outside the LLM and adds rule checks instead of inference calls, so it shortens the reaction to a shock instead of lengthening it. The remaining controls, such as source corroboration and pre-committed loss limits, follow professional risk-hedging practice, including portfolio-level risk budgeting, volatility-regime detection, and multi-leg hedging.
Sub-second latency is also not a binding constraint for strategies with daily or multi-day holding periods, such as \textit{swing trading}. For these strategies, the agent's edge lies in continuous, around-the-clock information ingestion. Human traders cannot maintain constant vigilance, whereas an autonomous agent monitors and reasons over information streams without attentional lapses. The trade-off between latency and returns across strategy horizons remains to be measured. Faster LLM decisions alone, however, do not make a population of agents safer. In our simulated market, 400 agents that decide every 60 seconds instead of every 240 deepen a 2010-sized crash to \sensOutSixty{} instead of \sensOutTwoForty{}, because they sell into the fall sooner (\S\ref{subsec:sim-controls}).

\noindent \textbf{Trading agents in the wild.}
Retail-built LLM trading agents are already operating with real capital. On X, Reddit, and Xiaohongshu (a major Chinese lifestyle and social-media platform), users share live returns yet never disclose architectures. Most of these agents are built through ``\textit{vibe coding}'' (lightly scaffolded, prompt-driven construction using tools such as Claude Code or Cursor) with minimal programming expertise, which lowers the deployment barrier while expanding the S1 attack surface. Because free data sources are limited in coverage, most no-code agents converge on a narrow set of feeds, which enables \textit{supply-side poisoning}. An adversary offering a free, high-quality-looking source could steer many agents simultaneously with \textit{zero knowledge of their internals}, risking correlated trades and flash-crash amplification~\cite{karppi2015hackcrash, itbrew2025flash}.
The same broad execution authority that makes autonomous trading attractive also means that, under a poisoned data feed, an agent can propagate aggressive erroneous trades at the same speed as legitimate ones. Trading agents therefore need a \textit{robustness} and \textit{security} codebook that holds under adversarial and stochastic market conditions, which is what \frameworkShort{} provides.

\noindent \textbf{Limitations.}
Our ratings describe 15 published designs, whose evidence comes from backtests, simulations, or case studies with no real capital, and they cannot replace penetration testing of running agents or of proprietary products such as RockAlpha~\cite{rockalpha2025} and Robinhood's agentic trading~\cite{robinhood2026agentic}. Because the indicators are read from papers, a \ding{109} means only that a control is not evidenced. The code audit shows that this gap runs both ways, covers only the \auditSchemes{} schemes with an official release, and depends on how partial evidence counts, since its agreement with the paper-level ratings drops from \auditMatch{} to \auditMatchStrict{} of \auditRatings{} ratings when partial indicators count as absent (Appendix~\ref{appendix:code-audit}). Finally, Sections~\ref{sec:beyond}--\ref{sec:open} and \S\ref{subsec:options} extend the systematization by analysis. The domain profiles are our assessment of the cited literature, no options agent was attacked, and the branching model assumes independent hops, so its thresholds describe the early phase of spread. The market experiment is a one-asset model driven by decisions measured on a grid of market states, so its agent counts describe a thinly traded stock and grow with depth (\S\ref{subsec:sim-robust}); calibrating the ABIDES simulator~\cite{byrd2020abides} to one stock's order book would give the counts for that stock.

\noindent\textbf{Potential harms.} A systematization of attack surfaces can guide attackers as well as defenders, and naming weaknesses in released code points to concrete targets. We limit this risk in three ways. We describe each surface at the level of design controls and release no attack code, payloads, or market-moving content. The weaknesses we report in released code concern research prototypes without order-execution paths, and the authors of all 15 schemes received our assessment before submission. The attacks we discuss, such as planting false news or coordinating trades to move prices, are forms of market manipulation that securities law already prohibits, so they add little for an attacker but show deployers which controls are missing. Finally, a partial or low-risk rating in Table~\ref{tab:combined-eval} certifies nothing about a deployed system, which adds execution paths that the design and the released code lack.

\section{Conclusion}
We rated 15 academic trading agents with one codebook and audited \auditSchemes{} code releases. No paper described a halt that bypasses the LLM, and no release enforced a loss limit outside it. In a simulated market, a few hundred agents that decided as three LLMs did caused and amplified flash crashes that the exchange pause missed, and a fast path and hard limits stopped both. These controls belong outside the model.

\section*{Ethical Considerations}
This work aims to help researchers and deployers build trading agents that are robust and secure. We assessed published designs and audited released code statically. We ran no attack against the authors' systems, live markets, or production trading infrastructure, and we release no attack code or market-moving content. We confirmed each code-level weakness that we report (\S\ref{subsec:code-audit}) by reading the source rather than by exploiting it.

\noindent\textbf{Responsible disclosure.} Before submission, we emailed the first authors of all 15 schemes with the ratings and justifications for their scheme. The first author of StockAgent replied, acknowledged that the design had not considered these issues, and said that the team would improve the system accordingly. We will reflect any further responses in the final version.

\section*{Open Science}
We are committed to open science and to enabling independent verification and reuse of our results. This systematization is built entirely on publicly available academic papers; the 15 evaluated financial LLM trading systems and their original sources are cited throughout the paper. To make our evaluation reproducible, we release a complete artifact package containing: (i)~the \frameworkShort{} codebook with all evidence-based indicator definitions and deterministic scoring rules (Appendix~\ref{appendix:codebook}); (ii)~the per-agent evaluation data, including every metric rating and its supporting justification; and (iii)~the structured prompts used for the LLM-assisted literature analysis. All artifacts are available at \url{https://anonymous.4open.science/r/FARSIGHT-266A/}. The repository link is anonymized for double-blind review, and we will release the artifacts under an open-source license upon acceptance.

\bibliographystyle{IEEEtran}
\bibliography{main}

\appendix

\begin{table*}[htbp]
\centering
\caption{Comparison between LLM-based analysis (\faLayerGroup) and human expert (\faUsers) review under the \frameworkShort{} framework.}
\label{tab:farsight-human-llm-metricside}
\footnotesize
\setlength{\tabcolsep}{2.5pt}
\begin{tabular}{l*{9}{cc}cc}
\toprule
\textbf{Agent System}
& \multicolumn{2}{c}{\textbf{R1}}
& \multicolumn{2}{c}{\textbf{R2}}
& \multicolumn{2}{c}{\textbf{R3}}
& \multicolumn{2}{c}{\textbf{R4}}
& \multicolumn{2}{c}{\textbf{S1}}
& \multicolumn{2}{c}{\textbf{S2.1}}
& \multicolumn{2}{c}{\textbf{S2.2}}
& \multicolumn{2}{c}{\textbf{S2.3}}
& \multicolumn{2}{c}{\textbf{S3}}
& \textbf{\#Diff} & \textbf{Ratio} \\
\cmidrule(lr){2-19}
& \faLayerGroup & \faUsers 
& \faLayerGroup & \faUsers 
& \faLayerGroup & \faUsers 
& \faLayerGroup & \faUsers 
& \faLayerGroup & \faUsers 
& \faLayerGroup & \faUsers 
& \faLayerGroup & \faUsers 
& \faLayerGroup & \faUsers 
& \faLayerGroup & \faUsers 
& & \\
\midrule

FinMem
& \ding{109} & \ding{117}
& \ding{117} & \ding{109}
& \ding{108} & \ding{117}
& \ding{117} & \ding{117}
& \ding{109} & \ding{109}
& \ding{109} & \ding{109}
& \ding{109} & \ding{117}
& \ding{109} & \ding{109}
& \ding{109} & \ding{109}
& 4 & 44.4\% \\

TradingAgents
& \ding{108} & \ding{117}
& \ding{108} & \ding{117}
& \ding{108} & \ding{117}
& \ding{108} & \ding{117}
& \ding{117} & \ding{117}
& \ding{117} & \ding{117}
& \ding{117} & \ding{117}
& \ding{117} & \ding{117}
& \ding{117} & \ding{109}
& 5 & 55.6\% \\

FinCon
& \ding{108} & \ding{109}
& \ding{108} & \ding{109}
& \ding{108} & \ding{109}
& \ding{108} & \ding{109}
& \ding{109} & \ding{109}
& \ding{117} & \ding{117}
& \ding{117} & \ding{117}
& \ding{117} & \ding{117}
& \ding{117} & \ding{117}
& 4 & 44.4\% \\

QuantAgent
& \ding{109} & \ding{117}
& \ding{109} & \ding{117}
& \ding{117} & \ding{117}
& \ding{108} & \ding{117}
& \ding{109} & \ding{117}
& \ding{117} & \ding{117}
& \ding{117} & \ding{117}
& \ding{109} & \ding{117}
& \ding{117} & \ding{109}
& 6 & 66.7\% \\

FLAG-Trader
& \ding{117} & \ding{109}
& \ding{108} & \ding{109}
& \ding{117} & \ding{109}
& \ding{117} & \ding{109}
& \ding{109} & \ding{117}
& \ding{117} & \ding{117}
& \ding{117} & \ding{117}
& \ding{117} & \ding{117}
& \ding{117} & \ding{117}
& 5 & 55.6\% \\
\midrule
\textbf{Total \#Diff / Ratio (per Metric)}
& \multicolumn{2}{c}{5 / 100\%}
& \multicolumn{2}{c}{5 / 100\%}
& \multicolumn{2}{c}{4 / 80\%}
& \multicolumn{2}{c}{4 / 80\%}
& \multicolumn{2}{c}{2 / 40\%}
& \multicolumn{2}{c}{0 / 0\%}
& \multicolumn{2}{c}{1 / 20\%}
& \multicolumn{2}{c}{1 / 20\%}
& \multicolumn{2}{c}{2 / 40\%}
& 24 & 53.3\% \\
\bottomrule
\end{tabular}

\vspace{3pt}
\ding{108} \textbf{(Fully Robust / Low Risk)} \quad
\ding{117} \textbf{(Partially Robust / Medium Risk)} \quad
\ding{109} \textbf{(Not Robust / High Risk)}
\end{table*}

\section{Metric Assignment Criteria}
\label{subsec:metric-assignment}

To maximize reproducibility, we adopt an \textit{evidence-based checklist} methodology for metric assignment.
For each metric, we define a set of \textit{evidence indicators}, concrete, binary (present/absent) features that evaluators identify in the agent's published paper, including described mechanisms, system diagrams, and reported experiments.
Three-level ratings (\ding{108}\,Fully Robust/Low Risk, \ding{117}\,Partial, \ding{109}\,Not Robust/High Risk) follow \textit{deterministic scoring rules} based on which indicators are satisfied, eliminating subjective judgment from the scoring step.
The complete indicator definitions and scoring rules for all nine evaluation metrics (R1--R4, S1, S2.1--S2.3, S3) are provided in the codebook (Appendix~\ref{appendix:codebook}, Table~\ref{tab:codebook_combined}). S2 is decomposed into three sub-metrics, S2.1 (prompt-injection), S2.2 (tool/API exploitation), and S2.3 (memory manipulation), each scored independently, yielding nine ratings per agent.

All 15 agents were independently scored by two evaluators. Evaluator~A is a domain expert with in-depth knowledge of financial LLM agent architectures (including hands-on experience reproducing and extending TradingAgents-class systems). Evaluator~B has a solid understanding of LLM security fundamentals and financial markets but is not a specialist in agent-level system design. Both evaluators applied the same codebook (Appendix~\ref{appendix:codebook}) without discussion during the scoring phase.
Across all 15 agents $\times$ 9 metrics (135 individual ratings), the two evaluators reached identical scores on 126 ratings, yielding a raw agreement rate of 93.3\% and a Cohen's $\kappa$ of 0.89, indicating near-perfect agreement~\cite{mchugh2012interrater}. The nine disagreements were concentrated in metrics R3, R4, and S2.2, where Evaluator~B occasionally rated an agent as \textit{Not Robust} (\ding{109}) while Evaluator~A assigned \textit{Partial} (\ding{117}). Root-cause analysis showed that these discrepancies stemmed from Evaluator~B's incomplete familiarity with certain agent-internal mechanisms, for example, implicit risk-adjustment loops in memory-based agents, or tool-permission scoping embedded in multi-agent communication protocols. Such mechanisms are not apparent from surface-level paper descriptions but are identifiable through deeper architectural reading. All disagreements were resolved through joint discussion, and final scores reflect the consensus ratings reported in Table~\ref{tab:combined-eval}.

\section{Codebook Used to Apply \frameworkShort{} Framework Metrics}
\label{appendix:codebook}

Following the approach of Akanda et al.~\cite{akanda2025sok}, we operationalize each \frameworkShort{} metric into an \textit{evidence-based checklist} of binary (present/absent) indicators, each anchored to concrete artifacts observable in a published paper (mechanism descriptions, system diagrams, or experimental results). The code-level audit (Appendix~\ref{appendix:code-audit}) applies the same indicators to source code.
Ratings are assigned by deterministic scoring rules, ensuring that two independent evaluators examining the same paper will check the same indicators and arrive at the same score.
Below we specify each indicator set and its scoring logic; Table~\ref{tab:codebook_combined} presents the full codebook in tabular form.

\vspace{3pt}
\noindent\textbf{Robustness Evidence Indicators.}

\noindent\textit{R1 (Crash Foresight).}\;
\textbf{E1}: Quantitative anomaly detection (e.g., VIX thresholds, statistical outlier tests, correlation-breakdown monitoring, PCA-based volatility clustering).\;
\textbf{E2}: Market-regime awareness via qualitative signals (e.g., sentiment profiling, macro-trend assessment, multi-source news synthesis).\;
\textbf{E3}: Automated threshold-based alert or trigger mechanism (programmatic, not dependent on LLM reasoning alone).\\
\textit{Rule}: \ding{108}\,=\,E1\,$\wedge$\,E3;\;\; \ding{117}\,=\,$\geq$1 of \{E1, E2, E3\} present but $\neg$(E1\,$\wedge$\,E3);\;\; \ding{109}\,=\,none.

\noindent\textit{R2 (Rapid Reaction).}\;
\textbf{E1}: Fast-path interrupt mechanism executable \textit{without} LLM inference (e.g., external rules engine, hardware-level circuit breaker, or safe-mode switch that fires before LLM dispatch).\;
\textbf{E2}: Automated order freeze or cancellation under detected shock (programmatic action that may still pass through the LLM-mediated reasoning chain).\;
\textbf{E3}: Risk-authority override protocol (risk agent can overrule other agents without multi-round debate).\\
\textit{Rule}: \ding{108}\,=\,E1 present;\;\; \ding{117}\,=\,$\geq$1 of \{E2, E3\} present but $\neg$E1;\;\; \ding{109}\,=\,none.

\noindent\textit{R3 (Loss Containment).}\;
\textbf{E1}: Quantitative stop-loss parameters defined (specific drawdown thresholds, position caps, or percentage limits).\;
\textbf{E2}: Risk-aware behavioral adjustment (exposure reduction, risk-mode switching, or loss-triggered strategy change).\;
\textbf{E3}: Automated enforcement (stop-loss executes programmatically, independent of LLM reasoning or human approval).\;
\textbf{E4}: Multi-layer or volatility-adjusted containment (position-level + portfolio-level, or dynamic calibration).\\
\textit{Rule}: \ding{108}\,=\,E1\,$\wedge$\,E3;\;\; \ding{117}\,=\,$\geq$1 of \{E1, E2, E3, E4\} present but $\neg$(E1\,$\wedge$\,E3);\;\; \ding{109}\,=\,none.

\noindent\textit{R4 (Recovery).}\;
\textbf{E1}: Crash-triggered recovery protocol (distinct recovery activated by tail events, not merely continuous learning).\;
\textbf{E2}: Structured post-mortem analysis (fault diagnosis, historical scenario comparison, or root-cause identification).\;
\textbf{E3}: Explicit parameter recalibration post-crash (reduced leverage, wider confidence thresholds, strategy downgrade).\;
\textbf{E4}: Experience replay or self-reflection mechanism that feeds past outcomes into future decisions.\\
\textit{Rule}: \ding{108}\,=\,E1\,$\wedge$\,(E2\,$\vee$\,E3\,$\vee$\,E4);\;\; \ding{117}\,=\,$\geq$1 of \{E1, E2, E3, E4\} present but $\neg$(E1\,$\wedge$\,(E2\,$\vee$\,E3\,$\vee$\,E4));\;\; \ding{109}\,=\,none.

\vspace{3pt}
\noindent\textbf{Security Evidence Indicators.}

\noindent\textit{S1 (Information-Source Attacks).}\;
\textbf{E1}: Source provenance tracking (source domain, timestamp, and authentication metadata are recorded).\;
\textbf{E2}: Cross-source corroboration required before trade execution ($\geq$2 independent sources must confirm a market-moving signal).\;
\textbf{E3}: Structured quantitative aggregation (voting, ranking, or policy learning) across independent analysis channels.\;
\textbf{E4}: Human curation or design-level scope restriction that limits raw external data exposure.\\
\textit{Rule}: \ding{108}\,=\,E1\,$\wedge$\,E2;\;\; \ding{117}\,=\,$\geq$1 of \{E1, E2, E3, E4\} present but $\neg$(E1\,$\wedge$\,E2);\;\; \ding{109}\,=\,none.

\smallskip
\noindent\textit{S2 Sub-metric Structure.}\;
S2 is decomposed into three independently scored sub-metrics, S2.1 (prompt-injection), S2.2 (tool/API exploitation), and S2.3 (memory or context manipulation). Each sub-metric uses three indicators with a uniform pattern: \textbf{E1} captures an \emph{active defensive mechanism} for that attack class, \textbf{E2} captures \emph{strict design-level} elimination of the corresponding attack surface, and \textbf{E3} captures \emph{partial design-level} restriction. The same scoring rule applies to all three sub-metrics.

\noindent\textit{S2.1 (Prompt-Injection and Instruction Override).}\;
\textbf{E1}: Active prompt-injection defense, system-prompt immutability with user/data-input separation, input sanitization or injection filtering, or multi-agent verification (debate, voting, role separation) that requires defeating multiple agents simultaneously.\;
\textbf{E2}: Strict design-level absence of open prompt channels. Only curated, authenticated, or research-environment inputs feed the agent; no end-user free-form prompts processed.\;
\textbf{E3}: Partial design-level restriction. Inputs are simulated, sandboxed, or otherwise constrained, but residual prompt-layer exposure remains.\\
\textit{Rule}: \ding{108}\,=\,E2;\;\; \ding{117}\,=\,$\geq$1 of \{E1, E3\} present but $\neg$E2;\;\; \ding{109}\,=\,none.

\noindent\textit{S2.2 (Tool-Use and API Exploitation).}\;
\textbf{E1}: Active tool/API protection, role-based access control (RBAC), capability scoping separating read from execution permissions, dual-authorization, rate limiting, or risk-management agents that gate trade execution.\;
\textbf{E2}: Strict design-level absence of live actuating tools. No trade-execution path exists, and the system operates as analysis-only or simulation-only.\;
\textbf{E3}: Partial design-level restriction. Live tools exist but are limited to read-only or sandboxed endpoints, with actuating tools off by default or available only under explicit human trigger.\\
\textit{Rule}: \ding{108}\,=\,E2;\;\; \ding{117}\,=\,$\geq$1 of \{E1, E3\} present but $\neg$E2;\;\; \ding{109}\,=\,none.

\noindent\textit{S2.3 (Memory or Context Manipulation).}\;
\textbf{E1}: Active memory protection, write-gating, quarantine memory, provenance-checked promotion, or feedback/risk-driven validation before observations enter persistent storage.\;
\textbf{E2}: Strict design-level absence of long-term memory. The agent has no persistent or self-evolving memory state.\;
\textbf{E3}: Partial design-level restriction. Memory is short-term, ephemeral across sessions, or otherwise limited in self-reinforcement.\\
\textit{Rule}: \ding{108}\,=\,E2;\;\; \ding{117}\,=\,$\geq$1 of \{E1, E3\} present but $\neg$E2;\;\; \ding{109}\,=\,none.

\noindent\textit{Note on S2.1--S2.3.}\; The \ding{108} criterion (E2) is intentionally reserved for advisory-only or offline-analysis systems that do not expose the corresponding surface at all; execution-capable agents with strong active defenses (E1) reach at most \ding{117} by construction. A \ding{117} on an S2 sub-metric therefore means that the surface is exposed, not that the agent is trivially exploitable.

\noindent\textit{S3 (Agent-as-Attacker).}\;
\textbf{E1}: Behavioral anomaly detection (monitors for sudden changes in order frequency, asset class, or trade volume).\;
\textbf{E2}: Agent isolation capability (can sandbox or disable a single compromised agent without system-wide shutdown).\;
\textbf{E3}: Inter-agent message validation (peer messages treated as untrusted inputs; policy engine validates before execution).\;
\textbf{E4}: Human hard-stop override channel that can immediately halt all agent actions.\;
\textbf{E5}: Strict design-level absence of autonomous market-action capability. The agent has no live execution path and operates as an analytical, advisory, or purely offline simulation-only framework. Multi-agent simulations with a shared, unisolated order book do not qualify.\;
\textbf{E6}: Partial design-level restriction of autonomous market-action capability. The agent can interface with markets but is not inherently autonomous (e.g., minimal live API exposure, requires explicit human triggers, or limited execution scope).\\
\textit{Rule}: \ding{108}\,=\,E5\,$\vee$\,(E1\,$\wedge$\,E2);\;\; \ding{117}\,=\,$\geq$1 of \{E1, E2, E3, E4, E6\} present but $\neg$E5\,$\wedge$\,$\neg$(E1\,$\wedge$\,E2);\;\; \ding{109}\,=\,none.

\begin{table*}[htbp]
\centering
\caption{Evidence-Based Codebook for \frameworkShort{} Robustness Metrics (security metrics in Table~\ref{tab:codebook_security}). Each metric is decomposed into binary evidence indicators. Ratings follow deterministic rules: \ding{108}\,=\,Fully Robust / Low Risk;\; \ding{117}\,=\,Partially Robust / Medium Risk;\; \ding{109}\,=\,Not Robust / High Risk.}
\label{tab:codebook_combined}
\renewcommand{\arraystretch}{1.05}
\footnotesize
\begin{tabular}{p{1.2cm} p{0.7cm} p{5.2cm} p{4cm} p{4cm}}
\toprule
\textbf{Metric} & \textbf{ID} & \textbf{Evidence Indicator (what evaluators look for)} & \textbf{Observable Artifact in Paper} & \textbf{Scoring Rule} \\
\midrule
\multicolumn{5}{l}{\textbf{Robustness Metrics}} \\
\midrule
\multirow{3}{*}{\makecell[l]{R1: Crash\\Foresight}}
  & E1 & Quantitative anomaly detection & VIX thresholds, statistical outlier tests, correlation monitoring, PCA clustering & \multirow{3}{*}{\makecell[l]{\ding{108} = E1 $\wedge$ E3\\ \ding{117} = $\geq$1 of \{E1,E2,E3\}\\\quad but $\neg$(E1$\wedge$E3)\\ \ding{109} = none}} \\
  & E2 & Market-regime awareness via qualitative signals & Sentiment profiling, macro-trend assessment, news synthesis & \\
  & E3 & Automated threshold-based trigger mechanism & Programmatic alerts not dependent on LLM alone & \\
\midrule
\multirow{3}{*}{\makecell[l]{R2: Rapid\\Reaction}}
  & E1 & Fast-path interrupt without LLM inference & External rules engine, hardware-level circuit breaker, or safe-mode switch firing before LLM dispatch & \multirow{3}{*}{\makecell[l]{\ding{108} = E1 present\\ \ding{117} = $\geq$1 of \{E2,E3\}\\\quad but $\neg$E1\\ \ding{109} = none}} \\
  & E2 & Automated order freeze/cancellation under shock & Programmatic cancel of resting orders (LLM-mediated allowed) & \\
  & E3 & Risk-authority override protocol & Risk agent overrules without debate & \\
\midrule
\multirow{4}{*}{\makecell[l]{R3: Loss\\Contain.}}
  & E1 & Quantitative stop-loss parameters defined & Drawdown thresholds, position caps, \% limits & \multirow{4}{*}{\makecell[l]{\ding{108} = E1 $\wedge$ E3\\ \ding{117} = $\geq$1 of \{E1..E4\}\\\quad but $\neg$(E1$\wedge$E3)\\ \ding{109} = none}} \\
  & E2 & Risk-aware behavioral adjustment & Exposure reduction, risk-mode switching & \\
  & E3 & Automated stop-loss enforcement & Executes without LLM/human approval & \\
  & E4 & Multi-layer or volatility-adjusted containment & Position + portfolio levels, or dynamic calibration & \\
\midrule
\multirow{4}{*}{\makecell[l]{R4:\\Recovery}}
  & E1 & Crash-triggered recovery protocol & Distinct from continuous learning; tail-event activated & \multirow{4}{*}{\makecell[l]{\ding{108} = E1 $\wedge$ (E2$\vee$E3$\vee$E4)\\ \ding{117} = $\geq$1 of \{E1..E4\}\\\quad but $\neg$(E1$\wedge$(E2$\vee$E3$\vee$E4))\\ \ding{109} = none}} \\
  & E2 & Structured post-mortem analysis & Fault diagnosis, scenario comparison & \\
  & E3 & Explicit parameter recalibration post-crash & Reduced leverage, wider thresholds & \\
  & E4 & Experience replay or self-reflection mechanism & Past outcomes feed future decisions & \\
\bottomrule
\end{tabular}
\end{table*}

\begin{table*}[htbp]
\centering
\caption{Evidence-based codebook for \frameworkShort{} security metrics (continued from Table~\ref{tab:codebook_combined}).}
\label{tab:codebook_security}
\renewcommand{\arraystretch}{1.05}
\footnotesize
\begin{tabular}{p{1.2cm} p{0.7cm} p{5.2cm} p{4cm} p{4cm}}
\toprule
\textbf{Metric} & \textbf{ID} & \textbf{Evidence Indicator (what evaluators look for)} & \textbf{Observable Artifact in Paper} & \textbf{Scoring Rule} \\
\midrule
\multicolumn{5}{l}{\textbf{Security Metrics}} \\
\midrule
\multirow{4}{*}{\makecell[l]{S1: Info-\\Source\\Attacks}}
  & E1 & Source provenance tracking & Domain, timestamp, authentication metadata & \multirow{4}{*}{\makecell[l]{\ding{108} = E1 $\wedge$ E2\\ \ding{117} = $\geq$1 of \{E1..E4\}\\\quad but $\neg$(E1$\wedge$E2)\\ \ding{109} = none}} \\
  & E2 & Cross-source corroboration before execution & $\geq$2 independent sources confirm signal & \\
  & E3 & Structured quantitative aggregation & Voting, ranking, or policy learning across channels & \\
  & E4 & Human curation or scope restriction & Curated input pipeline or restricted exposure & \\
\midrule
\multirow{3}{*}{\makecell[l]{S2.1:\\Prompt-\\Injection}}
  & E1 & Active prompt-injection defense & Immutable system prompt, input sanitization, multi-agent verification, role separation & \multirow{3}{*}{\makecell[l]{\ding{108} = E2\\ \ding{117} = $\geq$1 of \{E1,E3\}\\\quad but $\neg$E2\\ \ding{109} = none}} \\
  & E2 & Strict design-level scope restriction & No open prompt channels; curated/authenticated inputs only & \\
  & E3 & Partial design-level restriction & Simulated/sandboxed inputs with residual exposure & \\
\midrule
\multirow{3}{*}{\makecell[l]{S2.2:\\Tool/API\\Exploit.}}
  & E1 & Active tool/API protection & RBAC, capability scoping, dual-auth, rate limits, risk-gated execution & \multirow{3}{*}{\makecell[l]{\ding{108} = E2\\ \ding{117} = $\geq$1 of \{E1,E3\}\\\quad but $\neg$E2\\ \ding{109} = none}} \\
  & E2 & Strict design-level scope restriction & No live execution / actuating tools by design & \\
  & E3 & Partial design-level restriction & Read-only or sandboxed tools; actuating tools off by default & \\
\midrule
\multirow{3}{*}{\makecell[l]{S2.3:\\Memory/\\Context}}
  & E1 & Active memory protection & Write-gating, quarantine memory, provenance/risk-checked promotion & \multirow{3}{*}{\makecell[l]{\ding{108} = E2\\ \ding{117} = $\geq$1 of \{E1,E3\}\\\quad but $\neg$E2\\ \ding{109} = none}} \\
  & E2 & Strict design-level scope restriction & No persistent / self-evolving long-term memory & \\
  & E3 & Partial design-level restriction & Short-term/ephemeral memory only, no self-reinforcement & \\
\midrule
\multirow{6}{*}{\makecell[l]{S3: Agent\\as\\Attacker}}
  & E1 & Behavioral anomaly detection & Monitors order frequency, asset class, volume changes & \multirow{6}{*}{\makecell[l]{\ding{108} = E5 $\vee$ (E1$\wedge$E2)\\ \ding{117} = $\geq$1 of \{E1..E4,E6\}\\\quad but $\neg$E5 $\wedge$ $\neg$(E1$\wedge$E2)\\ \ding{109} = none}} \\
  & E2 & Agent isolation capability & Sandbox/disable one agent without system shutdown & \\
  & E3 & Inter-agent message validation & Peer messages untrusted; policy engine validates & \\
  & E4 & Human hard-stop override & Immediate halt of all agent actions & \\
  & E5 & Strict design-level absence of live execution & Analysis-only, advisory, or purely offline simulation (excludes multi-agent sims with a shared order book) & \\
  & E6 & Partial design-level restriction & Minimal live exposure or non-autonomous execution & \\
\bottomrule
\end{tabular}
\end{table*}

This evidence-based codebook provides \textit{structured, expert-applied} scoring rules: given the same published paper, two evaluators who identify the same indicators will arrive at the same rating. Consistent identification of the indicators themselves still requires domain expertise, as our LLM-application study (Appendix~\ref{appendix:llm_analysis}) confirms.
It also enables automated or hybrid evaluation workflows: as demonstrated in Appendix~\ref{appendix:llm_analysis}, we used large language models to assist in applying these indicators and to cross-validate human expert assessments.
Furthermore, the codebook is \textit{extensible}: new metrics require only defining additional indicator sets and scoring rules, and the modular design allows adaptation to other financial sectors (e.g., derivatives, cryptocurrency, international equities) or new risk dimensions (e.g., regulatory compliance, liquidity risk) without altering the existing evaluation logic.

\section{Per-Agent Metric Justifications}
\label{appendix:metric-justifications}

Below we provide justifications for partial (\ding{117}) and fully robust / low-risk (\ding{108}) scores in Table~\ref{tab:combined-eval}, following the codebook criteria in Appendix~\ref{appendix:codebook}. (Items not listed are not-robust / high-risk \ding{109} entries, for which no satisfied indicator was identified in the source paper.)

\noindent\textbf{R1 (Crash Foresight):}
\textit{FinMem}: sentiment-based risk profiling with adaptive memory switching satisfies E2, but no quantitative anomaly detection (no E1) and no programmatic threshold trigger (no E3).
\textit{Alpha-GPT 2.0}: human-in-the-loop alpha validation catches risky strategies offline (E2), but no continuous live surveillance (no E1, no E3).
\textit{QuantAgent}: dedicated risk monitoring agent provides qualitative regime awareness (E2), but lacks concrete quantitative metrics such as VIX tracking or correlation dislocation (no E1) and no programmatic threshold trigger (no E3).
\textit{TradingAgents}: Bull/Bear researcher agents assess market regime via semantic reasoning (E2), but rely on qualitative reasoning rather than statistical signals (no E1) and no threshold-based trigger (no E3).
\textit{TradingGroup}: multi-source news/sentiment synthesis (E2), but no threshold-based early warning (no E1, no E3).

\noindent\textbf{R2 (Rapid Reaction):}
\textit{QuantAgent}: risk-monitoring component triggers automated alerts under abnormal volatility (E2), but bounded by LLM inference latency, no LLM-independent fast path (no E1).
\textit{TradingAgents}: risk management team can override (E3), but acts through message-passing rather than algorithmic immediacy (no E1).
\textit{TradingGroup}: configurable stop-loss/take-profit thresholds trigger automated cancellation (E2), but static and user-defined without an LLM-independent fast path (no E1).

\noindent\textbf{R3 (Loss Containment):}
\textit{FinMem}: adaptive risk modes reduce exposure after performance declines (E2), but no defined drawdown thresholds (no E1) or programmatic liquidation triggers (no E3).
\textit{Alpha-GPT 2.0}: research-stage risk assessment (E2), but no execution-time enforcement (no E3).
\textit{QuantAgent}: position management functions within the risk agent (E2), but omits mathematical details on stop-loss parameters (no E1) and no programmatic enforcement (no E3).
\textit{TradingAgents}: coordinated loss containment via risk management team yields empirically smaller drawdowns (E2), but no transparent liquidation logic (no E3).
\textit{ContestTrade}: downranks poorly performing agents (E2), but adaptation occurs after losses materialize (no E1, no E3).
\textit{TradingGroup}: configurable stop-loss settings (E1), but without volatility-adjusted calibration (no E4) or fully autonomous programmatic enforcement (no E3).

\noindent\textbf{R4 (Recovery):}
\textit{FinMem}: self-evolution mechanism updates memory over time (E4), but no event-triggered recovery distinct from continuous learning (no E1).
\textit{FinAgent}: dual-level reflection framework feeds outcomes back into future decisions (E4), but no explicit crash-triggered reparameterization (no E1).
\textit{Alpha-GPT 2.0}: human researchers guide iterative improvements (E3), but no automated post-crash recovery protocol (no E1).
\textit{QuantAgent}: two-loop structure (inner optimization, outer testing) integrates outcomes (E4), but adaptation is slow and not crash-triggered (no E1).
\textit{TradingAgents}: continuous-learning loop (E4) without crash-specific recovery triggers (no E1).
\textit{ContestTrade}: evolutionary resilience promotes surviving agents (E4), but cannot handle simultaneous system-wide crashes (no E1).
\textit{TradingGroup}: self-reflection and continuous data synthesis (E4), but recovery is passive rather than crash-triggered (no E1).

\smallskip
\noindent\textbf{S1 (Information-Source Attacks):}
\textit{StockAgent}: simulated market environment limits raw external feed exposure (E4); no provenance tracking (no E1) or cross-source corroboration (no E2).
\textit{FinRobot}: structured aggregation across modular sub-agents (E3); no provenance or corroboration enforcement.
\textit{Alpha-GPT 2.0}: human-curated alpha-research inputs (E4); no provenance or cross-source corroboration before signal use.
\textit{QuantAgent}: structured fusion of technical features within the risk-monitoring agent (E3); no provenance or corroboration.
\textit{TradingAgents}: debate-and-synthesis aggregation across analyst/researcher teams (E3); no provenance or corroboration enforcement.
\textit{FLAG-Trader}: policy-learning aggregation over heterogeneous signals (E3); no provenance or corroboration.
\textit{ContestTrade}: real-time ranking and contest-based aggregation across parallel paths (E3); no provenance or corroboration.
\textit{TradingGroup}: collaborative aggregation across group communication channels (E3); no provenance or corroboration.

\smallskip
\noindent\textbf{S2.1 (Prompt-Injection \& Instruction Override):}
\textit{Alpha-GPT 2.0} (\ding{108}): operates on curated researcher inputs rather than open prompt channels, E2 (strict design-level absence of open prompt channels) satisfied.
\textit{MarketSenseAI, MarketSenseAI 2.0}: chain-of-thought analysis over curated SEC filings introduces multi-step structured reasoning (E1) and a partially constrained input pipeline (E3); no full design-level isolation (no E2).
\textit{StockAgent}: simulated event-driven inputs partially restrict prompt exposure (E3) plus role-decomposition across investor agents (E1); residual exposure remains (no E2).
\textit{FinCon, FinAgent, FinRobot, QuantAgent, TradExpert, TradingAgents, FLAG-Trader, ContestTrade, TradingGroup}: multi-agent verification, role separation, or chain-of-thought structure constrains direct prompt overrides (E1); none achieve full design-level absence of open prompt channels (no E2).

\smallskip
\noindent\textbf{S2.2 (Tool-Use \& API Exploitation):}
\textit{MarketSenseAI, MarketSenseAI 2.0} (\ding{108}): static document retrieval and analysis without live API-driven trade execution, E2 (strict design-level absence of actuating tools) satisfied.
\textit{Alpha-GPT 2.0} (\ding{108}): alpha-discovery research framework with no autonomous trade-execution path, E2 satisfied.
\textit{TradingGPT, FinMem, StockAgent}: limited tool exposure via single-agent or simulation context (E3); no full design-level execution isolation (no E2).
\textit{FinCon, FinAgent, FinRobot, QuantAgent, TradExpert, TradingAgents, FLAG-Trader, ContestTrade, TradingGroup}: risk-management gating, multi-agent veto, or capability scoping constrains execution (E1); actuating tools remain reachable (no E2).

\smallskip
\noindent\textbf{S2.3 (Memory or Context Manipulation):}
\textit{MarketSenseAI, MarketSenseAI 2.0} (\ding{108}): stateless analysis pipeline without persistent self-evolving memory, E2 satisfied.
\textit{TradExpert} (\ding{108}): mixture-of-experts produces predictions without persistent self-evolving memory state, E2 satisfied.
\textit{FinCon, FinAgent, FinRobot, Alpha-GPT 2.0, QuantAgent, TradingAgents, FLAG-Trader, ContestTrade}: feedback-driven, validated, or risk-gated memory updates (E1); memory remains long-term and modifiable (no E2).
\textit{StockAgent}: simulated environment limits memory poisoning surface (E3); long-term memory persists (no E2).

\smallskip
\noindent\textbf{S3 (Agent-as-Attacker):}
\textit{FinRobot} (\ding{108}): composable analytical platform whose design intentionally excludes any live execution or external trading connection, E5 (strict design-level absence of autonomous market-action capability) satisfied.
\textit{Alpha-GPT 2.0} (\ding{108}): alpha-discovery framework with no autonomous trade execution, E5 satisfied.
\textit{MarketSenseAI, MarketSenseAI 2.0}: minimal live API exposure with non-autonomous execution, E6 (partial design-level restriction) satisfied.
\textit{FinCon, FinAgent, TradExpert, FLAG-Trader, ContestTrade}: limited live capability with risk-management gating, contest filtering, or dual-control structure, E6 satisfied.

\section{Code-Level Audit Details}
\label{appendix:code-audit}

\begin{table*}[t]
\centering
\caption{Repositories and pinned commits of the code-level audit. Each repository was audited at the listed commit in three independent runs.}
\label{tab:audit-commits}
\small
\begin{tabular}{l l l l}
\toprule
\textbf{Scheme} & \textbf{Repository} & \textbf{Commit} & \textbf{Date} \\
\midrule
FinMem & \texttt{pipiku915/FinMem-LLM-StockTrading} & \texttt{be814aa} & 2024-08-17 \\
StockAgent & \texttt{MingyuJ666/Stockagent} & \texttt{e2a9c05} & 2026-06-16 \\
FinAgent & \texttt{DVampire/FinAgent} & \texttt{17248a0} & 2024-08-31 \\
FinRobot & \texttt{AI4Finance-Foundation/FinRobot} & \texttt{2717499} & 2026-09-28 \\
QuantAgent & \texttt{Y-Research-SBU/QuantHarness} & \texttt{2e64c7b} & 2026-08-18 \\
TradingAgents & \texttt{TauricResearch/TradingAgents} & \texttt{1394a3f} & 2026-10-03 \\
ContestTrade & \texttt{FinStep-AI/ContestTrade} & \texttt{22432f9} & 2025-12-22 \\
\bottomrule
\end{tabular}
\end{table*}

\noindent\textbf{Scope.} We count a release as official only if the scheme's paper or an author's page links to it. Seven schemes qualify: FinMem, StockAgent, FinAgent (linked from the first author's homepage), FinRobot, QuantAgent (since renamed QuantHarness), TradingAgents, and ContestTrade. FinCon links to a repository that contains only a README and a license. TradingGPT, MarketSenseAI, MarketSenseAI~2.0, Alpha-GPT~2.0, TradExpert, FLAG-Trader, and TradingGroup have no official release. Table~\ref{tab:audit-commits} lists the pinned commits. Two releases differ from the system in the paper. FinRobot's repository now holds several generations of the platform, and we audited the original package that the paper describes. TradingAgents published code only after its last arXiv version, so its audit reflects the released implementation rather than the system in the paper.

\noindent\textbf{Procedure.} The audit is static: we read the source at the pinned commit and did not run it, install it, or call any external service. For each repository, Claude Sonnet~5.5 received the codebook (Appendix~\ref{appendix:codebook}), our paper-level ratings and their justifications (Appendix~\ref{appendix:metric-justifications}), and a fixed protocol. It labeled each of the \auditIndPerScheme{} evidence indicators as present, partial, or absent with file-and-line evidence, and it answered five factual questions: whether the code can submit orders, whether memory persists across runs, which data sources are wired in, whether any risk limit is enforced outside the LLM, and which described mechanisms are missing. Each repository was audited in \auditRuns{} runs, each in a fresh clone and without access to the other runs. A script then re-derives the nine ratings from the indicator labels with the codebook rules, so the rating step itself is deterministic. A partial indicator supports a partial rating but never satisfies a requirement for a full rating. Under a stricter convention that treats partial indicators as absent, \auditMatchStrict{} of \auditRatings{} code-level ratings match the paper-level ones, compared with \auditMatch{} under the main convention.

\noindent\textbf{Pattern matching and claim verification.} Independently of the LLM runs, a script searches each release for five pattern families: order-submission and broker APIs, dynamic code execution, persistence of memory or state, risk limits, and input validation. The order-submission family has no match in any of the seven releases, which agrees with all \auditRuns{} runs. A separate verification pass, run with Claude Opus on local clones at the pinned commits and independent of the three auditor runs, checked each of the 14 code-level claims that the main text relies on against the cited source lines and confirmed all of them. The same pass ruled on each of the \auditDisagree{} cross-run disagreements by reading the code, and every ruling agreed with the two-of-three majority.

\noindent\textbf{Per-scheme findings.}
\textit{FinMem} is an offline, single-agent backtest whose position is a share counter. Its memory layers and agent state are pickled to disk and reloaded from checkpoints with no validation before writes, and error strings from failed output validation can be stored as reflections. The risk-averse branch of its self-adaptive risk character is commented out in the released prompt.
\textit{StockAgent} is a pure simulation in which all agents trade through one shared, unisolated order book with no price band. The last matched price becomes the market price, so a single agent can set it. Each agent's free-text forum post reaches every other agent's next-day prompt without filtering, which is the S3 propagation path in our threat model.
\textit{FinAgent} is a daily-bar backtest with no order path. Its memory is a FAISS index with a pickled key map (\nolinkurl{finagent/memory/faiss.py}), written without validation and reloadable from disk, and three data downloaders print the market-data API key to standard output.
\textit{FinRobot}'s paper-era package has no order path, and its backtests use a simulated broker. Its default assistants, however, execute model-written code on the host without a sandbox or human approval (\nolinkurl{finrobot/agents/workflow.py}), and the SEC filing tool returns the exception text, including the token-bearing request URL, to the model when a download fails (\nolinkurl{finrobot/data_source/sec_utils.py}). Its Chroma retrieval stores persist across runs in fixed directories.
\textit{QuantAgent}'s release contains Indicator, Pattern, and Trend agents and one decision node, but no risk agent, no stop-loss, and no memory. It analyzes user-selected candles, from one minute to one month, on demand. Its local web interface also renders request parameters without escaping and serves files without path checks, which are web-application issues outside the codebook.
\textit{TradingAgents} returns a five-tier rating and has no order or simulated-exchange path, although its README describes one. A persistent decision log, added after the paper, stores model-written text verbatim and splits entries on a fixed delimiter (\nolinkurl{tradingagents/memory/log.py}), so model output that reproduces the delimiter can forge a past entry. A stop-loss field is filled in by the model but never enforced.
\textit{ContestTrade}'s contest mechanism, which gives the system its name, is not wired into the released pipeline, which passes research signals straight to the output. Its default stock-selection tool for the Chinese market executes model-written Python with \texttt{exec} in the host process (\nolinkurl{contest_trade/tools/stock_selector_akshare.py}).

\noindent\textbf{Threats to validity.} Static reading can miss behavior that appears only at run time, and an LLM auditor can overlook code. Three independent runs, pattern matching, and manual verification reduce this risk but do not remove it. Releases evolve, so our results describe the pinned commits rather than later versions. The code-level ratings also depend on how partial indicators are counted, which is why we report both conventions. The auditor saw our paper-level ratings and their justifications, which helped it locate the mechanisms that the papers describe but may anchor its labels toward those ratings. Two observations limit this effect: the final code-level ratings still differ from the paper-level ones in \auditCodeBetter{} cases upward and \auditCodeWorse{} downward, and the pattern scan, which never saw the ratings, agrees with every run that no release can submit orders.

\section{Distinct Characteristics of Financial LLM Agents}
\label{appendix:distinct-characteristics}
As illustrated in Section \ref{subsec:trading-architectures} Figure~\ref{fig:attack_demos}, financial LLM agents possess attack characteristics distinct from those in domains like healthcare, law, or education.  
Their exposure to real-time market data, dependence on heterogeneous signals, and high-frequency autonomy make them highly sensitive to latency, trust, and reflexivity attacks.  
We summarize five defining traits below.

\noindent\textbf{Instant.}  
This trait describes the attack window, not the speed of the agent. Markets reprice within milliseconds, so an attacker can create a transient price dislocation and exit before human or automated risk systems respond. An agent that deliberates for minutes (Table~\ref{tab:latency-comparison}) widens this window rather than closing it, and such monetizable micro-windows have no counterpart in other domains.

\noindent\textbf{Inconspicuous.}  
Tiny quote or tick manipulations appear legitimate, prompting agents to alter trading behavior while users notice no anomaly.  
Because financial agents act on minute price deltas, even subtle perturbations can trigger hidden but impactful responses.

\noindent\textbf{Unverified.}  
Inputs such as fake news, spoofed APIs, or forged order messages are often processed without provenance checks due to speed-first system design.  
This lack of validation allows attackers to issue unauthorized actions or trigger auto-trades with minimal friction.

\noindent\textbf{Adaptive.}  
By coordinating multi-source inputs (news, sentiment, technical feeds), attackers dynamically tailor strategies to exploit signal inconsistencies.  
The agents’ heterogeneous integration logic grants adversaries flexibility unmatched in more structured data domains.

\noindent\textbf{Cascading.}
Minor perturbations, like brief rumors or small sell orders, can cascade through interconnected trading agents, causing flash crashes or liquidity shocks.  
Because markets are reflexive and interdependent, one agent’s output becomes another’s trigger, amplifying small manipulations into systemic events.

\noindent\textbf{Evidence for each trait.}
\textit{Instant}: markets reprice faster than the agents decide (Table~\ref{tab:latency-comparison}), and 12 of the 15 schemes receive \ding{109} on R2 (Table~\ref{tab:combined-eval}).
\textit{Inconspicuous}: the S2.1 and S2.3 rows of Table~\ref{tab:generic-vs-finance} show that a small numeric nudge or a slowly compounding memory bias suffices, and the resulting losses resemble ordinary market noise.
\textit{Unverified}: no scheme earns full credit on S1, and 7 of the 15 receive \ding{109} (Table~\ref{tab:combined-eval}).
\textit{Adaptive}: 13 of the 15 schemes combine three or more input types (Table~\ref{tab:llm-agents-full}), which gives an attacker several channels to choose from.
\textit{Cascading}: 6 schemes receive \ding{109} on S3, and the S3 row of Table~\ref{tab:generic-vs-finance} traces how harm reaches third parties through prices and herding.

\section{Comparison of Investment Schemes and Stock-Trading Agents}
\label{appendix:investment-schemes}

Passive-income schemes operate with minimal intervention: once capital is invested, returns are generated automatically without active management. Fixed-income assets such as \textit{bonds} offer the highest stability and predictable yields, with risk limited mainly to large-scale defaults or market collapse. 
Similarly, \textit{mutual funds}, \textit{index funds}, and \textit{ETFs} embody the passive philosophy, requiring little investor action. They exhibit strong robustness and security due to limited market interaction and low exposure to real-time trading risks, though their annualized returns remain moderate by design.  
In contrast, \textit{LLM-based stock-trading agents} actively engage with live markets to pursue higher returns, but this introduces latency, prediction errors, and exposure to adversarial manipulation.  
Passive schemes thus provide stable, secure, and steady returns, while active stock agents offer higher potential gains at significantly greater risk, requiring investors to balance stability against growth ambition.

\section{LLM Literature Analysis Implementation}  
\label{appendix:llm_analysis}  
\noindent\textbf{Agreement with human experts.} We evaluated the utility of LLM-based analysis under the \frameworkShort{} framework to assess how effectively it mirrors human expert judgment. Applying ChatGPT Deep Research with predefined metrics, definitions, and scope (see Appendix~\ref{appendix:llm_analysis} for additional details) to a 5-agent sub-sample (FinMem, TradingAgents, FinCon, QuantAgent, FLAG-Trader), we found that the model disagreed with human expert evaluations on 24 of 45 metric cells (a 53.3\% disagreement rate, i.e., 46.7\% agreement), while still identifying similar key financial LLM agents and repeatedly flagging the same recurring weaknesses. While its reasoning depth remains more limited than that of human experts, the approach offers high scalability, enabling rapid and consistent framework-guided evaluation as the literature continues to expand.

The disagreements concentrate on robustness. The model disagreed with the human ratings on all five schemes for R1 and R2, on four of five for R3 and R4, and on at most two of five for each security metric (Table~\ref{tab:farsight-human-llm-metricside}). A likely reason is that the security metrics build on attack classes with an established literature, such as prompt injection, tool misuse, and memory poisoning, whereas the robustness metrics ask for trading-specific controls, such as fast-path interrupts and crash-triggered recovery, that the papers seldom name. Relying solely on black-box LLM assessment is therefore least reliable on exactly the metrics that are specific to trading.

We conducted an automated literature analysis using ChatGPT Deep Research through a structured prompt derived from our \frameworkShort{} framework. The process included supplying the framework definitions for robustness (R1--R4) and security (S1--S3) metrics, restricting the search to academic papers published between 2023 and 2026 that focused on ``financial LLM agents'' or ``LLM trading systems,'' and requesting structured summaries highlighting robustness gaps, attack surfaces, and mitigation strategies. The LLM outputs were then compared with our manually reviewed dataset to assess alignment accuracy, as shown in Table~\ref{tab:farsight-human-llm-metricside}.  
In total, the LLM identified eight candidate papers labeled as ``financial LLM agents,'' but only five of them (\textit{FinMem, FinCon, QuantAgent, TradingAgents, and FLAG-Trader}) were genuinely relevant. The remaining three were benchmark, survey, or simulation studies without real agentic designs. While the model produced strong summaries and coherent overviews, its detailed framework-based analysis was unreliable: its metric-level ratings disagreed with the human ratings on 24 of 45 cells (53.3\%). We attribute this to the model’s inability to precisely locate and interpret framework-specific evidence within long, multi-section papers. In particular, the context window limitation caused the LLM to miss fine-grained details while maintaining fluent but shallow summaries. This observation reinforces that while LLMs are promising for rapid literature summarization, their analytical accuracy in structured scientific evaluation remains substantially lower than that of human expert review. Table~\ref{tab:farsight-human-llm-metricside} shows that these disagreements fall mostly on the robustness metrics.

\begin{figure}[htbp]
    \centering
    \includegraphics[width=0.9\linewidth]{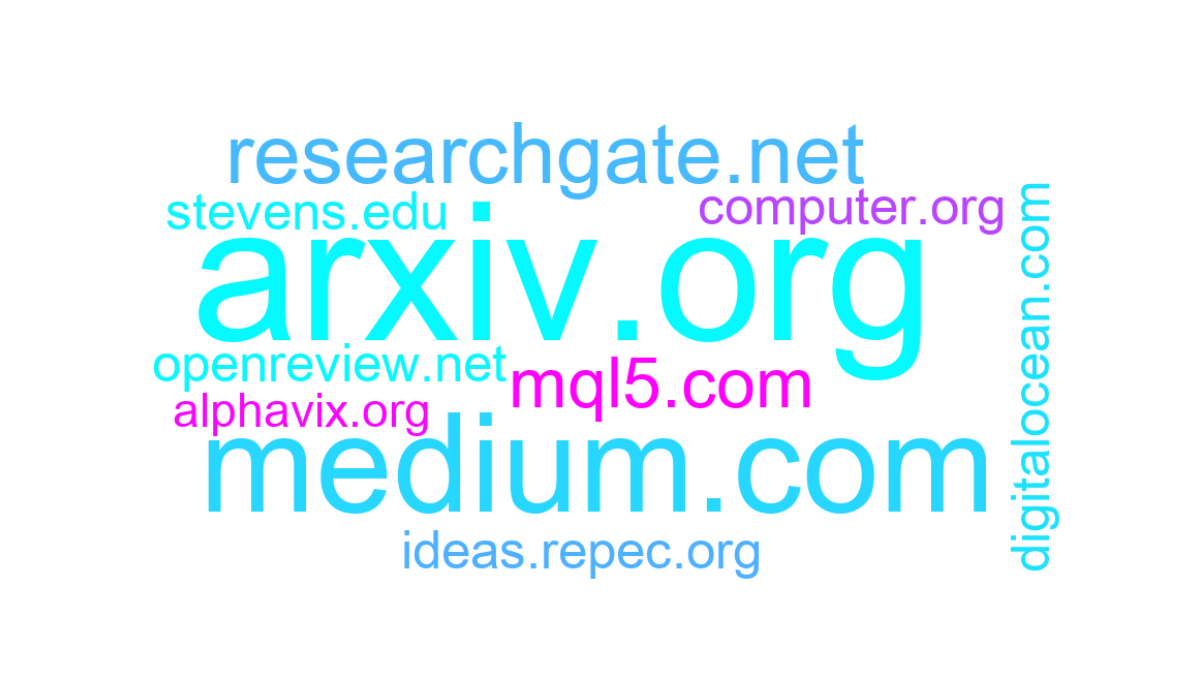}
    \caption{Source frequency distribution from ChatGPT Deep Research.}
    \label{fig:wordcloud-sources}
\end{figure}

We also examined the reasoning process and retrieved sources of the ChatGPT Deep Research system to better understand its literature analysis behavior. 
As shown in Figure~\ref{fig:wordcloud-sources}, the model queried a diverse set of sources, including \textit{arXiv}, \textit{ResearchGate}, \textit{RePEc}, and \textit{OpenReview}, but also accessed several non-academic domains such as \textit{Medium}, \textit{mql5.com}, and \textit{DigitalOcean}. 
Although the model could retrieve and summarize relevant papers, it frequently encountered \textit{reCAPTCHA} and rate-limit restrictions that reduced its search coverage compared to human experts. 
Additionally, it was easily distracted by unrelated technical blogs and community sites, resulting in lower focus and precision. 
This analysis demonstrates that while LLM-based search can efficiently explore new research areas, its reliability and selectivity remain inferior to those of human-guided exploration.
Overall, as shown in Table~\ref{tab:llm-capability}, our experiments indicate that while LLM-based literature reviews are effective for fast scanning and abstract summarization, they remain limited in deep reasoning, detailed analysis, and evidence-level verification.
\begin{table}[htbp]
\centering
\caption{Observed strengths and weaknesses of LLM-based review.}
\label{tab:llm-capability}
\footnotesize
\begin{tabular}{l l}
\toprule
\textbf{Strengths} & \textbf{Weaknesses} \\
\midrule
Fast scanning & Deep reasoning \\
Abstract summarization & Detailed analysis \\
Pattern discovery & Evidence verification \\
\bottomrule
\end{tabular}
\end{table}

\section{Domain Risk Profiles}
\label{appendix:domain-profiles}

This appendix gives the evidence for each cell of Figure~\ref{fig:domain-profiles}. Each property is rated from the perspective of a compromised or misled agent: \textit{high} means that the property raises the cost of a compromise in that domain. \textit{Irreversibility} asks whether the agent's actions can be undone. \textit{Adversary-authored inputs} asks whether outsiders routinely write what the agent reads. \textit{Speed of harm} compares the time from action to harm with the time a human reviewer needs. \textit{Coupling to other actors} asks whether the agent's outputs become other actors' inputs. \textit{Stealth of harm} asks whether harmful outcomes resemble normal variation. \textit{Reliance on in-model safety} asks whether any control that the agent cannot argue with sits between the agent and its irreversible actions.

\noindent\textbf{Trading (this paper).} \textit{Irreversibility} (high): An executed order settles before any human review, and losses scale with size and leverage~\cite{karppi2015hackcrash,kirilenko2017flash}. \textit{Adversary-authored inputs} (high): Anyone can write the news and social media that the agents read, and a forged headline moved the market in 2013~\cite{karppi2015hackcrash}. \textit{Speed of harm} (high): Markets reprice in milliseconds, while TradingAgents takes over 240 s per decision~\cite{kirilenko2017flash}. \textit{Coupling to other actors} (high): Every order changes the price that other agents, bots, and humans read (S3)~\cite{xia2026agentictradingllmagents}. \textit{Stealth of harm} (high): Losses from a slow memory bias or a numeric nudge look like ordinary market noise. \textit{Reliance on in-model safety} (high): No scheme earns full credit on R2 or R3, and no release enforces a loss limit outside the model.

\noindent\textbf{Healthcare.} \textit{Irreversibility} (high): An administered drug or a missed diagnosis cannot be undone. \textit{Adversary-authored inputs} (medium): Patients supply messages and images that can carry hidden instructions, and training data can be poisoned~\cite{clusmann2025prompt,alber2025medical}. \textit{Speed of harm} (medium): Orders take effect within minutes to hours, so human review is possible but compressed. \textit{Coupling to other actors} (medium): Agents write into a shared record that clinicians and other tools read and copy forward~\cite{tierney2024ambient}. \textit{Stealth of harm} (high): Poisoned medical models match clean ones on standard benchmarks, and clinical errors resemble the course of a disease~\cite{alber2025medical}. \textit{Reliance on in-model safety} (medium): Order-entry hard stops and pharmacist verification sit outside the agent, but they depend on attentive humans~\cite{kaushal2003effects,goddard2012automationbias}.

\noindent\textbf{Law.} \textit{Irreversibility} (medium): Filings can be corrected, but sanctions and missed deadlines stick~\cite{sdny2023mata}. \textit{Adversary-authored inputs} (high): The opposing party produces many of the documents the agent reads, and hidden prompts already appear in submitted manuscripts~\cite{lin2026hidden}. \textit{Speed of harm} (low): Court deadlines run in days, which leaves time for review. \textit{Coupling to other actors} (medium): Briefs, clause libraries, and citations are reused, so a fabricated case can be copied into later filings~\cite{magesh2024hallucination}. \textit{Stealth of harm} (medium): A fabricated citation is detectable when checked, yet commercial tools still produce them on 17--33\% of queries~\cite{magesh2024hallucination}. \textit{Reliance on in-model safety} (medium): The attorney's signature and duty to verify sit outside the model but rest on human diligence~\cite{sdny2023mata}.

\noindent\textbf{Software engineering.} \textit{Irreversibility} (medium): Deploys roll back, but deleted data and leaked credentials do not. \textit{Adversary-authored inputs} (high): Outsiders write the issues, pull requests, dependencies, and tool descriptions that coding agents read~\cite{hou2026context,halloran2026mcp}. \textit{Speed of harm} (medium): Automated pipelines ship changes within minutes. \textit{Coupling to other actors} (high): Other developers and their agents pull the packages and repositories that one agent writes. \textit{Stealth of harm} (high): A subtle backdoor reads like ordinary code. \textit{Reliance on in-model safety} (high): Coding agents often run with host privileges, and two audited trading releases execute model-written code on the host~\cite{guo2024redcode}.

\noindent\textbf{Crypto and DeFi.} \textit{Irreversibility} (high): On-chain transfers are final~\cite{patlan2025real}. \textit{Adversary-authored inputs} (high): Social posts and stored context can steer agents into unauthorized transfers~\cite{patlan2025real}. \textit{Speed of harm} (high): Blocks confirm within seconds, and front-running bots act within one block~\cite{daian2020flash}. \textit{Coupling to other actors} (high): Composable contracts and a shared mempool let one agent's transaction trigger others~\cite{daian2020flash}. \textit{Stealth of harm} (medium): The ledger is public, but counterparties are pseudonymous. \textit{Reliance on in-model safety} (high): Few controls sit between an agent's signing key and an immutable contract~\cite{patlan2025real}.

\noindent\textbf{Payments and commerce.} \textit{Irreversibility} (medium): Card payments can be reversed through chargebacks, but many transfers cannot~\cite{mao2026sokcommerce}. \textit{Adversary-authored inputs} (high): Counterparties write the merchant pages and listings that the agent reads~\cite{mao2026sokcommerce}. \textit{Speed of harm} (medium): Payments clear in seconds but settle later, which leaves a short window for review. \textit{Coupling to other actors} (medium): Agents negotiate with other agents over shared marketplaces~\cite{mao2026sokcommerce}. \textit{Stealth of harm} (medium): Fraud monitoring flags many abnormal payments. \textit{Reliance on in-model safety} (low): Transaction limits, authentication, and fraud scoring wrap every payer, human or agent.

\noindent\textbf{Scientific labs.} \textit{Irreversibility} (high): Synthesized compounds and physical hazards cannot be undone~\cite{boiko2023autonomous,mbran2024augmenting}. \textit{Adversary-authored inputs} (medium): Protocols and literature can be poisoned, but most inputs are curated~\cite{tang2025risks}. \textit{Speed of harm} (medium): Automated experiments run for minutes to hours~\cite{boiko2023autonomous}. \textit{Coupling to other actors} (low): Effects stay mostly within one lab until results are published. \textit{Stealth of harm} (medium): Failed or unsafe runs are usually visible to lab staff. \textit{Reliance on in-model safety} (medium): Lab safety protocols exist, but automated labs can run steps without a person at the bench~\cite{tang2025risks}.

\section{LLM Usage Considerations}

\noindent \textbf{Originality.}
The conceptual framework, the codebook, the paper-level ratings, and the conclusions are the authors' own. We used LLMs as tools in the ways listed below, and we manually verified that the collected links and references were legitimate and not fabricated.

\noindent \textbf{Transparency.}
We disclose all uses of LLMs in this work.
(1)~\textit{Literature analysis}: We used ChatGPT Deep Research to conduct automated literature analysis as described in Appendix~\ref{appendix:llm_analysis}, with results compared against human expert reviews.
(2)~\textit{Reference discovery}: ChatGPT was used to assist in discovering relevant online references (e.g., blog posts, industry reports). All discovered references were manually verified for legitimacy and relevance before inclusion.
(3)~\textit{Code audit}: Claude Sonnet~5.5 labeled codebook indicators in the released code of the open-source schemes, as described in Appendix~\ref{appendix:code-audit}. A script re-derives every rating from these labels, and a separate Claude Opus pass checked the cross-run disagreements and every code-level claim used in the main text.
(4)~\textit{Writing assistance}: Claude (Anthropic) drafted portions of the text under the authors' direction, including the related-work comparison, the code-audit sections, and Sections~\ref{sec:beyond}--\ref{sec:open} with their figures and tables. LLMs were also used for grammar checking and polishing. The authors take full responsibility for all content.
During this process, we identified several practical challenges, including reproducibility issues caused by closed-source models and inconsistent responses under identical prompts. A detailed discussion of these limitations and observations can be found in Appendix~\ref{appendix:llm_analysis}.

\noindent \textbf{Responsibility.}  
Our use of LLMs aims to advance community understanding of their role in SoK research, especially in analyzing financial and security implications of agentic LLM systems. We selected the GPT-based deep research tool for its performance and its stable access interface. All usage was limited to controlled academic evaluation, minimizing the number of queries and computational cost. No proprietary or sensitive data was used.

\end{document}